\documentclass[11pt]{article} 
\usepackage{microtype,ctable,subcaption,colortbl}
\DisableLigatures{encoding=*,family=*}
\usepackage{amsmath,amssymb,xfrac,framed,verbatim}
\usepackage{mathrsfs,accents,bbm,bm,amsthm}
\usepackage{slashed,mathtools}
\usepackage{graphicx,enumerate} 
\usepackage[unicode]{hyperref} 
\hypersetup{bookmarksnumbered=true,bookmarksopen=true,
  bookmarksopenlevel=2,breaklinks=true,citecolor=blue,
  colorlinks=true,linkcolor=red,linktoc=page,pdfborder={0 0 0},
  pdfkeywords={Notes, }{Spiked Instantons},pdfstartview=FitH,
  pdfauthor={NSP},pdfsubject={Spiked instantons},
  pdftitle={Spiked instantons},plainpages=false,unicode=true,urlcolor=blue}

\usepackage{makerobust}

\usepackage{geometry} 
\usepackage{fancyhdr} 
\usepackage{parskip} 
\usepackage{float}
\floatstyle{plaintop}
\restylefloat{table}
\usepackage[tableposition=bottom,justification=centering]{caption}

\makeatletter
\let\save@mathaccent\mathaccent
\newcommand*\if@single[3]{%
  \setbox0\hbox{${\mathaccent"0362{#1}}^H$}%
  \setbox2\hbox{${\mathaccent"0362{\kern0pt#1}}^H$}%
  \ifdim\ht0=\ht2 #3\else #2\fi
  }
\newcommand*\rel@kern[1]{\kern#1\dimexpr\macc@kerna}
\newcommand*\widebar[1]{\@ifnextchar^{{\wide@bar{#1}{0}}}{\wide@bar{#1}{1}}}
\newcommand*\wide@bar[2]{\if@single{#1}{\wide@bar@{#1}{#2}{1}}{\wide@bar@{#1}{#2}{2}}}
\newcommand*\wide@bar@[3]{%
  \begingroup
  \def\mathaccent##1##2{%
    \let\mathaccent\save@mathaccent
    \if#32 \let\macc@nucleus\first@char \fi
    \setbox\z@\hbox{$\macc@style{\macc@nucleus}_{}$}%
    \setbox\tw@\hbox{$\macc@style{\macc@nucleus}{}_{}$}%
    \dimen@\wd\tw@
    \advance\dimen@-\wd\z@
    \divide\dimen@ 3
    \@tempdima\wd\tw@
    \advance\@tempdima-\scriptspace
    \divide\@tempdima 10
    \advance\dimen@-\@tempdima
    \ifdim\dimen@>\z@ \dimen@0pt\fi
    \rel@kern{0.6}\kern-\dimen@
    \if#31
      \overline{\rel@kern{-0.6}\kern\dimen@\macc@nucleus\rel@kern{0.4}\kern\dimen@}%
      \advance\dimen@0.4\dimexpr\macc@kerna
      \let\final@kern#2%
      \ifdim\dimen@<\z@ \let\final@kern1\fi
      \if\final@kern1 \kern-\dimen@\fi
    \else
      \overline{\rel@kern{-0.6}\kern\dimen@#1}%
    \fi
  }%
  \macc@depth\@ne
  \let\math@bgroup\@empty \let\math@egroup\macc@set@skewchar
  \mathsurround\z@ \frozen@everymath{\mathgroup\macc@group\relax}%
  \macc@set@skewchar\relax
  \let\mathaccentV\macc@nested@a
  \if#31
    \macc@nested@a\relax111{#1}%
  \else
    \def\gobble@till@marker##1\endmarker{}%
    \futurelet\first@char\gobble@till@marker#1\endmarker
    \ifcat\noexpand\first@char A\else
      \def\first@char{}%
    \fi
    \macc@nested@a\relax111{\first@char}%
  \fi
  \endgroup
}
\makeatother

\makeatletter
\g@addto@macro\bfseries{\boldmath}
\makeatother

\newcommand{\w}[1]{\widebar{#1}} 
\newcommand{\bvert}{{\pmb{\rvert}}} 

\newcommand\Int{\mathbf{R}^{1,1}} 
\newcommand{\four}{\underline{\textbf 4}}
\newcommand{\six}{\underline{\textbf 6}}
\newcommand{\ud}{\text{d}}
\newcommand{\uD}{\mathrm{D}}
\newcommand{\uQ}{\mathrm{Q}}
\newcommand{\tr}{\text{Tr}}
\newcommand{\n}[1]{\mathcal{N}={(#1)}}

\newcommand{\cycm}{M}

\newcommand{\dt}[1]{\accentset{\mbox{\large\bfseries .}}{#1}}
\newcommand{\ndt}[1]{\accentset{\mbox{\phantom{\large\bfseries .}}}{#1}}

\newcommand{\rp}[1]{}
\newcommand{\pp}{{+\!\!\!+}}
\newcommand{\nmo}[1]{#1}

\newcommand\pmat[1] {\begin{pmatrix}#1\end{pmatrix}}
\newcommand{\wt}[1]{\widetilde{#1}}
\newcommand{\mc}[1]{\mathcal{#1}}
\newcommand{\mbb}[1]{\mathbf{#1}}
\newcommand{\mbbm}[1]{\mathbbm{#1}}
\newcommand{\tl}{\tilde}
\newcommand{\tp}[2]{\texorpdfstring{#1}{#2}}

\MakeRobustCommand\widebar
\MakeRobustCommand\dt
\MakeRobustCommand\ndt

\numberwithin{equation}{section} 
\makeatletter
\let\@fnsymbol\@arabic
\makeatother

\title{\textsc{Spiked instantons from intersecting D-branes}}
\author{Nikita Nekrasov \footnote{Simons Center for Geometry and
    Physics, \newline
    C.~N.~Yang Institute of Theoretical Physics, Stony Brook
    University, Stony Brook, NY 11794-3636.
    \newline e-mail: nikitastring@gmail.com} \and Naveen S. Prabhakar
  \footnote{C.~N.~Yang Institute of Theoretical Physics, Stony Brook
    University, Stony Brook, NY 11794-3840.\newline e-mail: naveen.s.prabhakar@gmail.com}}
\date{}

\begin{document}
\maketitle

\begin{abstract}
  The moduli space of spiked instantons that arises in the context of
  the BPS/CFT correspondence \cite{N1} is realised as the moduli space
  of classical vacua, i.e.~low-energy open string field
  configurations, of a certain stack of intersecting D1-branes and
  D5-branes in Type IIB string theory. The presence of a constant
  $B$-field induces an interesting dynamics involving the tachyon
  condensation.
  
\end{abstract}

\section{Introduction}
Gauge theories in four dimensions with $\mc{N} = 2$ supersymmetry
occupy a special place in a quantum field theorist's landscape. On the
one hand, the large amount of supersymmetry severely constrains the
dynamics in the form of non-renormalisation theorems, existence of
strong-weak dualities etc. On the other hand, it is weak enough to
permit interesting non-perturbative effects like confinement,
interacting infrared fixed points and so on. Such implications have
been explored extensively in the past decades, starting with the
seminal work of \cite{SW1, SW2, AD, G} to note a few, with connections
being uncovered to diverse areas such as quantum integrable systems
\cite{NS}, two dimensional conformal field theories \cite{LMN, N2, NO,
  N1, AGT} and so on.

One of the main features of $\mc{N} = 2$ supersymmetric gauge theories
is the existence of certain sectors in the space of observables which
do not receive perturbative quantum corrections beyond one loop, the
\emph{BPS observables}. It turns out that, due to $\mc{N} = 2$
supersymmetry, the full set of non-perturbative corrections to these
observables can be expressed as integrals over instanton moduli
space. Using the $\Omega$-background \cite{N2,
  NO}, one calculates these integrals directly by working
equivariantly with respect to various symmetries acting on the
instanton moduli space (as in \cite{MNS, LNS}), especially the rotational symmetry inherited
from four dimensional spacetime.

With the above technique at hand, one can study transitions in the
gauge theory between configurations of different instanton
number. Observables which encode information about such
non-perturbative transitions can be expressed in terms of integrals
over generalised instanton moduli spaces, known as the \emph{moduli
  space of spiked instantons}, first considered in
\cite{N3}. Considerations of the structure of these observables (the
\emph{$qq$-characters}) leads to far-reaching consequences such as the
BPS/CFT correspondence \cite{N1,N3,N4}.

String theory has provided more than one way of constructing large
classes of $\mc{N} = 2$ supersymmetric gauge theories \cite{DM, KKV,
  W4d}. One such class is the class of quiver gauge theories \cite{DM}
which can be engineered by considering the gauge theory on a stack of
D3-branes located at a singularity of ADE type. Instantons in this
gauge theory have an alternate description as D-instantons bound to
the D3-branes \cite{DBinB}. It is then very natural to consider a
state with arbitrary number of D-instantons some of which are bound to
the D3-branes and also processes in which the number of bound
D-instantons can change.

The simplest case of the non-perturbative transitions alluded to above
can be realised by adding an auxiliary stack of D3-branes intersecting
with the original stack at a point. This setup gives rise to the
moduli space of \emph{crossed instantons}. The transitions then
correspond to some D-instantons escaping from the worldvolume of one
stack of D3-branes onto the other stack. The full non-perturbative
partition function of the gauge theory on this second stack of
D3-branes is then interpreted as an observable in the original gauge
theory that encodes information about these transitions. This
observable can be written as the expectation value of a local operator
inserted at the point of intersection in the original gauge theory
\cite{N3}.

The analysis can be suitably generalised to include surface defects in
the original gauge theory which correspond to including further stacks
of D3-branes which intersect the original stack on codimension two
hypersurfaces. The moduli space of bound states of the D-instantons
with the various stacks of D3-branes is the moduli space of spiked
instantons. Calculating suitable integrals over this moduli space
would then provide us with a wealth of information regarding
non-perturbative transitions in $\mc{N} = 2$ gauge theories.

In the present paper, we obtain the moduli space of spiked instantons
by studying the low-energy field theory on the worldvolume of
D1-branes probing a configuration of intersecting D5-branes which
preserves $\n{0,2}$ supersymmetry on the common two dimensional
intersection (this is T-dual to the setup described above). The
precise configuration is described in Section 2 below. In order to
bind the D1-branes to the D5-branes, we turn on a small constant NSNS
$B$-field along the D5-branes. In Section 3, we study the amount of
supersymmetry preserved by the D-brane configuration in the presence
of such a background. The next step is to study the spectrum of open
strings stretching between various pairs of D-branes in order to
extract the low-energy theory on the D1-branes. This is done is
Section 4. The salient feature is the presence of tachyons for generic
values of the $B$-field.

In Section 5, we consider a stack of D1-branes two stacks of D5-branes
intersecting on along the D1-brane worldvolume and obtain the
low-energy theory on the D1-branes (the analogous setup without
$B$-field was considered in \cite{T}). We calculate the various
interaction terms in the low-energy effective action by considering
3-point open string amplitudes on the disk and obtain the moduli space
of crossed instantons as the classical vacuum moduli space of the
low-energy theory. In Section 6, we generalise to the spiked case. We
also provide two appendices, the first of which has details regarding
the propagation of open strings in a constant $B$-field and the vertex
operators for various states in the open string spectrum. Appendix B
has details about $\n{0,2}$ superspace that is relevant for the spiked
instanton story.

{\hfil\textbf{Acknowledgements}\hfil}\\ This research was supported
by the National Science Foundation under grant
no.~NSF-PHY/1404446. Any opinions, findings, and conclusions or
recommendations expressed in this material are those of the authors
and do not necessarily reflect the views of the National Science
Foundation. The authors thank Xinyu Zhang, Alex DiRe and Saebyeok
Jeong for discussions on various aspects of the problem. N.~N.~thanks
D.~Tong for discussions and for hospitality in Cambridge in the summer
of 2015. N.~P.~thanks Martin Ro\v cek for several
illuminating discussions on $\n{0,2}$ superspace.

\section{Preliminaries}
Write the ten dimensional spacetime
$\mbb R^{1,9} \simeq \mbb R^{1,1} \times \mbb R^8$ as
$\mbb{R}^{1,1} \times \mbb{C}^4$ by choosing a complex structure on
the $\mbb R^8$. Let $\four = \{1, 2, 3, 4\}$ be the set of coordinate
labels of the $\mbb{C}^4\, .$ There are six 2-planes $\mbb{C}^2_A$
that sit inside the $\mbb{C}^4$ labelled by an index
$\displaystyle A \in \six = \binom{\four}{2}$ i.e.~the set of
unordered pairs of numbers in $\four$. Explicitly,
$\six = \{(12), (13), (14), (23), (24), (34)\}$. We consider a system
of D-branes which consists of $k$ D1-branes spanning $\Int$ and $n_A$
D5-branes spanning $\Int \times \mbb{C}^2_A$ with $A \in \six$. Here
onwards, $\mathbf{R}^{1,1}$ refers to the common $1+1$ dimensional
intersection of the D-brane configuration and is taken to be along the
$x^0, x^9$ directions.
\begin{table}[htp!]\centering
\small
\renewcommand{\arraystretch}{1.5}
\begin{tabular}{c|cc|cc|cc|cc|cc}
\toprule
$\mbb R^{1,9}$ & 1 & 2 & 3 & 4 & 5 & 6 & 7 & 8 & 9 & 0 \tabularnewline
\cline{1-11}
$\mbb{C}^4 \times \Int$  & \multicolumn{2}{c|}{$z^1$}& \multicolumn{2}{c|}{$z^2$} & \multicolumn{2}{c|}{$z^3$} & \multicolumn{2}{c|}{$z^4$} & $x$ & $t$ \tabularnewline
\toprule
D1 & & & & & & & & & $\times$ & $\times$\tabularnewline
\hline
D5$_{(12)}$ & $\times$ & $\times$ & $\times$ & $\times$ & & & & & $\times$ & $\times$\tabularnewline
D5$_{(13)}$ & $\times$ & $\times$ & & & $\times$ & $\times$ & & & $\times$ & $\times$\tabularnewline
D5$_{(14)}$ & $\times$ & $\times$ & & & & & $\times$ & $\times$ & $\times$ & $\times$\tabularnewline
D5$_{(23)}$ & & & $\times$ & $\times$ & $\times$ & $\times$ & & & $\times$ & $\times$\tabularnewline
D5$_{(24)}$ & & & $\times$ & $\times$ & & & $\times$ & $\times$ & $\times$ & $\times$\tabularnewline
D5$_{(34)}$ & & & & & $\times$ & $\times$ & $\times$ & $\times$ & $\times$ & $\times$\tabularnewline
\bottomrule
\end{tabular}
\caption*{The D1-D5 system for spiked instantons.}
\end{table}

Type IIB string theory has two supersymmetry parameters $\epsilon$ and
$\tl\epsilon$ which are Majorana-Weyl spinors of the same chirality
(say left-handed). That is,
\begin{equation}
  \Gamma_c \epsilon = \epsilon\ ,\quad \Gamma_c\tl\epsilon = \tl\epsilon \quad \text{where}\quad
\Gamma_c = \Gamma^1\cdots\Gamma^9\Gamma^0\ \ \text{and}\ \ (\Gamma_c)^2 = \mbbm 1\ .
\end{equation}

The presence of a D$p$-brane gives the following constraint on the
supersymmetry parameters:
\begin{equation}
\tl\epsilon = \frac{1}{(p+1)!} \varepsilon_{\mu_0\mu_1\cdots\mu_p}\Gamma^{\mu_0\mu_1\cdots\mu_p}\epsilon\ .
\end{equation}
Here, $\mu_0,\ldots, \mu_p$ take $p+1$ values corresponding to the
spacetime extent of the D$p$-brane and
$\Gamma^{\mu_0\mu_1\cdots\mu_p}$ is the totally antisymmetrised
product of $p+1$ $\Gamma$-matrices. Suppose the spatial extent of the
D$p$-brane is along $\{x^{i_1}, \ldots,x^{i_p}\}$ with
$i_1 < i_2 < \cdots < i_p$. Then, the Levi-Civita symbol $\varepsilon$
is normalised such that $\varepsilon_{i_1i_2\cdots i_p 0} = +1$. In
the presence of D1-branes along $\Int$ and D5-branes along
$\Int \times \mbb C^2_{(12)}$, the constraints are
$\tl\epsilon = \Gamma^{90}\epsilon$ and
$\tl\epsilon = \Gamma^{123490}\epsilon$ which give
\begin{equation}
\Gamma^{1234}\epsilon = \epsilon\ .
\end{equation}
Since $\Gamma^{1234}$ squares to identity and is traceless, half of
the sixteen real components of $\epsilon$ are set to zero. This leaves
us with a total of eight independent supersymmetry parameters for the
D1-D5$_{(12)}$ system. Including all six stacks of D5-branes gives
us the following constraints on $\epsilon$:
\begin{equation}
\Gamma^{1234}\epsilon = \epsilon\ ,\ \ \Gamma^{1256}\epsilon = \epsilon\ ,\ \ \Gamma^{1278}\epsilon = \epsilon\ ,\ \ \Gamma^{3456}\epsilon = \epsilon\ ,\ \ \Gamma^{3478}\epsilon = \epsilon\ ,\ \ \Gamma^{5678}\epsilon = \epsilon\ .
\end{equation}
It is easy to check that these constraints are not all compatible with
each other. For example, combining the first and fourth constraints
above gives us (using that $(\Gamma^{34})^2 = -\mbbm 1$),
\begin{equation}
\Gamma^{1256}\epsilon = -\epsilon\ ,
\end{equation}
which is in contradiction with the second constraint in that that the
two respective eigenspaces of $\Gamma^{1256}$ do not overlap. Thus,
the above D-brane system is not supersymmetric. Is it stable?

The zero-point energy in the R sector is always zero due to worldsheet
supersymmetry. The NS sector zero-point energy $E_0$ for a D$p$-D$p'$
system
\begin{equation}
  E_{0} = -\frac{1}{2} + \frac{\nu}{8}\ ,
\end{equation}
where $\nu$ is the number of ND$+$DN boundary conditions among the
eight non-lightcone directions.
\begin{itemize}
\item For D1-D1 strings and D5$_A$-D5$_A$ strings, we have $\nu = 0$
  and $E_{0} = -\frac{1}{2}$ but this state is removed by
  GSO projection and one has a ground state with zero energy.

\item The D1-D5$_{A}$ system has $\nu = 4$ and a four-fold degenerate
  ground state with $E_{0} = 0$. The GSO projection removes
  half of the states.

\item The D5$_{(ac)}$-D5$_{(bc)}$ system with $a \neq b$ has $\nu = 4$
  and hence $E_{0} = 0$. Again, the GSO projected ground
  state is two-fold degenerate.

\item The D5$_A$-D5$_{\w A}$ system has $\nu = 8$ and
  $E_{0} = +\frac{1}{2}$. Here, $\w{A}$ refers to the
  complement of $A$ in the set $\four$:
  $\w{A} = \four \smallsetminus A$. For example, for $A = (12)$, we
  have $\w{A} = (34)$. The lowest energy state after GSO projection is
  eight-fold degenerate and has positive energy. These states do not
  survive in the low-energy limit and only the zero-energy states from
  the R sector survive. Such a fermionic ground state is allowed by
  chiral supersymmetry. Indeed, the above configuration preserves
  $\n{0,8}$ supersymmetry on the intersection. The massive states from
  the NS sector have superpartners which are first excited states in
  the R sector (see Section \ref{crossedbranes} below).

\end{itemize}
Thus, after GSO projection, the NS sector zero-point energies are
either zero or positive implying that the system is stable.

There is a different system of D1-branes and D5-branes which
does preserve some fraction of supersymmetry. We impose the maximal
set of constraints which are compatible with each other. One such set
is given by
\begin{equation}
  \Gamma^{1234}\epsilon = \epsilon\ ,\ \ \Gamma^{1256}\epsilon = -\epsilon\ ,\ \ \Gamma^{1278}\epsilon = \epsilon\ ,\ \ \Gamma^{3456}\epsilon = \epsilon\ ,\ \ \Gamma^{3478}\epsilon = -\epsilon\ ,\ \ \Gamma^{5678}\epsilon = \epsilon\ .
\end{equation}
The constraints with a negative sign correspond to
$\w{\text{D5}}$-branes. Only three of the above six constraints are
independent, preserving one-sixteenth of the 32 supercharges. Thus, a
configuration of D1-branes with four stacks of D5-branes and two
stacks of $\w{\text{D5}}$-branes in the above arrangement preserves
\textbf{two supercharges}. Next, we would like to study the effects of
turning on a constant NSNS $B$-field background.

\section{Supersymmetry in a constant \tp{$B$}{B}-field background}
Consider a constant NSNS $B$-field background of the form:
\begin{equation}\label{constB}
  2\pi\alpha' B_{12} = b_1\ ,\ \ 2\pi\alpha' B_{34} = b_2\ ,\ \ 2\pi\alpha' B_{56} = b_3\ ,\ \ 2\pi\alpha' B_{78} = b_4\ .
\end{equation}
This choice of $B$-field preserves the $\text{SO}(2)^4$ rotational
symmetry of the above intersecting D-brane system. Such a symmetry is
essential for eventually considering the generalisation to the
$\Omega$-background.

We first state our conventions and introduce some notation.
\begin{itemize}
\item Introduce the variables $v_a$, $a \in \four$ with
  \begin{equation}
    e^{2\pi i v_a} = \frac{1 + ib_a}{1 - ib_a}\ , \quad b_a = \tan \pi v_a\ ,\quad -\frac{1}{2} < v_a < \frac{1}{2}\ .
  \end{equation}
  The limits $v_a \to \pm \tfrac{1}{2}$ correspond to
  $b_a \to \pm\infty$.
  
\item For each $A \in \six$, let
  $\Gamma_{A} = \Gamma^{2a-1}\Gamma^{2a}\Gamma^{2b-1}\Gamma^{2b}$ for
  $A = (ab)$.
  
\item Choose the following representation for the
  $\Gamma$-matrices. This representation corresponds to a particular
  choice of the cocycle operators for open string vertex operators
  given in \cite{KLLSW}. See Appendix \ref{bfield} for more details.
  \begin{align*}
    \Gamma^1 & = \sigma_1 \otimes \mbbm{1} \otimes \mbbm{1} \otimes \mbbm{1} \otimes \mbbm{1}\ ,\qquad \phantom{-\sigma_2}\Gamma^7 = -\sigma_3 \otimes \sigma_3 \otimes \sigma_3 \otimes \sigma_2 \otimes \mbbm{1}\ ,\\
    \Gamma^2 & = \sigma_2 \otimes \mbbm{1} \otimes \mbbm{1} \otimes \mbbm{1} \otimes \mbbm{1}\ ,\qquad  \phantom{-\sigma_2}\Gamma^8 = \sigma_3 \otimes \sigma_3 \otimes \sigma_3 \otimes \sigma_1 \otimes \mbbm{1}\ ,\\
    \Gamma^3 & = \sigma_3 \otimes \sigma_2 \otimes \mbbm{1} \otimes \mbbm{1} \otimes \mbbm{1}\ ,\qquad \!\phantom{-\sigma_2} \Gamma^9 = \sigma_3 \otimes \sigma_3 \otimes \sigma_3 \otimes \sigma_3 \otimes \sigma_1\ ,\\
    \Gamma^4 & = -\sigma_3 \otimes \sigma_1 \otimes \mbbm{1} \otimes \mbbm{1} \otimes \mbbm{1}\ ,\qquad \!\phantom{\sigma_2}\Gamma^0 = \sigma_3 \otimes \sigma_3 \otimes \sigma_3 \otimes \sigma_3 \otimes (-i\sigma_2)\ ,\\
    \Gamma^5 & = -\sigma_3 \otimes \sigma_3 \otimes \sigma_1 \otimes \mbbm{1} \otimes \mbbm{1}\ ,\qquad \!\!\phantom{\sigma_2}\Gamma_c = \sigma_3 \otimes \sigma_3 \otimes \sigma_3 \otimes \sigma_3 \otimes \sigma_3\ ,\\
    \Gamma^6 & = -\sigma_3 \otimes \sigma_3 \otimes \sigma_2 \otimes \mbbm{1} \otimes \mbbm{1}\ ,\qquad  \!\!\phantom{\sigma_2}C_- = e^{3\pi i/4} \sigma_2 \otimes \sigma_1 \otimes \sigma_2 \otimes \sigma_1 \otimes \sigma_2\ .
  \end{align*}
  The chirality matrices in $\mbb{C}^2_{A}$ are chosen to be
  $\Gamma_c(\mbb{C}^2_{A}) = \Gamma_A$ where $\Gamma_A$ is defined above
  and the chirality matrix in $\Int$ is
  $\Gamma_c(\Int) = -\Gamma^0\Gamma^9$.
  
\item The $32_{\mbb C}$ dimensional spinor representation can then be
  constructed by considering simultaneous eigenvectors
  $|\pm,\pm,\pm,\pm,\pm\rangle$ of $-i\Gamma^{12}$, $-i\Gamma^{34}$,
  $-i\Gamma^{56}$, $-i\Gamma^{78}$ and $-\Gamma^{09}$, and using the
  linear combinations $-i\Gamma^1 \pm \Gamma^2$, \ldots,
  $-i\Gamma^7 \pm \Gamma^8$, $\Gamma^0 \pm \Gamma^9$ as raising and
  lowering operators respectively. The basis of the representation is
  then given by the 32 vectors $|\pm,\pm,\pm,\pm,\pm\rangle$. The
  left-handed (right-handed) $16_{\mbb C}$ subspace of $\Gamma_c$ is
  then spanned by the subset of the above with even (odd) number of
  negative signs.
\end{itemize}

Next, we study the amount of supersymmetry preserved in the presence
of a constant $B$-field for the two different D-brane systems
considered above.
\subsubsection*{No anti D5-branes}
In the presence of a constant $B$-field of the form \eqref{constB},
the constraint arising from the stack of D5$_{A}$ branes becomes
\begin{equation}
  \tl\epsilon = \Gamma^{90} R_{A}\,\epsilon\ ,
\end{equation}
where $R_{A}$ is given by
\begin{equation}
  R_{A} = \exp \left(\sum_{a\in A}\theta_a\Gamma^{2a-1}\Gamma^{2a}\right)\ ,\
\end{equation}
with $\theta_a := \pi (v_a + \frac{1}{2})$. Combining this with the
constraint $\tl\epsilon = \Gamma^{90}\epsilon$ from the D1-branes, we
get
\begin{equation}\label{susy}
  R_A\,\epsilon = \epsilon\ \ \text{for every}\ \  A \in \six\ .
\end{equation}

Let $r(\theta) := \exp ( i \sigma_3 \theta)$. Then, we have
\begin{align}
  R_{(12)} &= r(\theta_1) \otimes r( \theta_2) \otimes \mbbm 1  \otimes \mbbm 1 \otimes \mbbm 1\ ,\ \ R_{(13)} = r(\theta_1) \otimes \mbbm 1 \otimes r(\theta_3) \otimes \mbbm 1 \otimes \mbbm 1\ ,\nonumber\\
  R_{(14)} &=  r(\theta_1) \otimes \mbbm 1 \otimes \mbbm 1 \otimes r(\theta_4) \otimes \mbbm 1 \ ,\ \ R_{(23)} = \mbbm 1 \otimes r(\theta_2) \otimes r(\theta_3) \otimes \mbbm 1 \otimes \mbbm 1 \ ,\nonumber\\
  R_{(24)} &=  \mbbm 1 \otimes r(\theta_2) \otimes \mbbm 1 \otimes r(\theta_4)  \otimes \mbbm 1\ ,\ \ R_{(34)} = \mbbm 1 \otimes \mbbm 1 \otimes r(\theta_3) \otimes r(\theta_4) \otimes \mbbm 1 \ .
\end{align}
The equations $R_A\epsilon = \epsilon$ have a solution if, for some
choice of signs,
\begin{equation}
  \exp \left(\pm i\theta_a \pm i\theta_b\right) = 1\quad \text{with}\quad 0 \leq \theta_a \leq \pi\quad \forall\quad a \in \four\ .
\end{equation}
By inspection, it can be seen that the above set of equations has no
solution except when all the $\theta_a = 0$ or when all the $\theta_a
= \pi$. This corresponds to $v_a \to \pm\frac{1}{2}$, or equivalently,
$b_a \to \pm\infty$. Thus, turning on a finite $B$-field of the above
form does not make the brane configuration supersymmetric.

\subsubsection*{Four D5-branes and two anti D5-branes}
Next, let us consider the configuration of D5-branes and
$\w{\text{D5}}$-branes that preserves two supercharges. The constraints on
$\epsilon$ can then be succinctly written as
\begin{equation}
  \Gamma_A\epsilon =\varepsilon_{A\w A}\ \epsilon\ ,
\end{equation}
where $\w{A} = \four \smallsetminus A$. For example, when $A = (12)$, we
have $\w{A} = (34)$ and
$\varepsilon_{A\w A} = \varepsilon_{1234} = +1$ and when $A = (13)$ we
have $\varepsilon_{1324} = -1$. When we turn on the above constant
$B$-field, the equations become
\begin{equation}
  R_A \epsilon = \varepsilon_{A\w A}\ \epsilon\ .
\end{equation}
This gives the conditions
\begin{equation}
  \exp \left(\pm i\theta_a \pm i\theta_b\right) = \varepsilon_{ab\w a\w b}\quad\text{with}\quad A = (ab)\ ,\ \w{A} = (\w{a}\w{b}) \quad\text{for}\quad A \in \six\ .
\end{equation}
These equations have solutions corresponding to finite $B$ only when
$\displaystyle \theta_a = \frac \pi 2\ \text{for all}\ a \in \four$
with some choice of signs. This corresponds to $v_a = 0$ which is the
zero $B$-field point.

\subsubsection*{Stability}
The question of stability arises in this situation too. First we
observe that supersymmetry is completely lost about the original
vacuum for a non-zero finite value of the constant $B$-field. Thus,
stability is no longer guaranteed. Secondly, a constant $B$-field
background typically introduces instability in the form of tachyons in
the D-brane spectrum.

In some situations, e.~g.~the D1-D5 system, the effects of the
$B$-field can be accommodated by turning on a Fayet-Iliopoulos
parameter in the low energy effective action. The tachyon instability
leads to the system transitioning to a nearby vacuum at which point
supersymmetry is restored.

We shall see that something similar happens in the spiked scenario as
well, with some differences. To study the stability we need to derive
the spectrum of open strings in the presence of D-branes in a constant
$B$-field background.

\section{Spectrum of \tp{D$p$-D$p'$}{Dp-Dp'} strings}\label{constBstring}
The boundary conditions for an open string are modified in the
presence of a $B$-field. Let the worldsheet bosons and fermions along
$\mbb{C}^4$ be resp.~$Z^a(\sigma,\tau)$ and
$\Psi^{\pm,a}(\sigma,\tau)$, $a \in \four$. Neumann boundary
conditions along $\mbb{C}_a$ are modified to (cf. Appendix
\ref{bfield}):
\begin{align}
  \text{\textbf{Twisted(T):}}&\quad\quad \partial_{++} Z^a = e^{-2\pi i v_a}\partial_{--} Z^a\ ,\qquad \Psi^{a+} = e^{2\pi i v_a} \Psi^{a-}\ .
\end{align}
Neumann and Dirichlet boundary conditions are obtained by setting $v_a
= 0$ and $v_a \to \frac{1}{2}$ respectively. Sending one of the
$v_a$'s to $-\frac{1}{2}$ would give Dirichlet boundary conditions on
an anti D-brane. Consider the more general boundary conditions with
$-\frac{1}{2} \leq \mu, \nu \leq \frac{1}{2}$:
\begin{align}
  \partial_{++} Z &= e^{-2\pi i\nu} \partial_{--} Z\ ,\quad \Psi^{+} = e^{2\pi i \nu} \Psi^{-} \quad\text{at}\quad\sigma = 0\ ,\nonumber\\
  \partial_{++} Z &= e^{-2\pi i\mu}\partial_{--} Z\ ,\quad \Psi^{+} = \pm e^{2\pi i \mu} \Psi^{-} \quad\text{at}\quad\sigma = \pi\ ,
\end{align}
The low-energy spectrum for this system has been worked out in
Appendix \ref{bfield}. We summarise the results here.
\begin{enumerate}
\item \textbf{Non-integer modes:} The worldsheet boson $Z$ has moding
  $\mbb Z + \theta$ with $\theta = \mu - \nu$. The R sector fermions
  have the same moding as $Z$ due to rigid supersymmetry on the
  worldsheet and the NS sector fermions have moding
  $\mbb Z + \epsilon$ with
  $\epsilon = \theta + \frac{1}{2} = \mu - \nu + \frac{1}{2}$.
  
\item \textbf{Excitations:} The zero-point energy in the NS sector is
  given by
  \begin{equation}
    E_{0} = \frac{1}{8} - \frac{1}{2}\big||\theta| - \tfrac{1}{2}\big|\ .
  \end{equation}
  The first excited state in the NS sector has energy
  $E_{0} + |\epsilon|$ or $E_{0} + |1 - \epsilon|$
  when $-\frac{1}{2} \leq \epsilon \leq \frac{1}{2}$ and
  $\frac{1}{2} < \epsilon < \frac{3}{2}$ respectively.
  
  The zero-point energy in the R sector vanishes due to rigid
  supersymmetry on the worldsheet. The first excited state in the R
  sector has energy $|\theta|$ for $0 \leq |\theta| \leq \frac{1}{2}$
  and $1 - |\theta|$ for $\frac{1}{2} \leq |\theta| \leq 1$.
  
\item \textbf{Spectral flow:} When $\epsilon$ crosses the integer $s$
  from the left ($s = 0$ or 1), the state with energy $s - \epsilon$
  becomes negative and enters the Dirac sea and the state
  $\epsilon - s$ crosses into the positive energy region. The raising
  and lowering roles of the NS fermion operators $d_s$ and $d_s^\dag$
  are interchanged. Using $d_s^\dag d_s = -d_s d_s^\dag + 1$, we see
  that the number operator changes by one unit $N_d \to N_d + 1$. This
  changes the sign of the parity operator
  $(-)^{F_{\text{NS}}} := (-1)^{N_d}$ and the GSO projectors
  $\frac{1}{2}(1\pm(-)^{F_{\text{NS}}})$ are consequently
  interchanged. A similar phenomenon occurs in the R sector when
  $\theta$ crosses 0. \label{flip}
\end{enumerate}

\subsection{D1-D1 strings}
The open strings satisfy $\textbf{NN}$ boundary conditions along
$\Int$ and $\textbf{DD}$ boundary conditions along $\mbb{C}^4$. The
worldsheet bosons have momentum zero modes along $\Int$ and none along
$\mbb{C}^4$ and hence all the states are supported along $\Int$.

\textbf{NS sector:} There are no zero modes for the NS fermions and
the NS zero-point energy is $-\frac{1}{2}$. The NS fermion oscillators
${d_1^{\mu}}^\dag$, $\mu = 0,9$ and ${d_1^a}^\dag$, $a \in \four$
raise the energy by $\frac{1}{2}$. The oscillators ${d_1^{\mu}}^\dag$
gives rise to two states which are the components of a gauge field
$v_{\pm\pm}(x, t)$ while the four complex oscillators ${d_1^a}^\dag$
create four states in the adjoint of $\text{U}(k)$ corresponding to
complex scalars $B_a(x, t)$. Assigning the NS vacuum a fermion number
$F_{\text{NS}} = -1$, the GSO projection with projector
$\frac{1}{2}(1 + (-)^{F_\text{NS}})$ projects out the vacuum while
retaining the zero-energy states.

\textbf{R sector:} The R sector has ten zero modes thus giving a real
32 dimensional ground state transforming in the adjoint of
$\text{U}(k)$. The fermion parity $(-)^{F_{\text{R}}}$ on the zero
modes is then
$(-)^{F_{\text{R}}} = \Gamma^{1\cdots 8}\Gamma^{90} =
\Gamma_c(\mbb{R}^{1,9})$. The GSO projection with
$\frac{1}{2}(1 + (-)^{F_\text{R}})$ gives a left-handed fermion in
$1+9$ dimensions which splits up into eight right-handed and eight
left-handed fermions in $1+1$ dimensions.

We decompose the spacetime scalars and fermions into representations
of $\text{SO}(\mbb{C}^2_A) \times \text{SO}(\mbb{C}^2_{\w{A}})$ using
$\Gamma_c(\mbb{R}^{1,9}) =
\Gamma_c(\Int)\Gamma_c(\mbb{C}^2_A)\Gamma_c(\mbb{C}^2_{\w{A}})$. Writing
each $\text{SO}(4)$ as $\text{SU}(2) \times \text{SU}(2)$ with
$\alpha, \dt\alpha, \alpha', \dt\alpha'$ denoting the fundamentals of
the four $\text{SU}(2)$'s, we have
  \begin{align}
    \text{Scalars}:\quad  &X^{\alpha\dt\alpha} \oplus X^{\alpha'\dt\alpha'}\ ,\nonumber\\
    \text{Fermions}:\quad &\lambda_{-}^{\alpha\alpha'} \oplus \lambda_{-}^{\dt\alpha\dt\alpha'} \oplus \zeta_{\ndt +}^{\alpha\dt\alpha'} \oplus \zeta_{\ndt +}^{\alpha'\dt\alpha}\ ,
  \end{align}
  with reality conditions
  $\lambda_-^{\alpha\alpha'} =
  -\varepsilon^{\alpha\beta}\varepsilon^{\alpha'\beta'}\,\w{\lambda_-^{\beta\beta'}}$
  and so on for the fermions.

\subsection{D1-\tp{D5$_A$}{D5A} strings} 
For a D1-D5$_A$ string the boundary conditions are \textbf{DT} for $a
\in A$ and \textbf{DD} for $a \in \w A$. These boundary conditions
imply $\mbb Z + v_a - \frac{1}{2}$ moding for the bosons $Z^a$ with $a
\in A$ and $\mbb Z$ moding for $a \in \w A$. The R fermions have the
same moding as the bosons and the NS fermions have moding $\mbb Z +
v_a$ for $a \in A$ and $\mbb Z + \frac{1}{2}$ for $a \in \w A$. Since
the string is orientable, states from different orientations are
distinct and have to be combined together in order to form a CPT
invariant spectrum.

\textbf{NS sector:} Let $A = (ab)$. The NS zero-point energy is given
by $-\frac{1}{2}(|v_a| + |v_b|)$. For $v_a$ and $v_b$ close to zero,
the oscillators with lowest positive energy are from the NS fermions
and increase energy by $|v_a|$ and $|v_b|$. The first four states in
the NS sector have the energies
\begin{equation}
  \frac{1}{2}(\pm|v_a| \pm |v_b|)\ \ \text{or equivalently,}\ \ \frac{1}{2}(\pm v_a \pm v_b)\ .
\end{equation}
When either of $v_a$ and $v_b$ crosses zero, the sign of
$(-)^{F_{\text{NS}}}$ is flipped (cf. point \ref{flip} above). It is
then easy to see that states which have definite values of
$(-)^{F_{\text{NS}}}$ are $\frac{1}{2}(\pm v_a \pm v_b)$ rather than
$\frac{1}{2}(\pm|v_a| \pm|v_b|)$.

We assign $(-)^{F_{\text{NS}}} = -1$ to the state with energy
$-\frac{1}{2}(v_a + v_b)$ and choose the GSO projector to be
$\frac{1}{2}(1 + (-)^{F_{\text{NS}}})$. The states with energies
$\pm\frac{1}{2}(v_a + v_b)$ are projected out and the states that
remain are
\begin{equation}
  -\frac{1}{2}(v_a - v_b)\ ,\ \frac{1}{2}(v_a - v_b)\ .
\end{equation}
These states transform in the $(\bm{k}, \w{\bm{n}}_{\bm{A}})$ of
$\text{U}(k) \times \text{U}(n_A)$. The string with opposite
orientation furnishes two more states with the same energy and which
transform in the $(\w{\bm{k}}, \bm{n_A})$ of $\text{U}(k) \times
\text{U}(n_A)$. Thus, we get two complex scalars $\phi^1$ and
$\phi^2$ in the bifundamental of $\text{U}(k) \times \text{U}(n_A)$
with masses given by
\begin{equation}
  m^2 = \pm\frac{1}{2\alpha'}(v_a - v_b)\ .
\end{equation}
In the limit $v_a, v_b \to 0$, the two states become degenerate. We
also have $(-)^{F_\text{NS}} = \Gamma_A = \Gamma_c(\mbb{C}^2_A)$ which
implies that the above GSO projection results in a left-handed spinor
$\phi^\alpha$ in $\mbb{C}^2_A$. These constitute the two complex
scalars of a $\n{4,4}$ bifundamental hypermultiplet in $\Int$.

The two complex NS fermions along $\mbb{C}^2_{\w{A}}$ have energy
$\frac{1}{2}$. The oscillators associated to worldsheet bosons along
$\mbb{C}^2_{A}$ have energies $\frac{1}{2} \pm v_a$,
$\frac{1}{2} \pm v_b$.

\textbf{R sector:} The zero-point energy vanishes in the R
sector. There are six zero modes from fermions along
$\Int \times \mbb{C}^2_{\w{A}}$ which give an eight dimensional ground
state consisting of spinors $|\alpha',\pm\rangle$ and
$|\dt\alpha',\pm\rangle$ where $+(-)$ indicates left(right)-handed
spinors in $\Int$ and $\alpha'$($\dt\alpha'$) left(right)-handed
spinors in $\mbb{C}^2_{\w{A}}$. The fermion parity operator
$(-)^{F_\text{R}}$ is given by
$(-)^{F_\text{R}} = \Gamma_{\w{A}}\Gamma^{90} =
\Gamma_c(\Int)\Gamma_c(\mbb{C}^2_{\w{A}})$. The GSO projection with
$\frac{1}{2}(1 - (-)^{F_\text{R}})$ retains the states that satisfy
$\Gamma_c(\Int) = \pm 1$, $\Gamma_c(\mbb{C}^2_{\w{A}}) = \mp
1$. Together with the states from the oppositely oriented string, we
thus have spinors $\zeta^{\alpha'-} = -\zeta^{\alpha'}_+$ and
$\lambda^{\dt\alpha'+} = \lambda^{\dt\alpha'}_-$. They
transform in the $(\bm{k}, \w{\bm{n}}_{\bm{A}})$ of
$\text{U}(k) \times \text{U}(n_A)$ and constitute the fermionic part
of the $\n{4,4}$ bifundamental hypermultiplet in $\Int$.

The first few single-oscillator excitations in the R sector come from
fermionic and bosonic oscillators along $\mbb{C}^2_{(ab)}$ with energies
$\frac{1}{2} \pm v_{a}$, $\frac{1}{2} \pm v_{b}$. The oscillators
in the other directions raise the energy by $1$. After GSO projection,
the states from worldsheet bosons acting on $|\alpha',-\rangle$ and
$|\dt\alpha',+\rangle$ and states from R fermions acting on
$|\alpha',+\rangle$ and $|\dt\alpha',-\rangle$ are retained.

\textbf{D1-$\w{\text{D5}}_A$ strings:} The above analysis carries
through but the modings become $\mbb{Z} + \frac{1}{2} - v_a$,
$\mbb{Z}+ v_b - \frac{1}{2}$ for $Z_a$ and $Z_b$ respectively. Also,
the GSO projection is carried out with the opposite projectors. The NS
sector states remaining after GSO projection have energies
$\pm\frac{1}{2}(v_a + v_b)$ and in the limit $v_a, v_b \to 0$ form a
right-handed spinor in $\mbb{C}^2_{A}$. The R sector fermions which
are right(left)-handed in $\Int$ are right(left)-handed in
$\mbb{C}^2_{\w{A}}$.

\subsection{\tp{D5$_A$-D5$_{\w{A}}$}{D5A-D5(Abar)} strings}\label{crossedbranes}
The boundary conditions are \textbf{TD} for $a \in A$ and \textbf{DT}
for $a \in \w A$. These imply the following modings for the bosons and
R fermions:
\begin{equation}
  \mbb Z + \frac{1}{2} - v_a \ \,\text{for}\ \,a \in A\quad \text{and}\quad \mbb Z + v_{\w{a}} - \frac{1}{2}\ \,\text{for}\ \,\w{a} \in \w A\ .
\end{equation}
The NS fermions have $\mbb Z - v_a$ and $\mbb Z + v_{\w{a}}$ moding
respectively.

\textbf{NS sector:} The zero point energy in the NS sector is then
\begin{align}
&\frac{1}{2} - \frac{1}{2}\,\sum_{a\,\in\,\four}\,|v_a|\ .
\end{align}
The lowest excitation energies in the NS sector are $|v_a|$ for
$a \in \four$. The first few states are then
\begin{equation}
  \frac{1}{2}(1 \pm v_1 \pm v_2 \pm v_3 \pm v_4)\ .
\end{equation}
We assign $(-)^{F_{\text{NS}}} = -1$ to the state with energy
$\frac{1}{2}(1 -( v_1 + v_2 + v_3 + v_4))$. GSO projection with
$\frac{1}{2}(1 + (-)^{F_{\text{NS}}})$ removes states with an even
number of negative signs. The remaining states are
\begin{align}
  &\frac{1}{2}[1 \pm (v_1 - v_2 - v_3 - v_4)]\ ,\ \frac{1}{2}[1 \pm (v_1 + v_2 + v_3 - v_4)]\ ,\nonumber\\
  &\frac{1}{2}[1 \pm (v_1 + v_2 - v_3 + v_4)]\ ,\ \frac{1}{2}[1 \pm (v_1 - v_2 + v_3 + v_4)]\ .
\end{align}
For small enough $|v_a|$, the above energies are all positive: there
is no tachyon or massless state in the NS sector. There is another
copy of these states from the string with opposite
orientation. Together, they form eight massive complex scalars that
transform in the $(\bm{n_A}, \w{\bm{n}}_{\w{\bm{A}}})$ of
$\text{U}(n_A) \times \text{U}(n_{\w{A}})$.

\textbf{R sector:} The ground state energy in the R sector is zero as
always. The only zero modes are the ones along $\Int$ and we denote
them by $\Gamma^0$ and $\Gamma^9$. We have
$(-)^{F_{\text{R}}} = \Gamma^{90} = \Gamma_c(\Int)$. Assign
$(-)^{F_{\text{R}}} = -1$ for the ground state $|\text{R}\rangle$ and
define
\begin{equation}
  g = \frac{\Gamma^9 + \Gamma^0}{\sqrt{2}}\ ,\quad g^\dag = \frac{\Gamma^9 - \Gamma^0}{\sqrt{2}}\ .
\end{equation}
Acting on $|\text{R}\rangle$ with $g^\dag$ provides another state of
zero energy but with $(-)^{F_{\text{R}}} = +1$. The GSO projection
with $\frac{1}{2}(1 + (-)^{F_{\text{R}}})$ retains
$g^\dag|\text{R}\rangle$ which is a left-handed fermion. Together with
a similar state from the oppositely oriented string, this fermion
transforms in the bifundamental of
$\text{U}(n_A) \times \text{U}(n_{\w{A}})$.

For small $v_a$, the first two sets of single-oscillator excitations
for the worldsheet bosons and R fermions come from the $\mbb C^4$
directions and have energy $\frac{1}{2} \mp v_a$. The GSO projection
keeps the eight states obtained from the worldsheet bosons acting on
$g^\dag|\text{R}\rangle$ and the eight states from R fermions acting
on $|\text{R}\rangle$. Together with states from the oppositely
oriented string, they form four right- and left-moving fermions with
mass-squared $\frac{1}{2} + v_a$ and four right- and left-moving
fermions with mass-squared $\frac{1}{2} - v_a$.

In the limit $v_a \to 0$, the eight right-moving and eight left-moving
fermions become degenerate and the eight right-movers are in fact the
superpartners of the scalars from the NS sector.

\subsection{\tp{D5$_{(ca)}$-D5$_{(cb)}$}{D5(ca)-D5(cb)} strings}
Here $\mbb C^2_{(ca)}$ and $\mbb C^2_{(cb)}$ share a common
$\mbb C_c$. Let the remaining direction be $\mbb C_d$. The boundary
conditions are now \textbf{TT} for $Z^c$, \textbf{TD} for $Z^a$,
\textbf{DT} for $Z^b$ and \textbf{DD} for $Z^d$. The modings are
$\mbb Z$ for $Z^c$ and $Z^d$, $\mbb Z + \frac{1}{2} - v_a$ for $Z^a$
and $\mbb Z + v_b - \frac{1}{2}$ for $Z^b$. The R fermions have the
same modings and the NS fermions have the modings shifted by
$\frac{1}{2}$. The modings are the same as for a
D1-$\w{\text{D5}}_{(ab)}$ system. The worldsheet bosons have momentum
and position zero modes along $\Int \times \mbb{C}_c$. Hence all the
states will be supported on the four dimensional space
$\Int \times \mbb{C}_c$.

\textbf{NS sector:} The zero-point energy is $-\frac{1}{2}(|v_a| +
|v_b|)$ and the lowest-lying excitation energies are $|v_a|$ and
$|v_b|$. Thus, the lowest energy states are $\frac{1}{2}(\pm v_a \pm
v_b)$. We assign $(-)^{F_{\text{NS}}} = -1$ to the state
$-\frac{1}{2}(v_a + v_b)$ and perform GSO projection with
$\frac{1}{2}(1 - (-)^{F_{\text{NS}}})$ to get the states
\begin{equation}
  -\frac{1}{2}(v_a + v_b)\ ,\ \frac{1}{2}(v_a + v_b)\ .
\end{equation}
After including states from the oppositely oriented string, these give
two complex scalars $\sigma^1$, $\sigma^2$ which transform as
$(\bm{n_{(ca)}}, \w{\bm{n}}_{\bm{(cb)}})$ with masses $m^2 =
\pm\frac{1}{2\alpha'} (v_a + v_b)$. In the limit $v_a, v_b \to 0$, the
two scalars are massless and combine into a right-handed spinor in
$\mbb{C}^2_{(ab)}$ since $(-)^{F_{\text{NS}}} =
\Gamma_c(\mbb{C}^2_{(ab)})$. These constitute the bosonic part of a
$\mc{N} = 2$ hypermultiplet in $\Int \times \mbb{C}_c$.

\textbf{R sector:} The worldsheet fermions along
$\Int \times \mbb{C}^2_{(cd)}$ are integer moded, giving six zero
modes and an eight dimensional ground state. The fermion parity
operator is given by
$(-)^{F_{\text{R}}} = \Gamma_c(\mbb{C}^2_{(cd)})\Gamma^{90} =
\Gamma_c(\Int \times \mbb{C}_{c})\Gamma_c(\mbb{C}_{d})$ where
$\Gamma_c(\Int \times \mbb{C}_c) =
i\Gamma^{2c-1}\Gamma^{2c}\Gamma^{9}\Gamma^0$ and
$\Gamma_c(\mbb{C}_d) = -i\Gamma^{2d-1}\Gamma^{2d}$. We use the GSO
projector $\frac{1}{2}(1 + (-)^{F_{\text{R}}})$ to get a left-handed
fermion $\lambda$ and a right-handed fermion $\w\zeta$ in
$\Int \times \mbb{C}_c$ with $\Gamma_c(\mbb{C}_d) = \pm 1$
respectively. These constitute the fermionic part of a $\mc{N} = 2$
hypermultiplet.

\textbf{D5$_{(ca)}$-$\w{\text{D5}}_{(cb)}$ strings:} For this case,
the GSO projections are reversed. The NS sector states then have
energies $\pm\frac{1}{2}(v_a - v_b)$ and form a left-handed spinor in
$\mbb{C}^2_{(ab)}$ in the limit $v_a, v_b \to 0$. The R sector
fermions have opposite eigenvalues under $\Gamma_c(\mbb{C}_d)$.

\section{Crossed instantons}
We first consider the simpler configuration of \emph{crossed
  instantons}: $k$ D1-branes along $\Int$, $n$ D5-branes along
$\Int \times \mbb C^2_{(12)}$ and $n'$ D5-branes along
$\Int \times \mbb C^2_{(34)}$. This setup preserves four supercharges
organised into $\n{0,4}$ supersymmetry on the two dimensional
intersection $\Int$. This setup has been studied in the context of
$\mbb{AdS}_3$ holography by \cite{T, GMMS} and others. Another place
where $\n{0,4}$ supersymmetry appears is the ADHM sigma model
\cite{W3} which has a stringy realisation as a D1-D5-D9 brane system
\cite{Dgauge}. More recently, the authors in \cite{PSY} explore a
class of $\n{0,4}$ superconformal theories obtained by compactifying
M5-branes on four-manifolds of the form $\mbb{P}^1 \times \mc{C}$
where $\mc{C}$ is a Riemann surface with punctures.

We are interested in studying the bound states of D1-branes with the
crossed D5-branes above with the constant $B$-field background in
\eqref{constB}. As we have seen in the previous section, there are
generically tachyons in the spectrum and supersymmetry is broken.  We
are interested in the end point of tachyon condensation \cite{Sen, A,
  GS} and the all-important question: is supersymmetry restored at the
end point of the condensation?

We shall find that for a particular locus in the space of $B$-fields,
the supersymmetry breaking can be described by a Fayet-Iliopoulos term
in the low-energy theory. For small values of $B$-field, we can then
study the condensation of the tachyons in the low-energy effective
theory. The relevant low-energy degrees of freedom are those of a
supersymmetric $\text{U}(k)$ gauge theory interacting with various
matter multiplets supported on $\Int$. In particular, we freeze the
supersymmetric gauge degrees of freedom supported on the D5-branes to
their classical vacuum expectation values.

\textbf{Note:} The above D5-brane system without the D1-branes has
been studied in great detail by many authors, notably by
\cite{IKS}. There are chiral fermions (the field $\lambda$ below) in
the $1+1$ dimensional intersection arising from the strings stretching
between the two stacks of D5-branes. The chiral fermions render the
gauge theories on the intersection anomalous and the degrees of
freedom in the bulk of the D5-branes are necessary to cancel these
anomalies via the anomaly inflow mechanism. Since we have frozen these
gauge degrees of freedom, these issues are not immediately relevant to
our analysis below. In our case, the low-energy theory on the
intersection has $\text{U}(n) \times \text{U}(n')$ as rigid
symmetries.

The spacetime Lorentz group $\text{SO}(1,9)$ is broken down to
$\text{SO}(1,1) \times \text{SO}(4) \times \text{SO}(4)'$. The low
energy theory on $\Int$ has the internal rigid symmetry group
$\text{SO}(4) \times \text{SO}(4)' \times \text{U}(n) \times
\text{U}(n')$. \emph{It will be useful to write
}$\text{SO}(4) \times \text{SO}(4)' = \text{SU}(2)_{L} \times
\text{SU}(2)_{ R} \times \text{SU}(2)'_{ L} \times \text{SU}(2)'_{ R}$
\emph{with the indices $(\alpha, \dt\alpha, \alpha', \dt\alpha')$
  denoting the fundamental representations of the respective
  \emph{$\text{SU}(2)$}s.}

The sixteen components of the left-handed spinor $\epsilon$ can be
written in terms of spinors which have definite chirality under each
of $\text{SO}(1,1)$, $\text{SO}(4)$ and $\text{SO}(4)'$ as follows:
\begin{equation}
\epsilon = \eta_{\ndt L}^{\alpha\alpha'} \oplus \eta_{\ndt R}^{\alpha\dt\alpha'} \oplus \eta_{\ndt R}^{\dt\alpha\alpha'} \oplus \eta_{\ndt L}^{\dt\alpha\dt\alpha'}\ .
\end{equation}
The subscripts indicate chirality in $1+1$ dimensions. We see that the
product of the three chiralities is $+1$ which agrees with $\epsilon$
being left-handed in $9+1$ dimensions. There must also be a reality
condition on each of the $\eta$'s that arises from the Majorana
condition on $\epsilon$. Since the fundamental representation of
$\text{SU}(2)$ is pseudoreal, the $\eta$'s are in a real representation of the
corresponding $\text{SU}(2) \times \text{SU}(2)$. In other words, we have
\begin{equation}
\eta_{\ndt R}^{\alpha\alpha'} = -\varepsilon^{\alpha\beta}\varepsilon^{\alpha'\beta'}\,\w{\eta_{\ndt R}^{\beta\beta'}}\ \ \text{and so on.}
\end{equation}
(To check this, write
$\eta^{\alpha\alpha'} = \eta_m(\sigma^m)^{\alpha\alpha'}$ for some
dummy real 4-vector $\eta_m$ with
$\sigma^m = (\sigma^1, \sigma^2, \sigma^3, i\mbbm 1)$ ,
$\varepsilon^{12} = \varepsilon^{1'2'} = +1$ and
$\varepsilon^{\alpha\beta}\varepsilon_{\beta\gamma} =
-\delta^{\alpha}{}_{\gamma}$,
$\varepsilon^{\dt\alpha\dt\beta}\varepsilon_{\dt\beta\dt\gamma} =
-\delta^{\dt\alpha}{}_{\dt\gamma}$.)

The constraints on $\epsilon$ due to the above configuration of branes
are $\Gamma^{1234}\epsilon = \epsilon$ and
$\Gamma^{5678}\epsilon = \epsilon$ which means $\epsilon$ has to be
left-handed in both $\mbb C^2_{(12)}$ and $\mbb C^2_{(34)}$. Thus,
there are four real left-handed supersymmetry parameters
$\eta_{\ndt L}^{\alpha\alpha'}$ corresponding to supersymmetry in the
left-moving sector: \emph{we have $\n{0,4}$ supersymmetry in the $1+1$
  dimensional intersection $\Int$.} The R-symmetry of the $\n{0,4}$
supersymmetry algebra is $\text{SU}(2)_L \times \text{SU}(2)'_L$ and
the parameters $\eta_{\ndt L}^{\alpha\alpha'}$ transform as a bispinor
under this R-symmetry. We denote $\eta_{\ndt L}^{\alpha\alpha'}$ as
$\eta^{\alpha\alpha'+}$ or equivalently $\eta_{\ndt-}^{\alpha\alpha'}$
in the sequel.

\subsection{Low-energy spectrum and \tp{$\n{0,2}$}{N=(0,2)}
  decomposition}
We write the low-energy action in $\n{0,2}$ superspace by choosing a
particular $\n{0,2}$ subalgebra of the $\n{0,4}$ supersymmetry
algebra. See Appendix \ref{02app} for a description of $\n{0,2}$
superspace.

We choose the $\n{0,2}$ subalgebra generated by
$\eta^{11'+} := \eta^+$ and
$-\eta^{22'+} = \w{\eta^{11'+}} = \w\eta{}^+$ (this will be the
subalgebra preserved by the spiked instanton configuration). The
supercoordinates are $\theta^+$ and $\w\theta{}^{+}$. The R-symmetry
$U(1)_{\ell}$ of the left-moving supersymmetry is generated by
$F_{\ell} := F_L + F_R + F'_L + F'_R = F_{34} + F_{78}$ where
$F_L = \frac{1}{2}(-F_{12} + F_{34})$,
$F_R = \frac{1}{2}(F_{12} + F_{34})$,
$F'_L = \frac{1}{2}(-F_{56} + F_{78})$ and
$F'_R = \frac{1}{2}(F_{56} + F_{78})$. In our conventions,
$\eta^+ = \eta^{11'+}$ has charges $F_{12} = F_{56} = -\frac{1}{2}$
and $F_{34} = F_{78} = \frac{1}{2}$ giving $F_L = F'_L = +1/2$ and
$F_R = F'_R = 0$ and hence a charge of $+1$ under $U(1)_{\ell}$. The
$\n{0,2}$ content of the various multiplets from D$p$-D$p'$ strings
are summarised in Table \ref{crossedfields}. The various fields are
displayed with indices that indicate their
$\text{SO}(4)\times\text{SO}(4)'$ representations.

\textbf{Note:} In order to avoid too many indices on the fields, the
scalar component of a chiral multiplet $\Phi$ will be denoted by the
same letter and the right-handed spin-$\tfrac 12$ component by
$\zeta_{\Phi}$ in the sequel. Also, the left-handed spin-$\tfrac 12$
component of a Fermi superfield $\Lambda_{a}$ will be denoted as
$\lambda_{a}$ where $a$ is an index that runs over all Fermi
superfields in the theory. For example, the chiral multiplet $\wt{J}$
in Table \ref{d1d512} has components $\wt\phi{}^{\,2'}{}^\dag$ and
$\wt\zeta_+^{\,1}{}^\dag$ which will be alternatively referred to as
$\wt{J}$ and $\zeta_{\wt{J}}$ respectively. The left-moving fermionic
component Fermi superfield $\Lambda_{\wt J}$ will be denoted as
$\lambda_{\wt J}$.
\begin{table}\small
  \renewcommand{\arraystretch}{1.5}
  \begin{subtable}{\linewidth}\centering
    \caption{\textbf{D1-D1 strings}}\label{d1d1}
    \begin{tabular}{c|c|c}
      \toprule
      ${(0,4)}$ multiplet & Fields & ${(0,2)}$ multiplets   \tabularnewline
      \midrule
      Vector & $v_{--}\ ;\ \lambda_{-}^{\alpha\alpha'}$ & Vector $V = (v_{--}\,;\, \lambda_{-}^{11'})$,\ Fermi $\rp0\Lambda_2 = (\lambda_-^{12'})$\tabularnewline\hline
      Standard hyper & $X^{\alpha\dt\alpha}\ ;\ \zeta_{+}^{\alpha'\dt\alpha}$ & Chiral $\rp0\!B_1 = (X^{1\dt 1}\,;\,\zeta_{+}^{2'\dt 1})$,\ \ Chiral $\rp\pp\!B_2 = (X^{1\dt 2}\,;\,\zeta_{+}^{2'\dt 2})$ \tabularnewline\hline
      Twisted hyper & $X^{\alpha'\dt\alpha'}\ ;\ \zeta_{+}^{\alpha\dt\alpha'}$ & Chiral $\rp0\!B_3 = (X^{1'\dt 1'}\,;\, \zeta_{+}^{2\dt 1'})$,\ \ Chiral $\rp\pp\!B_4 = (X^{1'\dt 2'}\,;\, \zeta_{+}^{2\dt 2'})$\tabularnewline\hline
      Fermi & $\lambda_{-}^{\dt\alpha\dt\alpha'}$ & Fermi $\rp=\Lambda_3 = (\lambda_-^{\dt 1\dt 1'})$,\ \ Fermi $\rp0\Lambda_4 = (\lambda_-^{\dt 1\dt 2'})$\tabularnewline
      \bottomrule
    \end{tabular}
  \end{subtable}\vspace{20pt}
  \begin{subtable}{\linewidth}\centering
    \caption{\textbf{D1-D5$_{(12)}$ strings} \\ $I$, $\Lambda_I$ transform in the $(\bm{k},\w{\bm{n}})$  of $\text{U}(k) \times \text{U}(n)$ while $J$, $\Lambda_J$ transform in the $(\w{\bm{k}},\bm{n})$.}\label{d1d512}
    \begin{tabular}{c|c|c}
      \toprule
      ${(0,4)}$ multiplet & Fields & ${(0,2)}$ multiplets   \tabularnewline
      \midrule
      Standard hyper & $\phi^{\alpha}\ ;\ \zeta_{+}^{\alpha'}$ & Chiral $\rp{+} I = (\phi^{1}\,;\, \zeta_{+}^{2'})$,\ \ Chiral $\rp{+} J = (\phi^{2}{}^\dag\,;\, \zeta_{+}^{1'}{}^\dag)$ \tabularnewline\hline
      Fermi & $\lambda_{-}^{\dt\alpha'}$ & Fermi $\rp-\Lambda_I = (\lambda_-^{\dt 2'})$\ ,\ Fermi $\rp-\Lambda_J = (\lambda_-^{\dt 1'}{}^\dag)$ \tabularnewline
      \bottomrule
    \end{tabular}
  \end{subtable}\vspace{20pt}
  \begin{subtable}{\linewidth}\centering
    \caption{\textbf{D1-D5$_{(34)}$ strings} \\ $\wt{I}$, $\wt\Lambda_I$ transform in the $(\bm{k},\w{\bm{n}}\bm{'})$ of $\text{U}(k) \times \text{U}(n')$ while $\wt{J}$, $\wt\Lambda_J$ transforms in the $(\w{\bm{k}},\bm{n'})$.}\label{d1d534}
    \begin{tabular}{c|c|c}
      \toprule
      ${(0,4)}$ multiplet & Fields & ${(0,2)}$ multiplets   \tabularnewline
      \midrule
      Twisted hyper & $\wt\phi{}^{\,\alpha'}\ ;\ \wt\zeta_{+}^{\,\alpha}$ & Chiral $\rp+\wt{I} = (\wt\phi^{\,1'}\,;\, \wt\zeta_{+}^{\,2})$,\ \ Chiral $\rp+\wt{J} = (\wt\phi^{\,2'}{}^\dag\,;\, \wt\zeta_{+}^{\,1}{}^\dag)$ \tabularnewline\hline
      Fermi & $\wt\lambda_{-}^{\,\dt\alpha}$ & Fermi $\rp-\wt\Lambda_I = (\lambda_-^{\dt 2})$\ ,\ Fermi $\rp-\wt\Lambda_J = (\lambda_-^{\dt 1}{}^\dag)$ \tabularnewline
      \bottomrule
    \end{tabular}
  \end{subtable}\vspace{20pt}
  \begin{subtable}{\linewidth}\centering
    \caption{\textbf{D5$_{(12)}$-D5$_{(34)}$ strings} \\ $\Lambda$ transforms in the $(\bm{n},\w{\bm{n}}\bm{'})$  of $\text{U}(n) \times \text{U}(n')$.}\label{d5d5}
    \begin{tabular}{c|c|c}
      \toprule
      ${(0,4)}$ multiplet & Fields & ${(0,2)}$ multiplets  \tabularnewline
      \midrule
      Fermi & $\lambda_{-}$ & Fermi $\rp0\Lambda = (\lambda_-)$\tabularnewline
      \bottomrule
    \end{tabular}
  \end{subtable}
\caption{Various $\n{0,2}$ multiplets for the crossed instanton system.}\label{crossedfields}
\end{table}
\subsection{Tachyons and Fayet-Iliopoulos terms}
We are interested in generalising the above setup to one with a
constant $B$-field of the form \eqref{constB}. We have seen in the
analysis of the open string spectrum that there are tachyons of
mass-squared
\begin{equation}\label{crossedtach}
  \frac{1}{2\alpha'}|v_1 - v_2|\ ,\quad  \frac{1}{2\alpha'}|v_3 - v_4|\ ,
\end{equation}
in the D1-D5 and D1-D5$'$ spectra (the on-shell formula is
$m^2 = -k^2$). In our conventions these correspond to the fields $I$,
$\wt{I}$ for $(v_1 - v_2), (v_3 - v_4) > 0$ and $J$, $\wt{J}$ for
$(v_1 - v_2), (v_3 - v_4) < 0$. The system is no longer supersymmetric
about the original vacuum (where all the vacuum expectation values are
set to zero) due to the presence of tachyons. Can this supersymmetry
breaking be interpreted as an $F$-term or $D$-term breaking?

Let us study the simpler problem $k$ D1-branes along $\Int$ and $n$
D5-branes along $\Int \times \mbb{C}^2_A$. The low-energy effective
action for the D5-branes contains the following coupling to the
(pullback of the) 2-form RR gauge field $C_2$:
\begin{equation}
\frac{e_5}{2} \int_{\Int \times \mbb C^2_A}  C_{2} \wedge \tr\,(\mc{F} \wedge \mc{F})\ ,
\end{equation}
where $\mc{F} := 2\pi\alpha'(F - B)$ with $F$ the
$\text{U}(n)$ field strength on the stack of D5-branes and $B$ 
the (pullback of the) NSNS $B$-field. The charge quantum $e_5$ is
related to the D5-brane tension as $e_5 = T_5$ by virtue of its BPS
nature and is given by
\begin{equation}
  e_5 = \frac{1}{g_s \sqrt{\alpha'}\, (2\pi\sqrt{\alpha'}\,)^5}\ .
\end{equation}
Let us consider a situation where $C_2$ is supported along $\Int$, $F$
along $\mbb C^2_{A}$ and $B = 0$. Then, the above coupling becomes
\begin{equation}
\mc{S}_{\text{WZ},2} = e_1 k \int_{\Int} C_2\ ,\quad\text{with}\quad k = \frac{1}{8\pi^2}\int_{\mbb C^2_{A}} \tr\ F \wedge F\ .
\end{equation}
Here, $\displaystyle e_1 = (g_s 2\pi\alpha')^{-1}$ is the D1-brane
charge quantum. The coupling $k$ is the familiar instanton number of a
$\text{U}(n)$ instanton in $\mbb C^2_A$. The above form of coupling
implies that instantons of charge $k$ in the $\text{U}(n)$ gauge
theory on the D5-branes induce D1-brane charge $e_1k$ on the
worldvolume. This was first realised in \cite{DBinB}.

A constant $B$-field along $\mbb{C}^2_A$ does a similar job and
induces a D1-brane charge density
\begin{equation} 
  \mc{J}_1 = \frac{N e_1}{8\pi^2}\, B \wedge B =  \frac{N e_1}{8\pi^2}\ \frac{b_a b_b}{(2\pi\alpha')^2}\ \ud \text{Vol}(\mbb C^2_{A})\ ,
\end{equation}
The instability is qualitatively different for different ranges of the
$B$-field values \cite{SW3}. Let $\mbb{C}^2_A$ have the standard
orientation. When $v_a$ and $v_b$ have opposite signs, $\mc{J}_1$ is
negative and corresponds to induced anti D1-branes. The tachyon in the
NS sector then corresponds to the standard D1-$\w{\text{D1}}$
tachyon. The condensation of this tachyon results in the annihilation
of part of the $\w{\text{D1}}$ charge density and results in an
excited state of the D5-brane with excitation energy proportional to
the tachyon mass $m^2 = \frac{1}{2\alpha'}|v_a - v_b|$.

When $v_a$ and $v_b$ have the same sign, the charge density is
positive and corresponds to induced D1-branes. For $v_a \neq v_b$,
tachyon condensation corresponds to the external D1-brane dissolving
into the D5-brane and forming a bound state with the induced D1-branes
(the \emph{Higgs branch} of the D1-D5 system). The point with
$v_a = v_b \neq 0$ corresponds to a self-dual $B$-field in which case
the tachyon disappears and the D1-D5 system forms a bound state at
threshold.

In either of these scenarios, one can describe these tachyon masses as
arising from FI terms in the low energy effective action, at least for
small values of $v_a - v_b$.

In the present situation of crossed instantons, Fayet-Iliopoulos terms
arise as vacuum expectation values of auxiliary fields in the adjoint
representation of $\text{U}(k)$. We have one real auxiliary field $D$
and one complex auxiliary field $G_2$ in the $\n{0,4}$ vector
multiplet, two complex auxiliary fields $G_3$ and $G_4$ from the
$\n{0,4}$ Fermi multiplets $\Lambda_3$ and $\Lambda_4$. The FI terms
then correspond to the following $J$-terms in the $\n{0,2}$ action:
\begin{align}
  \mc S_{\text{FI}} &= -\frac{1}{\sqrt{2}}\text{Im}\int \ud^2x\,\uD_+\tr\left\{ -\sqrt{2} t \mc{F}_- + b_2 \Lambda_2 + b_3 \Lambda_3 + b_4 \Lambda_4\right\}\ ,\nonumber\\
                    &= \int \ud^2x\, \tr\left\{\frac{\theta}{2\pi}v_{01} + r D + \text{Re}(b_2 G_2 + b_3 G_3 + b_4 G_4)\right\}\ .
\end{align}
$\displaystyle t = \frac{\theta}{2\pi} + ir$ is the complexified
Fayet-Iliopoulos parameter where $\theta$ is the two dimensional
$\theta$-angle and $r$ is the real FI parameter. The components of the
field strength Fermi multiplet $\mc{F}_-$ are given by
\begin{equation}
\lambda^{11'}_- := -(\mc{F}_-)_\bvert\ ,\quad  D + iv_{01} := \left(\nabla_+\mc{F}_-\right)_\bvert\ .
\end{equation}
From the $\text{SO}(4) \times \text{SO}(4)'$ properties of the Fermi
multiplets in table \ref{d1d1}, it is easy to see that all FI terms
except $r$ break the $\text{SO}(2)^4$ rotational symmetry that is
preserved by the $B$-field in \eqref{constB}. Hence, only a non-zero
$r$ could possibly account for the effect of such a $B$-field. The
terms in the action involving $D$ are
\begin{equation}
\tr_k \left( \frac{1}{2g^2}D^2 - \sum_{a \in \four} [B_a, B_a^\dag] D - I D I^\dag + J^\dag D J - \wt{I} D \wt{I}^\dag + \wt{J}^\dag D \wt{J} + rD\right)\ ,
\end{equation}
which gives the field equation
\begin{equation}\label{Dfeq}
  \frac{1}{g^2}D = \sum_{a \in \four} [B_a, B_a^\dag] + I I^\dag - J^\dag J + \wt{I}\, \wt{I}^\dag - \wt{J}^\dag \wt{J} - r \cdot \mbbm 1_k\ .
\end{equation}
The contribution to the Lagrangian from the $D$-terms is
$-\frac{1}{2g^2}\tr_k D^2$ where $D$ substituted with its field
equation. There are various quartic interaction terms along with the
following mass terms for $I$, $\wt{I}$, $J$ and $\wt{J}$:
\begin{equation}
  -\frac{g^2}{2}\tr_k\left(-r I I^\dag - r \wt{I}\,\wt{I}^\dag + r J^\dag J + r \wt{J}^\dag \wt{J}\right)\ .
\end{equation}
As we can see, the mass-squared of $I$ and $\wt{I}$ are equal to
$-\frac{g^2}{2}r$ and those of $J$ and $\wt{J}$ are equal to
$+\frac{g^2}{2}r$. For small $B$-field, these answers could also be
obtained via three-point disk amplitudes \emph{a la} \cite{HK} with
two open string vertices (e.g.~$I$, $I^\dag$) and one closed string
vertex corresponding to $B_{\mu\nu}$. The computation involves
contributions from the boundary of the moduli space of the disk with
three punctures (two punctures on the boundary and one in the
bulk)\footnote{We thank Ashoke Sen for a discussion on this point.}
and will be reported in a forthcoming paper \cite{NP}.

Comparing this with \eqref{crossedtach}, we see
that the $B$-fields must be related to each other and $r$ as
\begin{equation}
  \frac{1}{2\pi g_s\alpha'{}^2} (v_1 - v_2) = \frac{1}{2\pi g_s\alpha'{}^2} (v_3 - v_4) = -r\ .
\end{equation}
Here, we have used that the coupling constant $g^2$ is given in terms
of $\alpha'$ and the closed string coupling $g_s$ as
$g^2 = 2\pi g_s \alpha'$. Thus, for the low-energy effective action to
be supersymmetric, the constant $B$-field must satisfy
\begin{equation}
v_1 - v_2 = v_3 - v_4\ .
\end{equation}
We restrict our attention to constant $B$-field backgrounds satisfying
the above constraint since these are amenable to an analysis using
standard ideas of $D$-type supersymmetry breaking.

\subsection{Yukawa couplings}
So far, we have determined the minimally coupled kinetic terms and the
masses coming from $D$-term interactions in the low energy effective
theory. The remaining terms describing the dynamics are the $E$-terms
and $J$-terms for the various Fermi multiplets. A simple way to obtain
these is to look at the Yukawa couplings in the theory. Recall from
Appendix \ref{02app} that Yukawa terms for a Fermi superfield $\Psi$
are of the general form
\begin{equation}\label{yukcross}
\mc{E}_\Psi = +\w\psi{}^a_{-}\, \frac{\partial E_a}{\partial \phi_j}\,\zeta_{j+}\ ,\quad\text{and}\quad \mc{J}^\Psi = -\frac{\partial J^a}{\partial \phi_j}\,\zeta_{j+}\,\psi_{a-}\ .
\end{equation}
We obtain these terms in the low-energy effective action by computing
3-point string amplitudes on the disk. The idea is to look for
non-zero amplitudes that involve only fields in the chiral multiplets
but not their complex conjugates i.e.~the fields in the chiral
multiplets displayed in Table \ref{crossedfields}.

A general open string vertex operator in a constant $B$-field
background has the form
\begin{equation}
  V_\lambda(k,z) = \omega(\lambda)\, c(z)\,\mc{B}(z)\, \nmo{e^{\lambda\cdot H(z)}\,e^{2ik\cdot X(z)}c_{\lambda}}\ .
\end{equation} 
Here, $\lambda$ is a weight in the covariant lattice
$D_2 \oplus D_2 \oplus \Gamma_{1,1}$ corresponding to the spacetime
symmetry $\text{SO}(1,1) \times SO(4) \times SO(4')$ and $c_\lambda$
is the associated cocycle operator. $\mc{B}(z)$ is the appropriate
product of boundary condition changing operators for the worldsheet
bosons. The weights for the various fields and the boundary condition
changing operators for the worldsheet bosons are derived in Appendix
\ref{bfield} and summarised in Tables \ref{d1vo} and \ref{d1d5vo}.

\begin{table}\small\centering
  \renewcommand{\arraystretch}{1.25}
  \caption{Covariant weights for the vertex operators arising from
    D1-D1 strings. In our conventions, a
    left-handed spinor $\psi^\alpha$ of $\text{SO}(4)$ is specified by
    the weights $\psi^{\alpha=1} = (-,+)$, $\psi^{\alpha=2} = (+,-)$
    and a right-handed spinor $\psi^{\dt\alpha}$ by
    $\psi^{\dt\alpha=1} = (-,-)$, $\psi^{\dt\alpha=2} =
    (+,+)$.} \label{d1vo}
    \begin{tabular}{c|c|c|c}
    \toprule
    State & Field & $\text{U}(1)_\ell$ & $ D_2 \oplus D_2 \oplus \Gamma_{1,1} $ weight   \tabularnewline
    \midrule
    D1-D1 vector & $v_{\pm\pm}$ & 0 & $ (0,0) \oplus (0,0) \oplus (\mp 1; -1)$ \tabularnewline\midrule
    D1-D1 scalars & $X^{1\dt 1}$, $B_1$ & 0 & $(-1,0) \oplus (0,0)\oplus (0;-1)$ \tabularnewline                       
    \phantom{Scalars} & $X^{1\dt 2}$, $B_2$ & 1 &  $ (0,1) \oplus (0,0)\oplus (0;-1)$ \tabularnewline                       
    \phantom{Scalars} & $X^{1'\dt 1'}$, $B_3$ & 0 & $ (0,0) \oplus (-1,0)\oplus (0;-1)$ \tabularnewline
    \phantom{Scalars} & $X^{1'\dt 2'}$, $B_4$ & 1 & $ (0,0) \oplus (0,1)\oplus (0;-1)$ \tabularnewline\midrule
    D1-D1 gauginos & $\lambda_-^{11'}$, $f$ & 1 & $ (-,+) \oplus (-,+) \oplus (+;-)$\tabularnewline
    \phantom{Gauginos} & $\lambda_-^{12'}$, $\lambda_2$ & 0 & $ (-,+) \oplus (+,-) \oplus (+;-)$\tabularnewline 
    \phantom{Gauginos} & $\lambda_-^{\dt 1 \dt 1'}$, $\lambda_3$ & -1 & $ (-,-) \oplus (-,-) \oplus (+;-)$\tabularnewline 
    \phantom{Gauginos} & $\lambda_-^{\dt 1 \dt 2'}$, $\lambda_4$ & 0 & $ (-,-) \oplus (+,+) \oplus (+;-)$\tabularnewline 
    \phantom{Gauginos} & $\zeta_+^{\dt 1 2'}$, $\zeta_1$ & -1 & $ (-,-) \oplus (+,-) \oplus (-;-)$\tabularnewline 
    \phantom{Gauginos} & $\zeta_+^{\dt 2 2'}$, $\zeta_2$ & 0 & $ (+,+) \oplus (+,-) \oplus (-;-)$\tabularnewline 
    \phantom{Gauginos} & $\zeta_+^{2\dt 1'}$, $\zeta_3$ & -1 & $ (+,-) \oplus (-,-) \oplus (-;-) $\tabularnewline
    \phantom{Gauginos} & $\zeta_+^{2\dt 2'}$, $\zeta_4$ & 0 & $ (+,-) \oplus (+,+) \oplus (-;-) $\tabularnewline \bottomrule
    \end{tabular}
\end{table}

\begin{table}\small\centering
  \renewcommand{\arraystretch}{1.25}
  \caption{Covariant weights for D1-D5$_{(12)}$, D1-D5$_{(34)}$ and D5$_{(12)}$-D5$_{(34)}$ strings.}\label{d1d5vo}
  \begin{tabular}{c|c|c|c}
    \toprule
    State & Field & $\text{U}(1)_\ell$ & $ D_2 \oplus D_2 \oplus \Gamma_{1,1} $ weight   \tabularnewline
    \midrule
    D1-D5$_{(12)}$ bosons & $\phi^1$, $I$ & $\frac{1}{2}-v_2$ & $ (-v_1-\frac{1}{2},-v_2+\frac{1}{2}) \oplus (0, 0) \oplus (0;-1) $\tabularnewline
      & $\phi^2{}^\dag$, $J$ & $\frac{1}{2}+v_2$ & $ (v_1-\frac{1}{2},v_2+\frac{1}{2}) \oplus (0, 0) \oplus (0;-1) $\tabularnewline\midrule
    D1-D5$_{(12)}$ fermions & $\zeta_+^{1'}{}^\dag$, $\zeta_J$ & $-\frac{1}{2}+v_2$ & $ (v_1,v_2) \oplus (+,-) \oplus (-;-) $\tabularnewline
     & $\zeta_+^{2'}$, $\zeta_I$ & $-\frac{1}{2} -v_2$ & $(-v_1,-v_2) \oplus (+,-) \oplus (-;-) $\tabularnewline
     & $\lambda_-^{\dt 1'}{}^\dag$, $\lambda_J$ & $\frac{1}{2}+v_2$ & $ (v_1,v_2) \oplus (+,+) \oplus (+;-) $\tabularnewline
     & $\lambda_-^{\dt 2'}$, $\lambda_I$ & $\frac{1}{2} -v_2$ & $ (-v_1,-v_2) \oplus (+,+) \oplus (+;-) $\tabularnewline \midrule
    D1-D5$_{(34)}$ bosons & $\wt\phi^{1'}$, $\wt{I}$ & $\frac{1}{2}-v_4$ & $ (0, 0) \oplus (-v_3-\frac{1}{2},-v_4+\frac{1}{2}) \oplus (0;-1) $\tabularnewline
      & $\wt\phi^{2'}{}^\dag$, $\wt{J}$ & $\frac{1}{2}+v_4$ & $ (0, 0)\oplus(v_3-\frac{1}{2},v_4+\frac{1}{2})\oplus(0;-1) $\tabularnewline\midrule
    D1-D5$_{(34)}$ fermions & $\wt\zeta_+^{1}{}^\dag$, $\wt\zeta_J$ & $-\frac{1}{2}+v_4$ & $ (+,-) \oplus (v_3,v_4) \oplus (-;-) $\tabularnewline
     & $\wt\zeta_+^{2}$, $\wt\zeta_I$ & $-\frac{1}{2}-v_4$ & $(+,-) \oplus (-v_3,-v_4) \oplus (-;-) $\tabularnewline
     & $\wt\lambda_-^{\dt 1}{}^\dag$, $\wt\lambda_J$ & $\frac{1}{2}+v_4$ & $(+,+) \oplus (v_3,v_4) \oplus  (+;-) $\tabularnewline
    & $\wt\lambda_-^{\dt 2}$, $\wt\lambda_I$ & $\frac{1}{2}-v_4$ & $(+,+) \oplus (-v_3,-v_4) \oplus (+;-) $\tabularnewline\midrule    
    D5$_{(12)}$-D5$_{(34)}$ fermions & $\lambda_-$, $\lambda$ & $v_2-v_4$ & $(v_1,v_2)\oplus(-v_3,-v_4)\oplus(+;-)$\tabularnewline
  \bottomrule
  \end{tabular}
\end{table}

The rest of the notation is quite standard: $c(z)$ is the coordinate
ghost, $H(z)$ is a 6-dimensional vector containing the five bosons
that bosonise the ten worldsheet fermions and the sixth boson being
the one that bosonises the superconformal ghosts, $k = (k^0, k^9)$ is
the $1+1$ dimensional momentum and $X = (X^0, X^9)$ are the worldsheet
bosons corresponding to the $1+1$ dimensional
intersection. $\omega(\lambda)$ is an \emph{a priori} undetermined
$c$-number phase.

The general structure of a 3-pt function with open string vertex
operators in the canonical ghost picture is given by
\begin{align}
\langle V_{\lambda_1}(k_1,x_1) V_{\lambda_2}(k_2,x_2)V_{\lambda_3}(k_3,x_3)\rangle\nonumber & = \omega(\lambda_1)\omega(\lambda_2)\omega(\lambda_3)\, \langle \mc{B}_1(x_1)\mc{B}_2(x_2)\mc{B}_3(x_3)\rangle\times \nonumber\\ &\quad \times\langle e^{\lambda_1\cdot H(x_1)}c_{\lambda_1}\, e^{\lambda_2\cdot H(x_2)}c_{\lambda_2}\, e^{\lambda_3\cdot H(x_3)}c_{\lambda_3}\rangle\nonumber\\ &\quad \times \langle c(x_1) c(x_2) c(x_3)\rangle\, \langle e^{ik_1\cdot X}(x_1) e^{ik_2\cdot X}(x_2) e^{ik_3\cdot X}(x_3)\rangle\nonumber\\
&= \omega(\lambda_1)\omega(\lambda_2)\omega(\lambda_3)\times\langle \mc{B}_1(x_1)\mc{B}_2(x_2)\mc{B}_3(x_3)\rangle\times\nonumber\\ &\quad \times\prod_{i < j} e^{i\pi\lambda_i\cdot \cycm\cdot \lambda_j} (x_i - x_j)^{1 + \lambda_i\cdot\lambda_j + 2\alpha' k_i\cdot k_j}\ .
\end{align}
A few comments are in order:
\begin{enumerate}
\item The phase prefactor $\prod_{i < j} e^{i\pi\lambda_i\cdot M\cdot\lambda_j}$
  in the last expression is due to the cocycle operators
  $c_{\lambda_i}$ commuting across the vertex operators
  $e^{\lambda_j\cdot H}$. Here, $M$ is a $6\times 6$ matrix whose form
  is given in Appendix \ref{bfield}. These phases are crucial for
  obtaining the correct low-energy Yukawa couplings.

\item For the case of crossed instantons, all the $E$-terms and
  $J$-terms turn out to be quadratic in the superfields. Looking at
  \eqref{yukcross}, it is easy to see that there will be two different
  amplitudes that arise from the same $E$- or $J$-term. We get
  relations between the phases $\omega(\lambda)$ by equating the
  coefficients of these two amplitudes.

\item The correlators are non-zero only when the spacetime momenta add
  up to zero, the $D_2 \oplus D_2 \oplus \Gamma_{1,1}$ weights add up
  to $(0,0,0,0,0;-2)$ with the first five entries signifying
  $\text{SO}(4)\times\text{SO}(4)'\times\text{SO}(1,1)$ invariance and
  the $-2$ indicating that the superconformal anomaly on the disk is
  soaked up.

\item When the correlators are non-zero, it can be shown that the
  different contributions to the exponent of $x_i - x_j$ coming from
  the coordinate ghosts, the BCC operators for the worldsheet bosons,
  the vertex operators for the worldsheet fermions and the vertex
  operators for the $\Int$ directions all add up to zero. This shows
  that the correlator is independent of the points of insertion of the
  vertex operators as it should be due to $\text{SL}(2,\mbb{R})$
  invariance.
\end{enumerate}

The $E$-term and $J$-term Yukawa couplings for the various Fermi
multiplets are as follows:
\underline{\textbf{The D1-D1 Fermi
    multiplets $\rp0\Lambda_2$, $\rp=\Lambda_3$, $\rp0\Lambda_4$}}
\begin{align}
\mc{J}^{\Lambda_2} &= \omega(B_3)\omega(\zeta_4)\omega(\lambda_2)\ e^{-\frac{\pi i}{4}}\ \tr_k\ ([B_3, \zeta_4]\lambda_2 + [\zeta_3, B_4]\lambda_2)\nonumber\\ &\quad +\omega(\wt{I})\omega(\wt\zeta_J)\omega(\lambda_2)\ e^{\frac{i\pi}{4}(-1-2v_3)}\ \tr_k\ (\wt{I}\,\wt\zeta_J\lambda_2 + \wt\zeta_I\wt{J}\lambda_2)\ ,\nonumber\\
\mc{E}_{\Lambda_2} &= \omega(\w\lambda_2) \omega(B_1)\omega(\zeta_2)\ i\ \tr_k\ (\w\lambda_2 [B_1, \zeta_2] + \w\lambda_2[\zeta_1, B_2])\nonumber\\ &\quad + \omega(\w\lambda_2)\omega(\zeta_I)\omega(J)\ e^{\frac{i\pi}{4}(-3+6v_1)}\ \tr_k\ (\w\lambda_2\zeta_I J + \w\lambda_2 I \zeta_J)\ .
\end{align}
\begin{align}
\mc{J}^{\Lambda_3} &= \omega(B_2)\omega(\zeta_4)\omega(\lambda_3)\ e^{-\frac{3\pi i}{4}}\ \tr_k\ ([B_2, \zeta_4]\lambda_3 + [\zeta_2, B_4]\lambda_3)\ ,\nonumber\\
\mc{E}_{\Lambda_3} &= \omega(\w\lambda_3) \omega(B_1)\omega(\zeta_3)\ i\ \tr_k\ (\w\lambda_3[B_1, \zeta_3] + \w\lambda_3 [\zeta_1, B_3])\ .
\end{align}
\begin{align}
\mc{J}^{\Lambda_4} &= \omega(B_2)\omega(\zeta_3)\omega(\lambda_4)\ e^{\frac{\pi i}{4}}\ \tr_k\ ([B_2, \zeta_3]\lambda_4 + [\zeta_2, B_3]\lambda_4)\ ,\nonumber\\
\mc{E}_{\Lambda_4} &= \omega(\w\lambda_4) \omega(B_1)\omega(\zeta_4)\ i\ \tr_k\ (\w\lambda_4[B_1, \zeta_4] + \w\lambda_4 [\zeta_1, B_4])\ .
\end{align}

Relations:
\begin{align}
 &\frac{\omega(B_3)\omega(\zeta_4)}{\omega(\zeta_3)\omega(B_4)} = \frac{\omega(B_1)\omega(\zeta_2)}{\omega(\zeta_1)\omega(B_2)} = i \ ,\ \  \frac{\omega(I)\omega(\zeta_J)}{\omega(\zeta_I)\omega(J)} = \frac{\omega(\wt{I})\omega(\wt\zeta_J)}{\omega(\wt\zeta_I)\omega(\wt{J})}  = e^{-\frac{i\pi}{4}}\ ,\nonumber\\
 &\frac{\omega(B_2)\omega(\zeta_4)}{\omega(\zeta_2)\omega(B_4)} = \frac{\omega(B_1)\omega(\zeta_3)}{\omega(\zeta_1)\omega(B_3)} = 1\ ,\ \ \frac{\omega(B_2)\omega(\zeta_3)}{\omega(\zeta_2)\omega(B_3)} = \frac{\omega(B_4)\omega(\zeta_1)}{\omega(\zeta_4)\omega(B_1)} = -i\ .
\end{align}

\underline{\textbf{The D1-D5$_{(12)}$ Fermi multiplets $\rp-\!\Lambda_I$, $\rp-\!\Lambda_J$}}
\begin{align}
\mc{J}^{\Lambda_J} &= \omega(B_3)\omega(\zeta_I)\omega(\lambda_J)\ i\ \tr_k (B_3\zeta_I\lambda_J + \zeta_3 I \lambda_J)\ ,\nonumber\\
\mc{E}_{\Lambda_J} &= \omega(\w\lambda_J)\omega(\zeta_J)\omega(B_4)\ e^{\frac{i\pi}{4}(4+8v_1)}\ \tr_k (\w\lambda_J \zeta_J B_4 + \w\lambda_J J \zeta_4)\ .
\end{align}
\begin{align}
\mc{J}^{\Lambda_I} &= \omega(\zeta_J)\omega(B_3)\omega(\lambda_I)\ e^{\frac{i\pi}{4}(-6+8v_1+8v_2)}\ \tr_k (\lambda_I \zeta_J B_3 + \lambda_I J \zeta_3  )\ ,\nonumber\\
\mc{E}_{\Lambda_I} &= \omega(\w\lambda_I)\omega(B_4)\omega(\zeta_I)\ e^{\frac{i\pi}{4}(8-8v_2)}\ \tr_k (B_4 \zeta_I \w\lambda_I + \zeta_4 I \w\lambda_I)\ .
\end{align}

Relations:
\begin{align}
 &\frac{\omega(I)\omega(\zeta_3)}{\omega(\zeta_I)\omega(B_3)} = e^{\frac{i\pi}{4}(-1 + 2v_1)}\ ,\ \ \frac{\omega(J)\omega(\zeta_4)}{\omega(\zeta_J)\omega(B_4)} = e^{\frac{i\pi}{4}(2 + 2v_1)}\ ,\nonumber\\
 &\frac{\omega(I)\omega(\zeta_4)}{\omega(\zeta_I)\omega(B_4)} = e^{\frac{i\pi}{4}(1 + 2v_1)}\ ,\ \ \frac{\omega(J)\omega(\zeta_3)}{\omega(\zeta_J)\omega(B_3)} = e^{\frac{i\pi}{4}(2v_1)}\ .
\end{align}

\underline{\textbf{The D1-D5$_{(34)}$ Fermi multiplets $\rp-\wt\Lambda_I$, $\rp-\wt\Lambda_J$}}
\begin{align}
\mc{J}^{\wt\Lambda_J} &= \omega(B_1)\omega(\wt\zeta_I)\omega(\wt\lambda_J)\ e^{\frac{i\pi}{4}(-6-4v_3-4v_4)}\ \tr_k (B_1\wt\zeta_I\wt\lambda_J + \zeta_1 \wt{I}\, \wt\lambda_J)\ ,\nonumber\\
\mc{E}_{\wt\Lambda_J} &= \omega(\w{\wt\lambda}_J)\omega(\wt\zeta_J)\omega(B_2)\ e^{\frac{i\pi}{4}(4v_3+4v_4)}\ \tr_k (\w{\wt\lambda}_J \wt\zeta_J B_2 + \w{\wt\lambda}_J \wt{J} \zeta_2)\ .
\end{align}
\begin{align}
\mc{J}^{\wt\Lambda_I} &= \omega(\wt\zeta_J)\omega(B_1)\omega(\wt\lambda_I)\ e^{\frac{i\pi}{4}(-6+4v_3+4v_4)}\ \tr_k (\wt\lambda_I \wt\zeta_J B_1 + \wt\lambda_I \wt{J} \zeta_1)\ ,\nonumber\\
\mc{E}_{\wt\Lambda_I} &= \omega(\w{\wt\lambda}_I)\omega(B_2)\omega(\wt\zeta_I)\ e^{\frac{i\pi}{4}(12-4v_3-4v_4)}\ \tr_k (B_2 \wt\zeta_I \w{\wt\lambda}_I + \zeta_2 \wt{I}\, \w{\wt\lambda}_I)\ .
\end{align}

Relations:
\begin{align}
 &\frac{\omega(\wt{I})\omega(\zeta_1)}{\omega(\wt\zeta_I)\omega(B_1)} = e^{\frac{i\pi}{4}(-1 + 2v_3)}\ ,\ \ \frac{\omega(\wt{J})\omega(\zeta_2)}{\omega(\wt\zeta_J)\omega(B_2)} = e^{\frac{i\pi}{4}(2 + 2v_3)}\ ,\nonumber\\
 &\frac{\omega(\wt{I})\omega(\zeta_2)}{\omega(\wt\zeta_I)\omega(B_2)} = e^{\frac{i\pi}{4}(1 + 2v_3)}\ ,\ \ \frac{\omega(\wt{J})\omega(\zeta_1)}{\omega(\wt\zeta_J)\omega(B_1)} = e^{\frac{i\pi}{4}(2v_3)}\ .
\end{align}

\underline{\textbf{The D5$_{(12)}$-D5$_{(34)}$ Fermi multiplet $\rp0\!\Lambda$}}
\begin{align}
\mc{J}^{\Lambda} &= \omega(\wt\zeta_J)\omega(I)\omega(\lambda)\ e^{\frac{i\pi}{4}(-2-2v_1-4v_2+4v_3+4v_4-4v_1v_2)}\ \tr_{k} ( I \lambda \wt\zeta_J + \zeta_I \lambda \wt{J})\ ,\nonumber\\
\mc{E}_{\Lambda} &= \omega(\w\lambda)\omega(J)\omega(\wt\zeta_I)\ e^{\frac{i\pi}{4}(2+6v_1+4v_2-4v_3-4v_4+4v_1v_2)}\ \tr_{k} (\wt\zeta_I \w\lambda J  +  \wt{I} \w\lambda \zeta_J)\ .
\end{align}

Relations:
\begin{align}
 &\frac{\omega(I)\omega(\wt\zeta_J)}{\omega(\zeta_I)\omega(\wt{J})} = e^{\frac{i\pi}{4}(-1 + 2v_1 - 2v_3)}\ ,\ \ \frac{\omega(J)\omega(\wt\zeta_I)}{\omega(\zeta_J)\omega(\wt{I})} = e^{\frac{i\pi}{4}(1+2v_1 -2v_3)}\ .
\end{align}

The above couplings can be massaged into a nicer form by using the
relations between the phases and choosing a convenient value for the
independent phases. In other words, we redefine the fields by
absorbing the independent phases into the corresponding fields. From
the relations, it is easy to see that there is only one independent
ratio of the form $\frac{\omega(\phi)}{\omega(\zeta_\phi)}$ where
$(\phi,\zeta_\phi)$ form a $\n{0,2}$ chiral superfield. The phases
corresponding to the Fermi superfields are not determined from the
above relations. Setting $\frac{\omega(B_1)}{\omega(\zeta_1)} = \w{c}$,
we have
\begin{align}
&\frac{\omega(B_1)}{\omega(\zeta_1)} = \frac{\omega(B_3)}{\omega(\zeta_3)} = \w{c}\ ,\quad \frac{\omega(B_2)}{\omega(\zeta_2)} = \frac{\omega(B_4)}{\omega(\zeta_4)} = -i\w{c}\ ,\nonumber\\ \frac{\omega(I)}{\omega(\zeta_I)} = e^{\frac{i\pi}{4}(-1+2v_1)} & \w{c}\ ,\quad  \frac{\omega(J)}{\omega(\zeta_J)} = e^{\frac{i\pi}{4}(2v_1)}\w{c}\ ,\quad
\frac{\omega(\wt{I})}{\omega(\wt\zeta_I)} = e^{\frac{i\pi}{4}(-1+2v_3)}\w{c}\ ,\quad  \frac{\omega(\wt{J})}{\omega(\wt\zeta_J)} = e^{\frac{i\pi}{4}(2v_3)}\w{c}\ .
\end{align}
We now make a particular choice for the independent phases:
\begin{align}
&c = e^{\frac{7i\pi}{8}}\ ,\ \omega(B_a) = 1\ ,\ \omega(I) = \omega(J) = e^{-\frac{i\pi}{4}(1 + 2v_1)}\ ,\ \omega(\wt{I}) = \omega(\wt{J}) = e^{\frac{i\pi}{4}(1 + 2v_3)}\ ,\nonumber\\
&\omega(\lambda_2) = \omega(\lambda_4) = e^{-\frac{i\pi}{8}}\ ,\ \omega(\lambda_3) = e^{\frac{3i\pi}{8}}\ ,\ \omega(\lambda_J) = e^{-\frac{3i\pi}{8}} e^{i\pi v_1}\ ,\ \omega(\lambda_I) = e^{-\frac{i\pi}{8}} e^{-i\pi(v_1 + 2v_2)}\ ,\nonumber\\
&\omega(\wt\lambda_J) = e^{\frac{i\pi}{8}} e^{i\pi (v_3 + v_4)}\ ,\quad \omega(\wt\lambda_I) = e^{\frac{3i\pi}{8}} e^{-i\pi(v_3 + v_4)}\ ,\ \omega(\lambda) = e^{-\frac{3i\pi}{8}} e^{i\pi(v_1 + v_2 - v_3 - v_4 + v_1v_2)}\ .
\end{align}
With the above choice of phases, the $J$- and $E$-terms are given by
\begin{align}
J^{\Lambda_2} &= [B_3, B_4] + \wt{I}\wt{J}\ ,\quad E_{\Lambda_2} = [B_1, B_2] + I J\ ,\nonumber\\ 
J^{\Lambda_3} &= [B_2, B_4]\ ,\quad E_{\Lambda_3} = -[B_1, B_3]\ ,\quad J^{\Lambda_4} = [B_2, B_3]\ ,\quad E_{\Lambda_4} = [B_1, B_4]\ ,\nonumber\\
J^{\Lambda_J} &= B_3 I\ ,\quad E_{\Lambda_J} = J B_4\ ,\quad J^{\Lambda_I} = - J B_3\ ,\quad E_{\Lambda_I} = B_4 I\ ,\nonumber\\
J^{\wt\Lambda_J} &= - B_1 \wt{I}\ ,\quad E_{\wt\Lambda_J} = - \wt{J} B_2\ ,\quad J^{\wt\Lambda_I} = \wt{J} B_1\ ,\quad E_{\wt\Lambda_I} = -B_2 \wt{I}\ ,\nonumber\\
J^{\Lambda} &= \wt{J} I\ ,\quad E_{\Lambda} = - J \wt{I}  \ .
\end{align}

\underline{\textbf{The identity $\tr_k\ J\cdot E = 0$:}}\\
We have
\begin{align}
&\tr_k\ \big\{([B_3, B_4] + \wt{I}\wt{J})([B_1, B_2] + I J) - [B_2, B_4][B_1, B_3] + [B_2, B_3][B_1, B_4] + \nonumber\\
&\qquad\ + B_3 I J B_4 - B_4 I J B_3 + B_1 \wt{I} \wt{J} B_2 - B_2 \wt{I} \wt{J} B_1 - I J \wt{I} \wt{J}\big\} = 0\ .
\end{align}
Thus, $\tr_k\ J\cdot E = 0$ is indeed satisfied and the low-energy
effective action is indeed $\n{0,2}$ supersymmetric. Since the action
is also covariant with respect to the
$\text{SU}(2)_L \times \text{SU}(2)'_L$ R-symmetry, it is $\n{0,4}$
supersymmetric as well. This is the same result that is obtained in
\cite{T} for the case of zero $B$-field.

\subsection{The crossed instanton moduli space}
The bosonic potential energy $U$ is
\begin{equation}
U = \frac{g^2}{2} \tr\,D^2  + \sum_a |E_a|^2 + \sum_a |J^a|^2\ ,
\end{equation}
with the auxiliary field $D$ substituted with its field equation in
\eqref{Dfeq}. The minima of the potential can be obtained by solving
the equations $D = 0$, $E_a = 0$ and $J^a = 0$. We relabel
$I, J \to I_{12}, J_{12}$ and $\wt{I}, \wt{J} \to I_{34}, J_{34}$ in
anticipation of the spiked instanton scenario. The vacuum moduli space
is then defined by the following equations upto a $\text{U}(k)$ gauge
transformation:
\begin{align}
  &\text{\underline{$D$-term:}}\\
  &\mu^{\mbb{R}} - r\cdot \mbbm 1_k = \sum_{a=1}^{4} [B_a, B_a^\dag] + I_{12} I_{12}^\dag - J_{12}^\dag J_{12} +  I_{34} I_{34}^\dag -  J_{34}^\dag J_{34} - r\cdot\mbbm 1_k = 0\ .\nonumber
\end{align}
\begin{align}
  &\text{\underline{$J$-terms:}}\\
  &\mu^{\mbb{C}}_{34} = [B_3, B_4] + I_{34} J_{34} = 0\ ,\quad \mu^{\mbb{C}}_{24} = [B_2, B_4] = 0\ ,\quad \mu^{\mbb{C}}_{23} = [B_2, B_3]= 0\ ,\nonumber\\
  &\sigma^{\mbb{C}}_{3,12} = B_3 I_{12} = 0\ ,\quad \tl\sigma^{\mbb{C}}_{3,12} = -J_{12} B_3 = 0\ ,\quad \sigma^{\mbb{C}}_{1,34} = -B_1 I_{34} = 0\ ,\nonumber\\
  &\tl\sigma^{\mbb{C}}_{1,34} = J_{34} B_1 = 0\ ,\quad \Upsilon^{\mbb{C}}_{12} = J_{34} I_{12} = 0 \ .\nonumber
\end{align}
\begin{align}
  &\text{\underline{$E$-terms:}}\\
  &\mu^{\mbb{C}}_{12} = [B_1, B_2] +  I_{12} J_{12} = 0\ ,\ \  \mu^{\mbb{C}}_{13} = -[B_1, B_3] = 0\ ,\ \  \mu^{\mbb{C}}_{14} = [B_1, B_4]= 0\ ,\nonumber\\
  &\sigma^{\mbb{C}}_{4,12} = B_4 I_{12} = 0\ ,\quad \tl\sigma^{\mbb{C}}_{4,12} = J_{12} B_4 = 0\ ,\quad \sigma^{\mbb{C}}_{2,34} = -B_2 I_{34} = 0\ ,\nonumber\\
  &\tl\sigma^{\mbb{C}}_{2,34} = -J_{34} B_2 = 0\ ,\quad \Upsilon^{\mbb{C}}_{34} = -J_{12} I_{34}  = 0\ .\nonumber
\end{align}

\textbf{Symmetries}\\ Note that the above equations are invariant
under $\text{U}(k) \times \text{U}(n) \times \text{U}(n')$
transformations. The crossed instanton moduli space is then defined by
the solutions of the above equations modulo $\text{U}(k)$ gauge
transformations. The group
$\text{P}\left(\text{U}(n) \times \text{U}(n')\right) \cong
\frac{{U}(n) \times \text{U}(n')}{\text{U}(1)_c}$, where
$\text{U}(1)_c$ is the common centre of
$\text{U}(n) \times \text{U}(n')$, remains a global symmetry on the
moduli space. These are the \emph{framing rotations} described in
\cite{N4}.

There are additional symmetries from the
$\text{SU}(2)_{L} \times \text{SU}(2)_{ R} \times \text{SU}(2)'_{ L}
\times \text{SU}(2)'_{ R}$ arising from rotations of the transverse
$\mbb{R}^8$. To see how many of these symmetries are preserved by the
vacuum moduli space, we first form real combinations of the
holomorphic equations above:
\begin{align}\label{realcrossed}
  s_A &:= \mu^\mbb{C}_A + \varepsilon_{A\w{A}} \left(\mu^\mbb{C}_{\w{A}}\right)^\dag = 0\ ,\quad\text{for}\quad A \in \six\ ,\nonumber\\
  \sigma_{\w{a}A} &:= \sigma^\mbb{C}_{\w{a}A} + \varepsilon_{\w{a}\w{b}A}\left(\tl\sigma^\mbb{C}_{\w{b}A}\right)^\dag = 0\ ,\quad\text{for}\quad A \in \six\ ,\quad \w{a} \in \w{A}\ ,\nonumber\\
  \Upsilon_A &:= \Upsilon^\mbb{C}_A  - \varepsilon_{A\w{A}}\left(\Upsilon^\mbb{C}_{\w{A}}\right)^\dag = 0\quad\text{for}\quad A \in \six\ .
\end{align}
Using the $\text{SO}(4) \times \text{SO}(4)'$ transformation
properties of the fields in Table \ref{crossedfields} it is easy to
see that the equations with $r = 0$ preserve a diagonal subgroup
$\text{SU}(2)_\Delta$ of the R-symmetry
$\text{SU}(2)_{L} \times \text{SU}(2)'_{ L}$. The equations
$\mu^\mbb{R}$, $s_{12}$ and $s_{34}$ form a triplet and the other real
equations are invariant under $\text{SU}(2)_\Delta$.

For $r \neq 0$, the subgroup $\text{SU}(2)_\Delta$ is broken down to
its maximal torus $\text{U}(1)_\Delta$ which is the R-symmetry
$\text{U}(1)_\ell$ of the $\n{0,2}$ subalgebra that was chosen
above. The factors $\text{SU}(2)_{ R} \times \text{SU}(2)'_{ R}$
survive as spectator symmetries. Hence, the total global symmetry on
the crossed instanton moduli space is
\begin{equation}
  \text{P}\left(\text{U}(n) \times \text{U}(n')\right) \times \text{SU}(2)_R \times \text{SU}(2)'_R \times \text{U}(1)_\Delta\ .
\end{equation}

{\textbf{Note:}} The vacuum moduli space for $r = 0$ splits up into
many distinct branches corresponding to the Coulomb branch, the two
Higgs branches (with the D1's binding to either of the D5-branes) and
mixed branches \cite{T}. Once a non-zero $r$ is introduced, the
D1-branes bind necessarily to some stack of D5-branes and the moduli
space becomes connected. Turning on $r$ also has the effect of
reducing the global symmetries as we saw above. It would be
interesting to repeat the R-charge analysis of \cite{T} in this case
and explore the infrared limit of this $\n{0,4}$ gauge theory and its
$\n{0,2}$ spiked generalisation along the lines of \cite{SiWi1,
  SiWi2}.

\section{Spiked instantons}
Consider the crossed instanton setup of D1-D5$_{(12)}$-D5$_{(34)}$
branes. Let us choose the $B$-field such that $v_1 v_2 \leq 0$ and
$v_3 v_4 \leq 0$. This ensures that the tachyons are of
D1-$\w{\text{D1}}$ type. In this region of the space of $B$-fields,
the tachyon mass can never be zero unless the $v$'s are zero.

Let us introduce a stack of D5$_{(23)}$-branes to the mix. In order to
realise a symmetric situation where the instability here is also of
D1-$\w{\text{D1}}$ type, we need $v_2 v_3 \leq 0$. This implies that
$v_1 v_3 \geq 0$ and $v_2 v_4 \geq 0$. Suppose we next add the two
stacks of five branes along $\Int \times \mbb{C}^2_{(13)}$ and
$\Int \times \mbb{C}^2_{(13)}$. The constraints $v_1 v_3 \geq 0$ and
$v_2 v_4 \geq 0$ and the requirement that the tachyons should be
D1-$\w{\text{D1}}$ tachyons automatically force these stacks to be
made of anti D5-branes!  We thus have the following configuration of
D5-branes and anti D5-branes:
\begin{equation}\label{spikedbranes}
\text{D5}_{(12)}\ ,\ \text{D5}_{(34)}\ ,\ \text{D5}_{(23)}\ ,\ \text{D5}_{(14)}\ ,\ \w{\text{D5}}_{(13)}\ ,\ \w{\text{D5}}_{(24)}\ . 
\end{equation}
This is the same configuration of six stacks of D5-branes which
preserves two supercharges when the $B$-field is dialled to
zero. Though two the six stacks of D5-branes are composed of
antibranes, the D1-branes bind to the various stacks of D5-branes and
$\w{\text{D5}}$-branes in a symmetric fashion in that all the
tachyonic instabilities are of the brane-antibrane type.

One may again enquire as to whether an FI term in the low-energy
effective action can accommodate the effect of the constant $B$-field
of the form \eqref{constB}. The masses of the tachyons for the various
D1-D5 (and D1-$\w{\text{D5}}$) strings can be read off from the
derivation of the open string spectrum in Section \ref{constBstring}:
\begin{align}\label{spikedtach}
  &\frac{1}{2\alpha'}|v_1 - v_2|\ ,\quad \frac{1}{2\alpha'}|v_3 - v_4|\ ,\quad \frac{1}{2\alpha'}|v_2 - v_3|\ ,\nonumber\\ &\frac{1}{2\alpha'}|v_1 - v_4|\ ,\quad \frac{1}{2\alpha'}|v_1 + v_3|\ ,\quad \frac{1}{2\alpha'}|v_2 + v_4|\ .
\end{align}
Repeating the analysis in the crossed case, we see that the field
equation for the auxiliary field $D$ becomes
\begin{equation}
  D = \sum_{a \in \four} [B_a, B_a^\dag] + \sum_{A \in \six}(I_A I_A^\dag - J_A^\dag J_A) - r \cdot \mbbm 1_k\ .
\end{equation}
giving rise to the same mass-squared $|r|$ to all the tachyons. Thus,
the $B$-field values must satisfy
\begin{equation}\label{spikedconstrt}
\boxed{  v_1 = -v_2 = v_3 = -v_4}\ ,
\end{equation}
in order to be accounted for by the real FI parameter in the
low-energy theory.

The presence of the extra four stacks of D5-branes gives rise to
additional terms in the $E$-terms and $J$-terms for the Fermi
multiplets $\Lambda_3$ and $\Lambda_4$. There are also additional
Fermi multiplets from the open strings stretching between D1-branes
and these stacks of D5-branes and $\w{\text{D5}}$-branes. Repeating
the disk amplitude calculation as above, one get the following
equations:
\begin{enumerate}
\item The real moment map:
  \begin{equation}
   \mu_{\mbb{R}} - r\cdot\mbbm 1_k :=  \sum_{a \in \four} [B_a, B_a^\dag] + \sum_{A \in \six}(I_A I_A^\dag - J_A^\dag J_A) - r \cdot \mbbm 1_k = 0\ .
  \end{equation}

\item For $A = (ab) \in \six$ with $a < b$,
  \begin{equation}\label{muA}
    \mu^\mbb{C}_A := [B_a, B_b] + I_A J_A = 0\ .
  \end{equation}

\item For $A \in \six$, $\w{A} = \four \smallsetminus A$ and $\w{a} \in \w{A}$,
  \begin{equation}\label{sigmaA}
    \sigma^\mbb{C}_{\w{a}A} := B_{\w{a}}I_A = 0\ ,\quad \tl\sigma^\mbb{C}_{\w{a}A} := J_A B_{\w{a}} = 0\ .
  \end{equation}

\item For $A \in \six$, $\w{A} = \four \smallsetminus A$,
  \begin{equation}\label{upsilonA}
    \Upsilon^\mbb{C}_A := J_{\w{A}} I_A = 0\ .
  \end{equation}
\end{enumerate}

\textbf{Symmetries}\\
The symmetries of the above equations can be obtained in a similar way
to the crossed instanton case. The total global symmetry is given by
\begin{equation}
  \text{P}\left(\bigtimes_{A \in \six} \text{U}(n_A)\right) \times \text{U}(1)^3\ ,
\end{equation}
where $\text{U}(1)^3$ is a maximal torus of $\text{SU}(4)$, the
isometry group of the transverse $\mbb{C}^4$ which preserves some
fraction of supersymmetry. 

\subsection*{Folded branes}
The above equations arise from considering D1-D1 strings, D1-D5$_A$
strings and D5$_A$-D5$_{\w{A}}$ strings. There are also additional
equations that result from the interaction of D1-branes with states
from open strings stretching between D5$_A$ and D5$_B$ with $A = (ac)$
and $B = (bc)$ i.e.~two stacks of D5-branes that have a line
$\mbb{C}_c$ in common. This is the setup of \emph{folded branes} and
once we throw in D1-branes, the classical moduli space of vacua is
called the moduli space of \emph{folded instantons}.

The open string spectrum for this case was analysed in Section 2 and
there we saw that there were tachyons in the NS sector with mass
$\frac{1}{2}|v_a \pm v_b|$ for D5-D5 strings and D5-$\w{\text{D5}}$
strings respectively. Note that this is opposite to the masses one
obtains for D1-D5 and D1-$\w{\text{D5}}$ strings. Thus, for the
configuration of branes in \eqref{spikedbranes} it is easy to see that
the spectrum of tachyon masses is precisely the same as in
\eqref{spikedtach}. With the constraint in \eqref{spikedconstrt}, all
tachyons have the same mass-squared which is equal to
$\frac{1}{\alpha'}|v_1|$.

All the states arising from such strings are supported over the four
dimensional subspace $\Int \times \mbb{C}_c$ with a constant $B$-field
$\tan \pi v_c$ along $\mbb{C}_c$ which makes the space
non-commutative. It has been conjectured in \cite{N3, N4} that the
interaction of these states with the states supported on $\Int$ gives
rise to an additional (infinite) set of equations of the form
\begin{equation}
\Upsilon_{A,B,j} = J_A (B_c)^{j - 1} I_B = 0\ ,\quad\text{for}\quad j = 1,2,\ldots
\end{equation}
The stringy derivation of these equations requires a deeper analysis
of the non-commutative theory on $\Int \times \mbb{C}_c$ and also the
calculation of $(j+2)$-point disk amplitudes. These calculations shall
be reported in a forthcoming paper \cite{NP}. $\hfill\square$

\newpage
\appendix
\section{Open strings in a constant \tp{$B$}{B}-field}\label{bfield}
We follow the treatment of background gauge fields in
\cite{ACNY}. Consider a open string propagating in flat spacetime in
the presence of a constant $B$-field with components only along
spatial directions. The $\n{1,1}$ supersymmetric worldsheet theory is
formulated in terms of the superfield $X^\mu$ with components
\begin{equation}
  X^\mu := X^\mu{}_\bvert\ ,\quad \psi^\mu_\pm := (\uD_\pm X^\mu)_\bvert\ ,\quad F^\mu = (\uD_+\uD_- X^\mu)_\bvert\ ,
\end{equation}
where the ${}_\bvert$ sets all the Grassmann coordinates to zero. Our
conventions are such that $\uD_\pm^2 = i\partial_{\pm\pm}$,
$\{\uD_+, \uD_-\} = 0$ and
$\partial_{\pm\pm} = \frac{1}{2}(\partial_\tau
\pm \partial_\sigma)$. The supersymmetric action is given by
\begin{align}\label{11action}
  \mc S &= \frac{1}{\pi\alpha'}\int\ud\tau\ud\sigma\, \uD_+\uD_-\left\{\left(G_{\mu\nu} + 2\pi\alpha'B_{\mu\nu}\right)\uD_+ X^\mu \uD_- X^\nu\right\}\ ,\nonumber\\
        & = \frac{1}{\pi\alpha'}\int\ud\tau\ud\sigma \left(G_{\mu\nu} + 2\pi\alpha'B_{\mu\nu}\right)\left(\partial_{++} X^\mu \partial_{--} X^\nu - F^\mu F^\nu + i\psi_-^\nu\partial_{++}\psi_-^\mu + i\psi_+^\mu\partial_{--}\psi_+^\nu\right)\ ,\nonumber\\
        & = \frac{1}{\pi\alpha'}\int\ud\tau\ud\sigma\, G_{\mu\nu}\left(\partial_{++} X^\mu \partial_{--} X^\nu - F^\mu F^\nu + i\psi_-^\mu\partial_{++}\psi_-^\nu + i\psi_+^\mu\partial_{--}\psi_+^\nu\right) + \nonumber\\
        &\quad -\frac{1}{\pi\alpha'}\int\ud\tau\, 2\pi\alpha'B_{\mu\nu} \Big[{2}(\partial_\tau X^\mu)X^\nu + i\psi_-^\mu\psi_-^\nu + i\psi_+^\mu\psi_+^\nu\Big]_{\sigma= 0}^{\sigma = \pi} + \text{total $\tau$-derivative}\ .
\end{align}
The boundary conditions that result from the Euler-Lagrange variation
are
\begin{align}
  &(G_{\mu\nu}\partial_\sigma X^\mu + 2\pi\alpha'B_{\mu\nu}\, \partial_\tau X^\mu)\delta X^\nu\Big|_{\sigma=0} = 0\ ,\nonumber\\
  &\delta\psi_-^\mu(G_{\mu\nu} - 2\pi\alpha'B_{\mu\nu}) \psi_-^\nu\Big|_{\sigma=0} = \delta\psi_+^\mu(G_{\mu\nu} + 2\pi\alpha'B_{\mu\nu}) \psi_+^\nu\Big|_{\sigma=0}\ ,
\end{align}
and the same ones for $\sigma = \pi$ as well. We assume
that the metric $G_{\mu\nu}$ is the standard Minkowski metric and
choose a coordinate system such that the constant $B$-field is in
block diagonal form:
\begin{equation}
  {2\pi\alpha'}B = \pmat{0 & b_1 & & & \\ -b_1 & 0 & & & \\ & & 0 & b_2 & \\ & & -b_2 & 0 & \\ & & & & \ddots}\ .
\end{equation}
If the metric contains off-diagonal components, it is in general not
possible to cast the $B$-field in the above form since the metric and
$B$-field preserve different subgroups of $\text{GL}(1,9)$. In such a
coordinate system, the above analysis reduces to that of an open
string in $\mbb{R}^2$ with a constant $B$-field
$\frac{b}{2\pi\alpha'}\, \ud x^1 \wedge \ud x^2$. Let us first study
the boundary conditions on the worldsheet bosons. Writing
$b = \tan \pi v$ and $Z := \frac{1}{\sqrt{2}}(X^1 + iX^2)$ the
boundary conditions become
\begin{align}
  (\partial_\sigma Z + i2\pi\alpha' B\, \partial_\tau Z)\delta \w{Z} = 0\ .
\end{align}
Thus, we can have two types of boundary conditions at each end:
\begin{align}
\textbf{Dirichlet (D)}&:\quad \delta Z = 0\ \ \text{i.e.}\ \ Z = z_0 \in \mbb{C}\ ,\nonumber\\
\textbf{Twisted (T)}&:\quad \partial_\sigma Z + i2\pi\alpha' B\, \partial_\tau Z = 0\ .
\end{align}
The twisted boundary conditions can also be written as
$\partial_{++} Z = e^{-2\pi i v} \partial_{--} Z$ with
$2\pi\alpha' B = \tan\pi v$. Dirichlet boundary conditions are
realised also by letting $v \to \infty$. In order to accommodate all
types of boundary conditions at both ends, we introduce the more
general boundary conditions
\begin{align} \label{btwisted}
  \partial_{++} Z &= e^{-2\pi i \nu} \partial_{--} Z\ , \quad \textrm{at }\sigma = 0\ ,\nonumber\\
  \partial_{++} Z &= e^{-2\pi i\mu}\partial_{--} Z\ , \quad \textrm{at }\sigma = \pi\ .
\end{align}
The boundary conditions with $B$-field can be realised by taking
$\nu = v$, $\mu = \frac{1}{2}$ for the {\bf TD} case and
$\nu = \frac{1}{2}$, $\mu = v$ for the {\bf DT} case. The solution to
the $Z$ field equation consists of independent left-moving and
right-moving waves:
\begin{equation}
Z(\tau,\sigma) = Z_L(\tau + \sigma) + Z_R(\tau - \sigma)\ ,
\end{equation}
with the mode expansions
\begin{align}
Z_L &= \tfrac{1}{2}z_L + \frac{\ell^2}{2}p_L (\tau + \sigma) + \frac{\ell}{2}\sum_{k\neq 0}\frac{\alpha_{L,k}}{k}\,
      e^{-ik(\tau + \sigma)}\
  ,\nonumber\\
Z_R &= \tfrac{1}{2}z_R + \frac{\ell^2}{2} p_R (\tau - \sigma) + \frac{\ell}{2}\sum_{k\neq 0}\frac{\alpha_{R,k}}{k}\,
      e^{-ik(\tau - \sigma)}\ .
\end{align}
Here, $\ell$ is the string length. The boundary conditions relate the
modes in $Z_L$ and $Z_R$ as
\begin{align}\label{bc1}
p_L &= e^{-2\pi i\nu} p_R\ ,\quad \alpha_{L,k} = e^{-2\pi i\nu}\alpha_{R,k}\ ,\nonumber\\
p_L &= e^{-2\pi i\mu} p_R\ ,\quad \alpha_{L,k}\, e^{-ik\pi} = e^{-2\pi i\mu} e^{ik\pi}\,\alpha_{R,k}\ .
\end{align}
For $\nu \neq \mu$ we get $p_L = p_R = 0$ and
\begin{equation}
\quad e^{2\pi i(k - \mu + \nu)} = 1 \Longrightarrow k  \in \mbb{Z} + \mu - \nu \ .
\end{equation}
Let $z_0 = \tfrac{1}{2}(z_L + z_R)$, $\theta = \mu - \nu$ and $\theta_n
= n + \theta$. The mode expansion for $Z$ becomes
\begin{align}\label{Zmode}
Z(\tau, \sigma) &= z_0\ +\ \ell\left[\sum_{m = 1}^{\infty}\frac{\alpha_{m}}{\theta_m}\ f_{m}(\tau,\sigma) + \sum_{n=0}^{\infty}\frac{\beta_{n}^\dag}{\theta_{-n}}\ f_{-n}(\tau,\sigma)\right]\ ,\nonumber\\
&\text{with}\quad f_n(\tau,\sigma) = e^{-i\pi\nu}e^{-{i\theta_n\tau}} \cos[\theta_n\sigma + \pi\nu]\ .
\end{align}
\textbf{Note:} For $\theta = 0$, there will be no $\beta^\dag_0$ term
above but there will be a momentum zero-mode $\ell^2p_R
e^{-i\pi\nu}\left(\tau\cos\pi\nu - i\sigma\sin\pi\nu\right) =: \ell^2
\w{p} (\tau - i b\sigma)$ where $b = \tan \pi\nu$. The mode expansion
becomes
\begin{align*}
  Z(\tau, \sigma) &= z_0 + \ell^2 \w{p} (\tau - ib\sigma) + \ell\sum_{m = 1}^{\infty}\left[\frac{\alpha_{m}}{m}\ f_{m}(\tau,\sigma) - \frac{\beta_{m}^\dag}{m}\ f_{-m}(\tau,\sigma)\right]\ ,\quad\text{for}\quad\theta = 0\ .
\end{align*}

We focus on the $\theta \neq 0$ case. The functions
$\varphi_{n}(\sigma) := \cos [\theta_n\sigma +\pi\nu]$ satisfy the
completeness relation:
\begin{equation}
  \int_{0}^{\pi}\,\ud\sigma\ \big[(\theta_m + \theta_n) + b\ \delta(\sigma) - b'\ \delta(\pi-\sigma)\big]\varphi_m(\sigma) \varphi_n(\sigma) = \pi\theta_m \delta_{mn}\ .
\end{equation}
These imply the following completeness relation for the functions
$f_n(\tau,\sigma) = e^{-i\theta_n\tau}\varphi_n(\sigma)$:
\begin{equation}
  \int_{0}^{\pi}\,\ud\sigma\ \w f_{m}\big[i\overset{\text{\tiny$\bm\longleftrightarrow$}}{\partial_{\tau}} + b\ \delta(\sigma) -b'\ \delta(\pi-\sigma)\big]f_n = \pi\theta_m\delta_{mn}\ ,
\end{equation}
and for the constant mode $z_0$:
\begin{equation}
  \int_{0}^{\pi}\,\ud\sigma\ \big[i\overset{\text{\tiny$\bm\longleftrightarrow$}}{\partial_{\tau}} + b\ \delta(\sigma) -b'\ \delta(\pi-\sigma)\big]f_n = 0\ .
\end{equation}
Using the above relations one can invert the formula for $Z$ to obtain
\begin{align}
z_0 & = \frac{1}{b - b'}\int {\ud\sigma} \left[ i \partial_\tau Z + \left(b\,\delta(\sigma) - b'\,\delta(\pi-\sigma)\right) Z\right]\ ,\nonumber\\
\ell\, \alpha_{m} &= \int \frac{\ud\sigma}{\pi} \left[ i\w f_m \partial_\tau Z + \left(\theta_m + b\,\delta(\sigma) - b'\,\delta(\pi-\sigma)\right)\w f_m Z\right]\ ,\nonumber\\
\ell\, \beta_{n}^\dag &= \int \frac{\ud\sigma}{\pi} \left[ i\w f_{-n} \partial_\tau Z + \left(\theta_{-m} + b\,\delta(\sigma) - b'\,\delta(\pi-\sigma)\right)\w f_{-n} Z\right]\ .
\end{align}
To quantise the system we impose the following equal-time
commutation relations:
\begin{equation}
[P(\tau,\sigma)\, ,Z(\tau,\sigma')] = -i\hbar\delta(\sigma-\sigma')\ \ ,\quad  [\w P(\tau,\sigma)\, ,\w Z(\tau,\sigma')] = -i\hbar\delta(\sigma-\sigma')\ .
\end{equation}
The conjugate momentum $P(\tau, \sigma)$ is given by
\begin{align}
P(\tau,\sigma) &= \frac{\partial \mc L}{\partial(\partial_\tau Z(\tau,\sigma))} = \frac{1}{2\pi\alpha'}\left[\partial_\tau \w Z(\tau,\sigma) - \frac{ib'}{2}\,\w Z(\tau,\pi) + \frac{ib}{2}\,\w Z(\tau,0)\right]\ .
\end{align}
In terms of $Z(\tau,\sigma)$ and $P(\tau,\sigma)$ the zero mode and
oscillators are given by
\begin{align}
z_0 & = \frac{1}{b - b'}\int {\ud\sigma} \left[ 2\pi i\alpha'\,\w P + \left(\tfrac{b}{2}\,\delta(\sigma) - \tfrac{b'}{2}\,\delta(\pi-\sigma)\right) Z\right]\ ,\nonumber\\
\ell\, \alpha_{m} &= \int \frac{\ud\sigma}{\pi} \left[2\pi i\alpha'\w f_m\, \w P + \left(\theta_m + \tfrac{b}{2}\,\delta(\sigma) - \tfrac{b'}{2}\,\delta(\pi-\sigma)\right)\w f_m Z\right]\ ,\nonumber\\
\ell\, \beta_{n}^\dag &= \int \frac{\ud\sigma}{\pi} \left[2\pi i\alpha'\w f_{-n}\, \w P + \left(\theta_{-n} + \tfrac{b}{2}\,\delta(\sigma) - \tfrac{b'}{2}\,\delta(\pi-\sigma)\right)\w f_{-n} Z\right]\ .
\end{align}
Setting $2\alpha' = \ell^2$ and using the above completeness relations, we get
\begin{equation}
\boxed{[z_0, \w z_0] = \frac{\ell^2}{b - b'}\ ,\quad [\alpha_{m}\, , \alpha_{m'}^\dag] = (m + \theta) \delta_{mm'}\ ,\quad [\beta_{n}\, , \beta_{n'}^\dag] = (n-\theta) \delta_{nn'}\ .}
\end{equation}

\subsubsection*{Worldsheet fermions}
In 1+1 dimensions, right(left)-handed spinors are left(right)-moving
on-shell and supersymmetry relates left-movers to left-movers and
right-movers to right-movers:
\begin{align}
\delta Z &=  -i\epsilon^+ \Psi^- + i\epsilon^- \Psi^+\ ,\quad \delta \Psi^+ = -2\partial_{--}Z \epsilon^- + \mc F \epsilon^-\ ,\quad\delta \Psi^- = 2\partial_{++}Z \epsilon^+ + \mc F\epsilon^+\ ,
\end{align}
where we have introduced the complex combinations
$\Psi^{\pm} = \frac{1}{\sqrt{2}}(\psi^{1\pm} + i\psi^{2\pm})$ and
$\mc{F} = \frac{1}{\sqrt{2}}(F^1 + iF^2)$. The presence of a boundary
reduces the supersymmetry by half by imposing a relation between the
parameters: $\epsilon^{+} = \pm\epsilon^-$. One can always impose
$\epsilon^+ = -\epsilon^-$ at one end, say $\sigma = 0$. On the other
end, two choices are possible and they correspond to the R and NS
sectors:
\begin{align}
\sigma = \pi:\qquad \left\{\ \begin{array}{rl} \epsilon^{+} = -\epsilon^- &\quad \text{Ramond}\ ,\\
\epsilon^{+} = \epsilon^- &\quad \text{Neveu-Schwarz}\ . \end{array} \right.
\end{align}
It is evident that rigid supersymmetry is present only in the R sector
and that it has only one parameter
$\epsilon = \epsilon^+ = -\epsilon^-$. The boundary condition on
$\Psi^{\pm}$ corresponding to
$\partial_{++}Z = e^{-2\pi i\nu}\partial_{--}Z$ at $\sigma = 0$ is
given by
\begin{equation}\label{fbc0}
\Psi^+ = e^{2\pi i\nu}\,\Psi^-\quad \text{at}\quad \sigma = 0\ .
\end{equation}
Similarly, the boundary condition at $\sigma = \pi$ is
\begin{equation}\label{fbcpi}
\text{At }\sigma = \pi:\quad \left\{\ \begin{array}{ll} \Psi^+ = e^{2\pi i\mu}\,\Psi^- &\quad \text{R sector}\ ,\\
\Psi^+ = -e^{2\pi i\mu}\,\Psi^- &\quad \text{NS sector}\ . \end{array} \right.
\end{equation}
In order to write down the mode expansions, we combine
$\Psi^+(\tau-\sigma)$ and $\Psi^-(\tau+\sigma)$ on $0 \leq \sigma \leq
\pi$ into one field $\Psi$ on the double interval $-\pi \leq \sigma
\leq \pi$ such that
\begin{equation}
  \Psi(\tau + \sigma) = \left\{\ \begin{array}{ll} \Psi^+(\tau + \sigma)  & -\pi \leq \sigma \leq 0\ ,\\ e^{2\pi i\nu}\, \Psi^-(\tau + \sigma) & \phantom{-}0 \leq \sigma \leq \pi\ . \end{array} \right.
\end{equation}
Treating $-\pi \leq \sigma \leq \pi$ as an angular variable we see
that $\Psi$ is continuous at $\sigma = 0$ by virtue of \eqref{fbc0}
and twisted-periodic across $\sigma = \pi$ due to \eqref{fbcpi}:
$\Psi(\pi) = \pm e^{2\pi i(\nu - \mu)}\Psi(-\pi)$. The mode expansion
for $\Psi(\tau + \sigma)$ in the R sector is
\begin{align}
\Psi_{\text{R}}(\tau + \sigma) &= \frac{\ell}{2}\left[\sum_{m = 1}^{\infty}\,a_{m}\, e^{-i\theta_m (\tau + \sigma)} + \sum_{n=0}^{\infty}\,b_{n}^\dag\,  e^{-i\theta_{-n}\,(\tau + \sigma)}\right]\ ,
\end{align}
and in the NS sector is
\begin{align}
\Psi_{\text{NS}}(\tau + \sigma) &= \frac{\ell}{2}\left[\sum_{r=1}^{\infty}\,c_{r}\, e^{-i\epsilon_r (\tau + \sigma)} + \sum_{s=0}^{\infty}\,d_{s}^\dag\,  e^{-i\epsilon_{-s}\,(\tau + \sigma)}\right]\ ,
\end{align}
where $\epsilon = \theta + \frac{1}{2} = \mu - \nu + \frac{1}{2}$ and
$\epsilon_n = \epsilon + n$. The action for the doubled
$\Psi(\tau + \sigma)$ is
\begin{equation}
\mc S[\Psi] = \frac{2i}{\pi\alpha'}\int \ud\tau \int_{-\pi}^{\pi}\ud\sigma\ \w \Psi\, \partial_{--}\Psi\ .
\end{equation}
The boundary term for the fermions in \eqref{11action} measures the
jump in $\w\Psi{}^+\Psi^+ + \w\Psi{}^-\Psi^-$ between two boundary
components ($\sigma = 0,\pi$) of the worldsheet. Since $\w\Psi\Psi$ is
periodic on the double interval, such a boundary term is absent in the
above action. The conjugate momentum is then
\begin{equation}
\Pi(\tau + \sigma) = \frac{\partial\mc L}{\partial\dot{\Psi}} = -\frac{2i}{\pi\alpha'}\, \w \Psi(\tau + \sigma)\ .
\end{equation}
The correct equal-time anticommutation relation follows from
Dirac's constrained Hamiltonian formalism:
\begin{equation}
\{\Pi(\tau + \sigma)\, , \Psi(\tau + \sigma')\} = -\frac{i}{2} \delta(\sigma - \sigma')\ \Longrightarrow\ 
\{\w \Psi(\tau + \sigma)\, , \Psi(\tau + \sigma')\} = \frac{\pi\alpha'}{4} \delta(\sigma - \sigma')\ .
\end{equation}
Using the completeness relations and $2\alpha' = \ell^2$, we get
\begin{align}
\{a_{m}\, ,a_{m'}^\dag\} &= \delta_{mm'}\ ,\quad \{b_{n}\, , b_{n'}^\dag\} = \delta_{nn'}\ ,\quad
\{c_{r}\, ,c_{r'}^\dag\} = \delta_{rr'}\ ,\quad \{d_{s}\, , d_{s'}^\dag\} = \delta_{ss'}\ .
\end{align}
The expression for $L_0$ in the R and NS sectors is given by
\begin{align}
&L_0^{(\text{R})} = \sum_{m=1}^{\infty} \Big[\alpha_m^\dag \alpha_m + (m+\theta) a_m^\dag a_m\Big] + \sum_{n=0}^{\infty}\Big[\beta_n^\dag \beta_n + (n-\theta) b_n^\dag b_n\Big]\ ,\nonumber\\   
&L_0^{(\text{NS})} - E_{0} = \sum_{m=1}^{\infty}\alpha_m^\dag \alpha_m + \sum_{r=1}^{\infty} (r+\epsilon) c_r^\dag c_r + \sum_{n=0}^{\infty}\beta_n^\dag \beta_n + \sum_{s=0}^{\infty} (s-\epsilon) d_s^\dag d_s\ .
\end{align}

Recall that $\epsilon = \mu - \nu + \frac{1}{2}$. Since $|\mu|, |\nu|
< \frac{1}{2}$, we have $-\frac{1}{2} < \epsilon < \frac{3}{2}$. The
first few states of the spectrum in the NS sector for different ranges
of $\epsilon$ are as in Table \ref{NStable}.
\begin{table}[ht!]\centering\small
  \renewcommand{\arraystretch}{1.5}
  \begin{subtable}{0.23\linewidth}
    \caption{$-\frac{1}{2} < \epsilon < 0$}
    \begin{tabular}{c|c}
      \toprule
      $E - E_0$ &$\quad \text{NS}\quad $  \tabularnewline\hline\hline
      $-\epsilon$ & $d_0^\dag$\tabularnewline\hline
      $\epsilon + 1$ & $c_1^\dag$\tabularnewline\hline
      $-\epsilon + {1}$ & $d_1^\dag$\tabularnewline\hline
      $\epsilon + {2}$ & $c_2^\dag$\tabularnewline\hline
    \end{tabular}
  \end{subtable}\quad\ 
  \begin{subtable}{0.23\linewidth}
    \caption{$0 < \epsilon < \frac{1}{2}$}
    \begin{tabular}{c|c}
      \toprule
      $E - E_0$ &$\quad \text{NS}\quad $\tabularnewline\hline\hline
      $\epsilon$ & $d_0$\tabularnewline\hline
      $-\epsilon + {1}$ & $d_1^\dag$\tabularnewline\hline                                    
      $\epsilon + 1$ & $c_1^\dag$\tabularnewline\hline
      $-\epsilon + 2$ & $d_2^\dag$\tabularnewline\hline
    \end{tabular}
  \end{subtable}\quad\ 
  \begin{subtable}{0.22\linewidth}
    \caption{$\frac{1}{2} < \epsilon < 1$}
    \begin{tabular}{c|c}
      \toprule
      $E - E_0$ &$\quad \text{NS}\quad $ \tabularnewline\hline\hline
      $-\epsilon + {1}$ & $d_1^\dag$\tabularnewline\hline
      $\epsilon$ & $d_0$\tabularnewline\hline
      $-\epsilon + 2$ & $d_2^\dag$\tabularnewline\hline
      $\epsilon + 1$ & $c_1^\dag$\tabularnewline\hline
    \end{tabular}
  \end{subtable}\quad\ 
  \begin{subtable}{0.22\linewidth}
    \caption{$1 < \epsilon < \frac{3}{2}$}
    \begin{tabular}{c|c}
      \toprule
      $E - E_0$ &$\quad \text{NS}\quad $\tabularnewline\hline\hline
      $\epsilon - 1$ & $d_1$\tabularnewline\hline
      $-\epsilon + 2$ & $d_2^\dag$\tabularnewline\hline
      $\epsilon$ & $d_0$\tabularnewline\hline
      $-\epsilon + 3$ & $d_3^\dag$\tabularnewline\hline
    \end{tabular}
  \end{subtable}%
  \caption{Spectral flow in the NS sector}\label{NStable}
\end{table}
Observe that as we dial up $\epsilon$, negative energy states from the
Dirac sea cross the zero-point energy and become positive energy
states. The first excited state in the NS sector has energy
$|\epsilon|$ or $|1 - \epsilon|$ depending on whether $-\frac{1}{2} <
\epsilon < \frac{1}{2}$ or $\frac{1}{2} < \epsilon < \frac{3}{2}$. A
similar analysis can be made for the R sector and the results are in
Table \ref{Rtable}.
\begin{table}[ht!]\centering\small
  \renewcommand{\arraystretch}{1.5}
  \begin{subtable}{0.23\linewidth}
    \centering    
    \caption{$-1 < \theta < -\frac{1}{2}$}
    \begin{tabular}{c|c}
      \toprule
      $E$ &$\quad \text{ R}\quad $  \tabularnewline\hline\hline
      $\theta + 1$ & $a_1^\dag$\tabularnewline\hline
      $-\theta$ & $b_0^\dag$\tabularnewline\hline
      $\theta + {2}$ & $a_2^\dag$\tabularnewline\hline
      $-\theta + {1}$ & $b_1^\dag$\tabularnewline\hline
    \end{tabular}
  \end{subtable}\quad\ 
  \begin{subtable}{0.23\linewidth}
    \caption{$-\frac{1}{2} < \theta < 0$}
    \begin{tabular}{c|c}
      \toprule
      $E$ &$\quad \text{ R}\quad $  \tabularnewline\hline\hline
      $-\theta$ & $b_0^\dag$\tabularnewline\hline
      $\theta + 1$ & $a_1^\dag$\tabularnewline\hline
      $-\theta + {1}$ & $b_1^\dag$\tabularnewline\hline
      $\theta + {2}$ & $a_2^\dag$\tabularnewline\hline
    \end{tabular}
  \end{subtable}\quad\ 
  \begin{subtable}{0.22\linewidth}
    \caption{$0 < \theta < \frac{1}{2}$}
    \begin{tabular}{c|c}
      \toprule
      $E$ &$\quad \text{ R}\quad $  \tabularnewline\hline\hline
      $\theta$ & $b_0$\tabularnewline\hline
      $-\theta + {1}$ & $b_1^\dag$\tabularnewline\hline
      $\theta + 1$ & $a_1^\dag$\tabularnewline\hline
      $-\theta + 2$ & $b_2^\dag$\tabularnewline\hline
    \end{tabular}
  \end{subtable}\quad\ 
  \begin{subtable}{0.22\linewidth}
    \caption{$\frac{1}{2} < \theta < 1$}
    \begin{tabular}{c|c}
      \toprule
      $E$ &$\quad \text{ R}\quad $  \tabularnewline\hline\hline
      $-\theta + {1}$ & $b_1^\dag$\tabularnewline\hline
      $\theta$ & $b_0$\tabularnewline\hline
      $-\theta + {2}$ & $b_2^\dag$\tabularnewline\hline
      $\theta + 1$ & $a_1^\dag$\tabularnewline\hline
    \end{tabular}
  \end{subtable}%
  \caption{Spectral flow in the R sector}\label{Rtable}
\end{table}

The zero-point energies for a complex boson and a complex fermion with
moding $\mbb Z + v + \frac{1}{2}$, $|v| \leq \frac{1}{2}$ are
\begin{equation}
\frac{1}{24} - \frac{v^2}{2}\ ,\ \ -\frac{1}{24} + \frac{v^2}{2}\quad \text{respectively}\ .
\end{equation}

The complex boson $Z$ has moding $\mbb Z + \theta$ and so do the
fermions in the R sector. This is a consequence of rigid supersymmetry
on the worldsheet in the R sector. Thus, the zero-point energy in the
R sector vanishes. The fermions in the NS sector have moding $\mbb Z +
\epsilon$ and the total zero-point energy in the NS sector is given by
\begin{align}
E_{0} &= \frac{1}{24} - \frac{\left(|\theta| -\tfrac{1}{2}\right)^2}{2} - \frac{1}{24} + \frac{\left(\big||\epsilon-\frac{1}{2}| - \frac{1}{2} \big|-\tfrac{1}{2}\right)^2}{2}\ ,\nonumber\\
                &= \frac{1}{8} - \frac{1}{2}\big||\theta| - \tfrac{1}{2}\big|\ .
\end{align}

Since $[z\, ,\w z] = \frac{\ell^2}{b-b'}$ and $[z\, , L_0] = 0$, we
can build an infinite tower of states (given that the $Z$ direction is
non-compact) from each $L_0$ eigenstate. Thus, each state in the
spectrum is infinitely degenerate and furnishes a representation of
the Heisenberg algebra of $z$ and $\w z$.

\subsection{Boundary condition changing operators}
We map the strip $-\infty < \tau < \infty$, $0 \leq \sigma \leq \pi$
to the upper half plane $\mbb{H} = \{z \in \mbb{C}\ |\ \text{Im}(z) >
0\}$ by first Wick-rotating $\tau = -it$ and using the map $z = \exp
(t + i\sigma)$. In particular, the boundary at $\sigma = 0, \pi$ is
mapped to $z = \w{z} > 0$ and $< 0$ respectively. Consider a complex
boson $Z$ with $\textbf{TT}\bm{'}$ boundary conditions. Using
$\partial_{++} = iz\partial$ and $\partial_{--} = i\w{z}\w\partial$,
we can write the corresponding boundary conditions on $\mbb{H}$:
\begin{equation}
  \partial Z = e^{-2\pi i\nu}\,\w\partial Z\quad\text{for}\quad z = \w{z} > 0\ ,\qquad \partial Z = e^{-2\pi i\mu}\,\w\partial Z\quad\text{for}\quad z = e^{2\pi i}\w{z} < 0\ .
\end{equation}
with $-\frac{1}{2} \leq \mu, \nu \leq \frac{1}{2}$. We define bulk
chiral currents $J(z) = i\partial Z$, $\w{J}(\w{z}) = i\w\partial Z$
using the modes in \eqref{Zmode}:
\begin{align}
J(z) = i\partial Z(z) &=  -\frac{i\ell}{2}\sum_{n = 1}^\infty \alpha_n\,  e^{-2\pi i\nu} z^{-1-\theta_n}  - \frac{i\ell}{2}\sum_{m = 0}^\infty \beta_m^\dag\, e^{-2\pi i\nu} z^{-1-\theta_{-m}}\ ,\nonumber\\
\w{J}(\w{z}) = i\w\partial Z(\w{z}) &=  -\frac{i\ell}{2}\sum_{n = 1}^\infty \alpha_n\,  \w{z}^{-1-\theta_n}  - \frac{i\ell}{2}\sum_{m = 0}^\infty \beta_m^\dag\, \w{z}^{-1-\theta_{-m}}\ ,
\end{align}
where $\theta = \mu - \nu$ and $\theta_n = \theta + n$. Since the
modes are not integers, we need to specify a branch cut: we choose it
to be at $ -\infty < z \leq 0$. We also define the hermitian conjugate
currents
\begin{equation}
  J^*(z) := z^{-1} \w{J(\w{z}^{-1})}\ ,\quad \w{J}{}^*(\w{z}) := \w{z}^{-1} \w{\w{J}(z^{-1})}\ .
\end{equation}
The gluing conditions for the currents are then:
\begin{align}
J(z) &= e^{-2\pi i\nu}\w{J}(\w{z})\quad\text{for}\quad z = \w{z} > 0\ ,\qquad\ \ \, J(z) = e^{2\pi i\mu} \w{J}(\w{z})\quad\text{for}\quad z = e^{2\pi i}\w{z} < 0\ ,\nonumber\\
J^*(z) &= e^{2\pi i\nu}\w{J}{}^*(\w{z})\quad\text{for}\quad z = \w{z} > 0\ ,\qquad J^*(z) = e^{-2\pi i\mu} \w{J}{}^*(\w{z})\quad\text{for}\quad z = e^{2\pi i}\w{z} < 0\ .
\end{align}
The gluing conditions allow us to extend the domain of definition of
the currents to the full $z$-plane by employing the \emph{doubling
  trick}:
\begin{align}
\mathsf{J}(z) &= -\frac{i\ell}{2}\sum_{n \geq 1} \alpha_{m}\,e^{-2\pi i\nu} z^{-1-\theta_n} - \frac{i\ell}{2} \sum_{m \geq 0} \beta_{m}^\dag\,e^{-2\pi i\nu} z^{-1-\theta_{-m}}\ ,\nonumber\\
\mathsf{J}^*(z) &= \frac{i\ell}{2} \sum_{n \geq 1} \alpha_{n}^\dag\, e^{2\pi i\nu} z^{-1+\theta_n} +  \frac{i\ell}{2} \sum_{m \geq 0} \beta_{m}\, e^{2\pi i\nu}z^{-1+\theta_{-m}} \ ,
\end{align}
The doubled stress tensor $\mathsf{T}(z)$ is given by
\begin{equation}\label{suga}
\mathsf{T}(z) = \lim_{w \to z}\,\frac{4}{\ell^2}\left( \mathsf{J}(w) \mathsf{J}^*(z) - \frac{\ell^2}{4(w-z)^2}\right)\ .
\end{equation}
The change in boundary conditions from $\mu$ to $\nu$ at $z = 0$ and
vice-versa at $z = \infty$ can be interpreted as there being present
\emph{boundary condition changing operators} (BCC) $\sigma(0)$ and
$\sigma^+(\infty)$ where $\sigma^+$ is the operator conjugate to
$\sigma$. The conformal dimension of $\sigma$ is obtained from the
one-point function of $\mathsf{T}(z)$. Following the treatment in
\cite{DFMS, FGRS} we first define $\mathsf{J} = \mathsf{J}_{\bm{>}} +
\mathsf{J}_{\bm{<}}$ where $\mathsf{J}_{\bm{>}}$ contains only
annihilation operators. We have, for $0 < \theta < 1$,
\begin{align}
  \mathsf{J}_{\bm{>}}(w) &= -\frac{i\ell}{2}\sum_{n \geq 1} \alpha_{n}\,e^{-2\pi i\nu} w^{-1-\theta_n} - \frac{i\ell}{2}\, \beta_0^\dag\,e^{-2\pi i\nu} w^{-\theta-1}\ ,
\end{align}
and for $-1 < \theta < 0$ the last term is absent. Next, we
compute
\begin{align}
\mathsf{J}(w)\mathsf{J}^*(z) - \frac{\ell^2}{4(w-z)^2} &= \mathsf{J}_{\bm{<}}(w)\mathsf{J}^*(z) + \mathsf{J}^*(z)\mathsf{J}_{\bm{>}}(w) + [\mathsf{J}_{\bm{>}}(w), \mathsf{J}^*(z)] - \frac{\ell^2}{4(w-z)^2}\ ,\nonumber\\
  &= \mathsf{J}_{\bm{<}}(w)\mathsf{J}^*(z) + \mathsf{J}^*(z)\mathsf{J}_{\bm{>}}(w) + \frac{\ell^2}{4}\ \partial_z\!\left[\left(\frac{z}{w}\right)^{(\star)}\!\frac{1}{w-z}\right]\ ,
\end{align}
where the exponent $(\star)$ is $\theta$ for $0 < \theta < 1$
and $1+\theta$ for $-1 < \theta < 0$. This finally gives
\begin{align}
  \mathsf{T}(z) &= \frac{|\theta|\left(1 - |\theta|\right)}{2 z^2} + \frac{4}{\ell^2}\mathsf{J}_{\bm{<}}(z)\mathsf{J}^*(z) + \frac{4}{\ell^2}\mathsf{J}^*(z)\mathsf{J}_{\bm{>}}(z)\ ,
\end{align}
which gives the one-point function
\begin{equation}
\langle \mathsf{T}(z) \rangle = \frac{|\theta|\left(1 - |\theta|\right)}{2 z^2}\ .
\end{equation}
This can be interpreted as there being two BCC operators $\sigma$,
$\sigma^+$ inserted resp.~at $z = 0$ and $z = \infty$ with
conformal weight $h_{\sigma} = \frac{|\theta|\left(1 -
    |\theta|\right)}{2}$. Their two-point function is
\begin{equation}
\langle \sigma(x_1) \sigma^+(x_2)\rangle = \frac{1}{|x_1 - x_2|^{2h_\sigma}}\ .
\end{equation}

We next describe BCC operators for the worldsheet fermions
$\Psi^\pm$. Since the fermions have conformal dimension
$\frac{1}{2}$, we should include a Jacobian factor $z^{-1/2}$
while mapping them from the strip to the upper half-plane. We employ
the doubling trick to directly write the R and NS fermions on the full
$z$-plane in the R and NS sectors:
\begin{align}
\Psi_{\text{R}}(z) &= \frac{\ell}{{2}}\sum_{n \geq 1}\,a_{n}\, z^{-\epsilon_n} + \frac{\ell}{{2}}\sum_{m \geq 0}\,b_{m}^\dag\,  z^{-\epsilon_{-m}}\ ,\nonumber\\
\Psi_{\text{NS}}(z) &= \frac{\ell}{{2}}\sum_{r \geq 1}\,c_{r}\, z^{-\theta_r -1} + \frac{\ell}{{2}}\sum_{s \geq 0}\,d_{s}^\dag\,  z^{-\theta_{-s} -1}\ .
\end{align}
with $\epsilon = \theta + \frac{1}{2}$. In order to describe the BCC
operators, we first bosonise $\Psi(z)$ by introducing an antihermitian
scalar $H(z)$:
\begin{equation}
\Psi(z) = \nmo{e^{H(z)}}\ ,\ \ \Psi^*(z) = \nmo{e^{-H(z)}} \quad\text{with}\quad \langle H(w) H(z) \rangle = \log (w-z)\ .
\end{equation}
The normal ordering symbol ${\bf :\ :}$ is omitted in the above
definition for the sake of brevity. Now, consider the OPE of the
operator $\nmo{e^{-\theta H(x)}}$ with $\Psi$:
\begin{align}
&\Psi(z) \nmo{e^{-\theta H(0)}}\ \sim z^{-\theta} \nmo{e^{(1-\theta)H(0)}} + \cdots\ .
\end{align}
First, we notice that as $z \to e^{2\pi i} z$, $\Psi$ picks up a phase
$e^{- 2\pi i\theta}$, which matches with the monodromy of
$\Psi_{\text{NS}}$. Further, we observe that the right-hand side of
the first OPE is regular as $z \to 0$ for $\theta < 0$. This requires
that $\Psi_{\text{NS}}(z)$ annihilate the state $\nmo{e^{-\theta
    H(0)}}|0\rangle$ in the limit $z \to 0$ where the state
$|0\rangle$ is the $\text{SL}(2,\mbb{R})$-invariant vacuum. From Table
\ref{NStable}, we see that the NS vacuum $|\textsc{ns}\rangle$ has the
same properties for $-\frac{1}{2} < \theta < 0$. Thus, we can
identify the state $\nmo{e^{-\theta H(0)}}|0\rangle$ with the NS
vacuum for this range of $\theta$:
\begin{equation}
\nmo{e^{-\theta H(0)}}|0\rangle = |\textsc{ns}\rangle\quad\text{for}\quad -\frac{1}{2} < \theta < 0\ .
\end{equation}
For the range $-1 < \theta < -\frac{1}{2}$, we see from Table
\ref{NStable} that $\Psi_{\text{NS}}$ annihilates the excited state
$d_0^\dag|\textsc{ns}\rangle$. Thus, for this range of $\theta$ one
has to identify $\nmo{e^{-\theta H(0)}}|0\rangle$ with the first
excited state:
\begin{equation}
\nmo{e^{-\theta H(0)}}|0\rangle = d_0^\dag|\textsc{ns}\rangle\quad\text{for}\quad -1 < \theta < -\frac{1}{2}\ .
\end{equation}
The NS vacuum is obtained by applying $d_0$ to the above and $d_0$ is
contained in the Hermitian conjugate field $\Psi^*(z)$, defined as
$$ \Psi^*(z) := z^{-1}\ \w{\Psi(\w{z}{}^{-1})}\ .$$
This gives
\begin{equation}
\Psi_{\text{NS}}^*(z) := \frac{\ell}{{2}}\sum_{r \geq 1}\,c_{r}^\dag\, z^{\theta_r} + \frac{\ell}{{2}}\sum_{s \geq 0}\,d_{s}\,  z^{\theta_{-s}}\ .
\end{equation}
Thus, the operator corresponding to the NS vacuum for this range of
$\theta$ is obtained by fusing $\Psi^*$ with $\nmo{e^{-\theta H(x)}}$:
\begin{equation}
\Psi^*(z) \nmo{e^{-\theta H(0)}}\ \sim z^{\theta}\nmo{e^{-(1+\theta)H(0)}} + \cdots\ .
\end{equation}
Thus the NS vacuum is to be identified with the operator on the right
hand side:
\begin{equation}
\nmo{e^{-(1+\theta)H(0)}}|0\rangle = |\textsc{ns}\rangle\quad\text{for}\quad -1 < \theta < -\frac{1}{2}\ .
\end{equation}
Similarly, for $\theta > 0$ there are two cases $0 < \theta <
\frac{1}{2}$ and $\frac{1}{2} < \theta < 1$ for which the operators
corresponding to $|\textsc{ns}\rangle$ are $\nmo{e^{-\theta H(x)}}$
and $\nmo{e^{(1-\theta)H(x)}}$ respectively. The same analysis can be
made for the R sector as well. We summarise the results in Table
\ref{vobfield}. We designate the operator corresponding to the NS and
R vacua as $s^{\textsc{ns}}(x)$ and $s^{\textsc{r}}(x)$
respectively. These shall be the BCC operators for the respective
sectors.
\begin{table}[ht!]\centering
  \small
  \renewcommand{\arraystretch}{1.5}
  \begin{tabular}{c|c|c|c|c}
    \toprule
    Ground BCC operator & $-1 < \theta < -\frac{1}{2}$ & $-\frac{1}{2} < \theta < 0$ & $0 < \theta < \frac{1}{2}$ & $\frac{1}{2} < \theta < 1$ \tabularnewline\midrule
    $s^{\textsc{ns}}(x)$ & $\nmo{e^{- (1 + \theta) H(x)}}$ & $\nmo{e^{- \theta H(x)}}$ & $\nmo{e^{- \theta H(x)}}$ & $\nmo{e^{(1 - \theta) H(x)}}$ \tabularnewline\midrule
    $s^{\textsc{r}}(x)$ & $\nmo{e^{- \epsilon H(x)}}$ & $\nmo{e^{- \epsilon H(x)}}$ & $\nmo{e^{(1-\epsilon) H(x)}}$ & $\nmo{e^{(1 - \epsilon) H(x)}}$ \tabularnewline
    \bottomrule
  \end{tabular}
  \caption{Ground BCC operators for the NS and R sectors}\label{vobfield}
\end{table}
Also observe that the operators $s^{\textsc{ns,r}}$ always have the
smallest conformal dimension in each range of $\theta$. The operators
corresponding to the excited states can be inferred in a similar
fashion and are summarised in Table \ref{evobfield}.
\begin{table}[ht!]\centering
  \small
  \renewcommand{\arraystretch}{1.5}
  \begin{tabular}{c|c|c|c|c}
    \toprule
    Excited BCC operator & $-1 < \theta < -\frac{1}{2}$ & $-\frac{1}{2} < \theta < 0$ & $0 < \theta < \frac{1}{2}$ & $\frac{1}{2} < \theta < 1$ \tabularnewline\midrule
    $t^{\textsc{ns}}(x)$ & $\nmo{e^{- \theta H(x)}}$ & $\nmo{e^{-(1+\theta) H(x)}}$ & $\nmo{e^{(1-\theta)H(x)}}$ & $\nmo{e^{-\theta H(x)}}$ \tabularnewline
    $\tilde{t}^{\textsc{ns}}(x)$ & $\nmo{e^{-(2+\theta)H(x)}}$ & $\nmo{e^{(1-\theta) H(x)}}$ & $\nmo{e^{-(1+\theta)H(x)}}$ & $\nmo{e^{(2-\theta) H(x)}}$ \tabularnewline\midrule
    $t^{\textsc{r}}(x)$ & $\nmo{e^{-(1+\epsilon) H(x)}}$ & $\nmo{e^{(1-\epsilon)H(x)}}$ & $\nmo{e^{-\epsilon H(x)}}$ & $\nmo{e^{(2-\epsilon) H(x)}}$ \tabularnewline
    $\tilde{t}^{\textsc{r}}(x)$ & $\nmo{e^{(1-\epsilon) H(x)}}$ & $\nmo{e^{-(1+\epsilon)H(x)}}$ & $\nmo{e^{(2-\epsilon) H(x)}}$ & $\nmo{e^{-\epsilon H(x)}}$ \tabularnewline
    \bottomrule
  \end{tabular}
  \caption{Excited BCC operators for the NS and R sectors}\label{evobfield}
\end{table}
Let us study the limiting cases of \textbf{NN}, \textbf{DD},
\textbf{DN}. For both \textbf{NN} and \textbf{DD} we have $\theta =
0$. From Table \ref{vobfield}, we see that for either of the two
limits $\theta \to 0^+$ or $\theta \to 0^-$, the NS vacuum is the
$\text{SL}(2,\mbb{R})$-invariant vacuum $|0\rangle$. The first two
excited states corresponding to $\nmo{e^{\pm H}}$ are
degenerate. In the R sector, for $\theta \to 0^\pm$, the vacuum
corresponds to $\nmo{e^{\pm H/2}}$ and the first excited state to
$\nmo{e^{\mp H/2}}$. The two states are degenerate, so either limit
gives the same spectrum.

For \textbf{DN} boundary conditions, we have $\mu = 0$ and $\nu =
\frac{1}{2}$ giving $\theta =-\frac{1}{2}$ and $\epsilon = 0$. In the
NS sector, the ground state and the first excited state are
degenerate, corresponding to the operators $\nmo{e^{\pm H/2}}$. In the
R sector, the ground state is the $\text{SL}(2,\mbb{R})$-invariant
vacuum and the first two excited states corresponding to $\nmo{e^{\pm
    H}}$ are degenerate.

For \textbf{ND} boundary conditions, we have $\mu = \frac{1}{2}$ and
$\nu = 0$ giving $\theta =\frac{1}{2}$ and $\epsilon = 1$. The
discussion on the NS and R sector states is identical to the
\textbf{DN} case.

For the \textbf{TD} case, we have $\mu = \frac{1}{2}$ and $\nu = v$
giving $\theta =\frac{1}{2} - v$ and $\epsilon = 1 - v$. The range of
$\theta$ is $0 \leq \theta \leq 1$, giving the ground BCC operator
$\nmo{e^{(1-\epsilon)H(x)}}$ in the R sector and $\nmo{e^{- \theta
    H(x)}}$, $\nmo{e^{(1 - \theta) H(x)}}$ in the NS sector for $0
\leq \theta \leq \frac{1}{2}$ and $\frac{1}{2} \leq \theta \leq 1$
respectively.

For the \textbf{DT} case, we have $\mu = v$ and $\nu = \frac{1}{2}$
giving $\theta =v - \frac{1}{2}$ and $\epsilon = v$. The ground BCC
operators are $\nmo{e^{-\epsilon H(x)}}$ in the R sector and
$\nmo{e^{-(1+\theta) H(x)}}$, $\nmo{e^{-\theta H(x)}}$ in the NS sector
for $-1 \leq \theta \leq -\frac{1}{2}$ and $-\frac{1}{2} \leq \theta
\leq 0$ respectively. 

\subsection{The covariant lattice}
Consider the following linear combinations of the holomorphic
(left-moving) part of the worldsheet
fermions:
\begin{equation}
  \Psi^{\pm e_a} = \frac{\psi^{2a-1} \pm i\psi^{2a}}{\sqrt{2}}\ ,\ a \in \four\ ,\quad \Psi^{\pm e_5} = \frac{\psi^9 \pm \psi^0}{\sqrt{2}}\ . 
\end{equation}
Here $e_m$, $m = 1,\ldots, 5$, are unit vectors $(e_m)_i =
\delta_{im}$ of the $D_5$ weight lattice. Along with their negatives,
they form the weights of the vector representation of
$\text{so}(1,9)$. Under complex conjugation the fermions behave as
follows:
\begin{equation}
  (\Psi^{+e_a})^* = \Psi^{-e_a}\ ,\quad (\Psi^{\pm e_5})^* = \Psi^{\pm e_5}\ .
\end{equation}
In order to bosonize these with the correct properties under complex
conjugation, we introduce antihermitian scalars $H_a(z)$ and a
hermitian scalar $H_5(z)$ which satisfy $\langle H_m(z) H_n(w)\rangle
= \delta_{mn} \log (z - w)$, $m,n = 1,\ldots, 5$. The bosonised
versions of the fermions are
\begin{align}
&\Psi^{\pm e_m}(z) :=\, \nmo{e^{\pm H_m(z)}c_{\pm e_m}}\ ,\quad m = 1, \ldots, 5\ .
\end{align}
The object $c_{\pm e_m}$ is a \emph{cocycle operator} which is defined
\cite{KLLSW} in terms of the fermion number operators $N_m$ as
\begin{equation}
c_{\pm e_m} := (-)^{N_1 + \cdots + N_{m-1}}\ .
\end{equation}
These ensure that the fermions $\Psi^{\pm e_m}$, $\Psi^{\pm e_n}$ for
$m \neq n$ anticommute after bosonisation. In a broader context, these
fermions are used to construct the currents whose modes satisfy the
commutation relations of the affine Ka\v{c}-Moody algebra
$\hat{D}_5$. The commutation relations between the generators
corresponding to the roots of the Lie algebra involve certain
2-cocycles. In order to obtain these 2-cocycles correctly via
bosonised vertex operators, we need to include the above \emph{cocycle
  operators} in the definition of the vertex operators. These
2-cocycles were first treated by Bardakci and Halpern \cite{BH} in the
physics literature and by Frenkel and Ka\v{c} \cite{FK}, and Segal
\cite{Se} in the mathematics literature.

In terms of the bosons $H_m$, the number operators are given by
$N_m := (\partial H_m)_0$ where $(\partial H_m)_0$ are the zero modes
of $\partial H_m$. The bosons $H_m$ have the following mode expansion
\cite{PR}:
\begin{equation}
  H_m(z) = h_m + N_m \log z + \sum_{k \neq 0} \frac{\alpha_{m,k}}{k} z^{-k}\ .
\end{equation}
The Hermitian conjugate field $H^*_m$ is defined as follows:
\begin{equation}
  H^*_m(z) := \w{H_m(\w{z}{}^{-1})}\ .
\end{equation}
Since $H_a$ are antihermitian and $H_5$ is hermitian,
the modes satisfy
\begin{align}
  &(h_a){}^\dag = -h_a\ ,\quad (N_a)^\dag = N_a\ ,\quad (\alpha_{a,k})^\dag = \alpha_{a,k}\ ,\ a \in \four\ ,\nonumber\\
  &(h_5){}^\dag = h_5\ ,\quad (N_5)^\dag = -N_5\ ,\quad (\alpha_{5,k})^\dag = -\alpha_{5,k}\ .
\end{align}
These properties will be required in the discussion on the cocycle
operators.  Next, we discuss superconformal ghosts. The contribution
due to these have to be included appropriately in each vertex operator
to ensure that operator products are mutually local. One bosonises the
superconformal ghosts $\beta$, $\gamma$ using a hermitian scalar field
$\varphi$ with $\langle \varphi(z)\varphi(w) \rangle = -\log (z - w)$
and two fermions $\xi(z)$ and $\eta(z)$ with $\langle \xi(z)\zeta(w)
\rangle = (z-w)^{-1}$:
\begin{equation}
\beta(z) :=\,\nmo{e^{-\varphi(z)}\partial\xi(z)}\ ,\quad \gamma(z) :=\,\nmo{\eta(z) e^{\varphi(z)}}\ .
\end{equation}
The fermions $\xi$ and $\eta$ are further bosonised as
\begin{equation}
\xi(z) = e^{\zeta(z)}\ ,\quad \eta(z) = e^{-\zeta(z)}\ .
\end{equation}
with $\zeta(z)$ a hermitian scalar with $\langle \zeta(z)\zeta(w)\rangle
= \log (z-w)$. The superghost picture number operator $N_6$ is given
by the zero mode of $\partial\zeta - \partial\varphi$. Under
conjugation, it satisfies
\begin{equation}
(N_6)^\dag = -N_6 - Q = -N_6 -2\ ,
\end{equation}
where $Q = 2$ is the background charge of the $\beta\gamma$ CFT. The
picture charge of $\nmo{e^{q\varphi(z)}}$ is $q$ and its conformal
dimension is $-\frac{1}{2}q(q + Q)$. The operator conjugate to
$\nmo{e^{q\varphi}}$ is $\nmo{e^{-(q + Q)\varphi}}$ and it also has
conformal dimension $-\frac{1}{2}q(q + Q)$.

In the canonical ghost picture, vertex operators in the NS sector
acquire a factor of $e^{-\varphi}$ and those in the R sector a factor
of $e^{-\varphi/2}$. The integer and half-integer exponents are
correlated with the integer and half-integer modes for the NS and R
fermions on the doubled plane. The integer and half-integer ghost
numbers can be interpreted as belonging to a $D_1$ weight lattice
which can then be combined with the spacetime $D_5$ weights to get a
\emph{covariant lattice} $\Gamma_{5,1}$. The lattice $\Gamma_{5,1}$ is
Lorentzian since $\langle \varphi(z) \varphi(w)\rangle = -\log(z - w)$
as opposed to $\langle {H_m}(z) {H_n}(w)\rangle = \delta_{mn}\log(z -
w)$. Writing $H_6 := -\varphi$, we have $\langle H_\mu(z) H_\nu(w)
\rangle = \eta_{\mu\nu} \log(z-w)$ with $\eta_{66} = -1$, $\eta_{6m} =
0$ and $\eta_{mn} = \delta_{mn}$.  A general vertex operator in the
(worldsheet) fermionic sector is then given by
\begin{equation}\label{vog51}
  V_\lambda(z) =\ \nmo{e^{\lambda\cdot H(z)}c_{\lambda}}\ .
\end{equation} 
where $\lambda$ is a weight in the covariant lattice $\Gamma_{5,1}$,
$c_\lambda$ is the cocycle operator corresponding to $\lambda$ and the
dot product $\lambda\cdot H$ is with respect to the Lorentzian metric
$\eta_{\mu\nu}$. We give a formula for $c_\lambda$ below. The
$\Gamma_{5,1}$ weights $\lambda$ with $\lambda_6 = -1, -\frac{1}{2},
-\frac{3}{2}$ directly correspond to physical states whose
mass-squared is given by
\begin{equation}
  \alpha' m^2 = \frac{1}{2}\lambda^2 + \lambda\cdot e_\varphi - 1\ ,
\end{equation}
where $e_\varphi$ is the unit vector $(0,0,0,0,0;1)$. The term
$\lambda\cdot e_\varphi$ arises due to the background charge of the
$\beta\gamma$ CFT. The other $\Gamma_{5,1}$ weights do not correspond
directly to physical states but linear combinations of the
corresponding vertex operators correspond to physical operators with
picture charge different from the canonical ones.

\subsection{The D1-D5\tp{$_A$}{A}-D5\tp{$_{\w{A}}$}{Abar} system}
Consider the D1-D5$_A$-D5$_{\w{A}}$ system. The spacetime Lorentz
symmetry $\text{SO}(1,9)$ is broken down to
$ \text{SO}(4) \times \text{SO}(4)' \times \text{SO}(1,1)$ with
spacetime now being the $1+1$ dimensional intersection $\Int$ of the
D-branes. The worldsheet fermion contribution to the total vertex
operator can thus be described by \eqref{vog51} where $\lambda$ is now
a weight in the covariant lattice
$D_2 \oplus D_2 \oplus \Gamma_{1,1}$. In the presence of a constant
$B$-field, the weights $\lambda$ have to be generalised to include
entries which are neither integral nor half-integral. The precise
weights can be obtained by following the procedure outlined in the
previous sections.

An (unintegrated) open string vertex operator with only fermionic
oscillators then has the form
\begin{equation}\label{finalvop}
  V_\lambda(k,z) = \omega(\lambda)\, c(z)\,\mc{B}(z)\, \nmo{e^{\lambda\cdot H(z)}\,e^{2ik\cdot X(z)}c_{\lambda}}\,\ ,
\end{equation} 
where $\omega(\lambda)$ is an arbitrary $c$-number phase, $c(z)$ is
the coordinate ghost, $\mc{B}(z)$ is the product of the appropriate
BCC operators for the worldsheet bosons and $k$ is the $1+1$
dimensional spacetime momentum. We have suppressed Chan-Paton
factors. The mass formula for a state with weight $\lambda$ becomes
\begin{equation}
  \alpha' m^2(\lambda) =  -\alpha' k^2 = \frac{1}{2}\lambda^2 + \lambda\cdot e_\varphi - 1 + \sum_{\sigma | \mc{B}}h_\sigma\ .
\end{equation}
The notation $\sum_{\sigma | \mc{B}}$ indicates the summation of the
conformal dimensions $h_\sigma$ of bosonic BCC operators $\sigma(z)$
present in $\Sigma(z)$ above. 

The weights of the vertex operators for the case with $A = (12)$,
$\w{A} = (34)$ are recorded in Tables \ref{d1vo} and \ref{d1d5vo} in
the main text. Let us calculate the masses of some of the states using
the above machinery. For the D1-D5$_{(12)}$ states $(\phi^1, \phi^2)$ we have
BCC operators for the worldsheet bosons $Z^1$ and $Z^2$ with conformal
dimensions $\frac{1}{2}\left(\frac{1}{4} - v_1^2\right)$ and
$\frac{1}{2}\left(\frac{1}{4} - v_2^2\right)$ respectively. We thus
get
\begin{align}
  \alpha' m^2(\phi^1) &= \frac{1}{2}\Big[(v_1 + \tfrac{1}{2})^2 + (v_2 - \tfrac{1}{2})^2 - 1\Big] + 1 - 1 + \frac{1}{2}\Big[(\tfrac{1}{4} - v_1^2) + (\tfrac{1}{4} - v_2^2)\Big]\nonumber\\ 
                      &= \frac{1}{2}(v_1 - v_2)\ .\\
  \alpha' m^2(\phi^2) &= \frac{1}{2}\Big[(v_1 - \tfrac{1}{2})^2 + (\tfrac{1}{2} + v_2)^2 - 1\Big] + 1 - 1 + \frac{1}{2}\Big[(\tfrac{1}{4} - v_1^2) + (\tfrac{1}{4} - v_2^2)\Big]\nonumber\\ 
                      &= -\frac{1}{2}(v_1 - v_2)\ .
\end{align}
Observe that as $v_a \to 0$, both the states become massless and
combine into a left-handed spinor $\phi^\alpha$. Lastly, let us
calculate the mass of the D5$_{(12)}$-D5$_{(34)}$ fermion:
\begin{align}
  \alpha' m^2(\lambda) &= \frac{1}{2}\Big[\frac{1}{4} +  \sum_{a \in \four}( v_a^2) - \frac{1}{4}\Big] + \frac{1}{2} - 1 + \frac{1}{2}\sum_{a \in \four} (\tfrac{1}{4} - v_a^2) = 0\ .
\end{align}
These formulas match with the worldsheet oscillator derivation in
Section \ref{constBstring}.

\subsection{Cocycle operators}
We follow the treatment in \cite{KLLSW} and write the cocycle
operators $c_{\lambda}$ as follows:
\begin{equation}
c_{\lambda} := \exp\big(i\pi\cycm_{\rho\sigma}\lambda^\rho  N^\sigma\big)\ ,
\end{equation}
where $N_\sigma$ is the vector of number operators $(N_1,\ldots,N_6)$ and
$\cycm_{\mu\nu}$ is the matrix
\begin{equation}
\cycm_{\mu\nu} = \pmat{0 & 0 & 0 & 0 & 0 & 0 \\ +1 & 0 & 0 & 0 & 0 & 0 \\ +1 & +1 & 0 & 0 & 0 & 0 \\ -1 & +1 & -1 & 0 & 0 & 0 \\ +1 & +1 & +1 & +1 & 0 & 0 \\ -1 & -1 & -1 & -1 & +1 & 0}\ .
\end{equation}
The indices $\sigma,\rho$ are raised and lowered using the indefinite
metric $\eta_{\mu\nu}$. The OPE between two vertex operators
$V_\lambda(z)$ and $V_{\lambda'}(z)$ then becomes
\begin{equation}
V_\lambda(z) V_{\lambda'}(w) \sim (z - w)^{\lambda\cdot\lambda'} e^{i\pi\lambda\cdot \cycm\cdot \lambda'} V_{\lambda + \lambda'}(w) + \cdots\ .
\end{equation} 

The signs in the matrix $\cycm$ are chosen to obtain the correct charge
conjugation matrices in the OPEs
\begin{align}
S^A(z) S^B(w) \sim (z-w)^{-1} C^{AB} + \cdots\ ,\quad S^{\dt A}(z) S^{\dt B}(w) \sim -i(z-w)^{-1} C^{\dt A\dt B} + \cdots\ ,
\end{align}
where $S^A$, $S^{\dt B}$ are the $9+1$ dimensional left- and
right-handed spinor vertex operators in the canonical ghost
picture. They are given by the $D_{5,1}$ weights
$(\pm\frac{1}{2},\pm\frac{1}{2},\pm\frac{1}{2},\pm\frac{1}{2},\pm\frac{1}{2};-\frac{1}{2})$
with even and odd number of minus signs respectively. The
corresponding $\Gamma$-matrices are obtained from the OPE
\begin{equation}
\psi^\mu(z) S^A(w) \sim (z-w)^{-1} (\Gamma^{\mu})^A{}_{\dt B}\, S_{(-3/2)}^{\dt B}(w) + \cdots\ ,
\end{equation}
where $\psi^\mu(z)$ is the $9+1$ dimensional vector vertex operator
from the NS sector with $D_{5,1}$ weight $(0,\ldots,0,\pm
1,0,\ldots,0; -1)$ and $S^{\dt B}_{(-3/2)}$ is the operator that is
conjugate to the operator $S^{\dt B}_{(-1/2)}$ in the canonical
ghost picture. We obtain the following helicity representation for the
$\Gamma$-matrices and the charge conjugation matrix from the above
OPEs \cite{KLLSW}:
\begin{align*}
\Gamma^1 & = \sigma_1 \otimes \mbbm{1} \otimes \mbbm{1} \otimes \mbbm{1} \otimes \mbbm{1}\ ,\qquad \phantom{-\sigma_2}\Gamma^7 = -\sigma_3 \otimes \sigma_3 \otimes \sigma_3 \otimes \sigma_2 \otimes \mbbm{1}\ ,\\
\Gamma^2 & = \sigma_2 \otimes \mbbm{1} \otimes \mbbm{1} \otimes \mbbm{1} \otimes \mbbm{1}\ ,\qquad  \phantom{-\sigma_2}\Gamma^8 = \sigma_3 \otimes \sigma_3 \otimes \sigma_3 \otimes \sigma_1 \otimes \mbbm{1}\ ,\\
\Gamma^3 & = \sigma_3 \otimes \sigma_2 \otimes \mbbm{1} \otimes \mbbm{1} \otimes \mbbm{1}\ ,\qquad \!\phantom{-\sigma_2} \Gamma^9 = \sigma_3 \otimes \sigma_3 \otimes \sigma_3 \otimes \sigma_3 \otimes \sigma_1\ ,\\
\Gamma^4 & = -\sigma_3 \otimes \sigma_1 \otimes \mbbm{1} \otimes \mbbm{1} \otimes \mbbm{1}\ ,\qquad \!\phantom{\sigma_2}\Gamma^0 = \sigma_3 \otimes \sigma_3 \otimes \sigma_3 \otimes \sigma_3 \otimes (-i\sigma_2)\ ,\\
\Gamma^5 & = -\sigma_3 \otimes \sigma_3 \otimes \sigma_1 \otimes \mbbm{1} \otimes \mbbm{1}\ ,\qquad \!\!\phantom{\sigma_2}\Gamma_c = \sigma_3 \otimes \sigma_3 \otimes \sigma_3 \otimes \sigma_3 \otimes \sigma_3\ ,\\
\Gamma^6 & = -\sigma_3 \otimes \sigma_3 \otimes \sigma_2 \otimes \mbbm{1} \otimes \mbbm{1}\ ,\qquad  \!\!\phantom{\sigma_2}C_- = e^{3\pi i/4} \sigma_2 \otimes \sigma_1 \otimes \sigma_2 \otimes \sigma_1 \otimes \sigma_2\ .
\end{align*}

\subsection{CPT conjugate vertex operators}
In the calculation of the Yukawa couplings arising from the various
$E$-terms and $J$-terms, the $J$-term couplings involve the
right-moving fermions while the $E$-term couplings involve the
conjugate right-moving fermions. Hence we need vertex operators for
CPT conjugate states. The transformation of the cocycle operators
under CPT are quite intricate and must be handled with care.

Recall that the $H_a$ are antihermitian and $H_5$, $H_6$ are
hermitian. The number operators satisfy:
\begin{equation}
(N_a)^\dag = N_a\ ,\ a \in \four\ , \quad (N_5)^\dag = -N_5\ ,\quad (N_6)^\dag = -N_6 - 2\ .
\end{equation}
Spacetime CPT is implemented as Hermitian conjugation on the vertex
operators. The operator $e^{\lambda\cdot H}$ thus transforms as
\begin{equation}
(e^{\lambda\cdot H})^\dag = (e^{\lambda_a H_a + \lambda_5 H_5 -\lambda_6 H_6})^\dag = e^{-\lambda_a H_a + \lambda_5 H_5 - \lambda_6 H_6} =: e^{\lambda^\star\cdot H}\ ,
\end{equation}
where we have defined $\lambda^\star := (-\lambda_a, \lambda_5;
\lambda_6)$ to be the CPT conjugate weight. The cocycle operator
$c_\lambda$ transforms as
\begin{align}
(c_\lambda)^\dag &= \exp\left[-i\pi\lambda\cdot \cycm\cdot N^\dag\right] = \exp\left[-i\pi(\lambda\cdot \cycm)_b N_b + i\pi(\lambda\cdot \cycm)_5 N_5\right]\ ,\nonumber\\
                 &= \exp\left[i\pi(-\lambda_a \cycm_{ab} N_b - \lambda_5 \cycm_{5a} N_a + \lambda_6 \cycm_{6a} N_a - \lambda_6 \cycm_{65} N_5)\right]\ ,\nonumber\\
                 &= \exp\left[i\pi(\lambda^\star\cdot \cycm)_a N_a + i\pi (\lambda^\star\cdot\cycm)_5 N_5\right]\times\exp\left[-2\pi i(\lambda_5 \cycm_{5a} N_a - \lambda_6 \cycm_{6a} N_a)\right]\ ,\nonumber\\
                 &= c_{\lambda^{\star}}\ e^{-\pi i\left[2(\lambda_5 + \lambda_6)\sum_b N_b\right]} =: c_{\lambda'}\ ,
\end{align}
where we have defined
$\lambda'=(-\lambda_a, -\lambda_5 - 2\lambda_6; \lambda_6) =
\lambda^\star -2(\lambda_5 + \lambda_6)e_5$. Thus, the CPT conjugate
of the operator $V_{\lambda}(k,z)$ in \eqref{finalvop} is given by
\begin{align}
  \wt{V}_\lambda(k,z) & = \omega(\lambda)^\dag\, c(z)\,\mc{B}^\dag(z)\, (c_{\lambda})^\dag\,(e^{\lambda\cdot H(z)})^\dag\, e^{-2ik\cdot X(z)}\nonumber\\
                   & = \omega(\lambda)^\dag\,e^{i\pi \lambda'\cdot\cycm\cdot\lambda^\star}\,c(z)\,\mc{B}^\dag(z)\, e^{\lambda^\star\cdot H(z)} c_{\lambda'}\,e^{-2ik\cdot X(z)}\ .
\end{align} 
$\hfill\square$

\section{\tp{$\n{0,2}$}{N=(0,2)} superspace}\label{02app}
A $1+1$ dimensional theory with $\n{0,2}$ supersymmetry has two
supercharges $(\uQ_+, \w\uQ_+)$ in the left-moving sector. The
corresponding supersymmetry parameters are left-handed spinors
$(\epsilon^+, \w\epsilon{}^+)$.

The Dirac equation in $1+1$ dimensions imposes that left(right)-handed
fermions are right(left)-movers on-shell. A scalar has both left- and
right-moving parts. The left-moving part of the scalar will then have
a superpartner fermion which is left-moving on-shell and hence
right-handed. Thus, a scalar multiplet has a scalar and a right-handed
fermion as its on-shell degrees of freedom. A fermion which is
right-moving on-shell (and hence left-handed) can form a multiplet on
its own under such a supersymmetry. We next describe these multiplets
and their gauged versions in superspace.

$\n{0,2}$ superspace is described by coordinates $(x^{\pm\pm},
\theta^+, \w\theta{}^+)$ where $\theta^+$ and $\w\theta{}^+$ are
left-handed spinors. The corresponding supercovariant derivatives are
denoted by $(\partial_{++}, \uD_{+}, \w{\uD}_+)$. They satisfy the algebra
\begin{equation}
  \uD_+^2 = \w{\uD}{}^2_+ = 0\ ,\quad \{\uD_+\, ,\w{\uD}_+\} = 2i\partial_{++}\ .
\end{equation}
We would like to study constrained superfields of the form
$\w\uD_+(\cdot) = 0$. The natural complex structure of the
supercovariant derivatives then imposes a complex structure on the
space of constrained superfields. There are three kinds of $\n{0,2}$
superfields that will be important for us: Vector, Chiral,
Fermi. Before we study these representations, let us briefly discuss
the representation theory of SO(1,1): the Lorentz group in 1+1
dimensions.

\subsection{Representations of \tp{$\text{SO}(1,1)$}{SO(1,1)}}
The group $\text{SO}(1,1)$ is abelian and has a single boost generator
$T_{01}$ with group element $$g(\lambda) = \exp(\lambda T_{01})\ ,\ \
\lambda \in \mbb{R}\ .$$ All irreducible representations are one
dimensional. The representation on coordinates $x^\mu$, $\mu = 0,1$,
is given by $T_{01} = -x_0 \partial_1 + x_1 \partial_0$. In terms of
lightcone coordinates $x^{\pm\pm} = x^0 \pm x^1$ where
$x^{++}$ is left-moving and $x^{--}$ is right-moving, we have
\begin{equation}
g(\lambda) \cdot x^{\pm\pm} = e^{\pm\lambda} x^{\pm\pm}\ .
\end{equation}
The vector representation mimics the above transformation rule:
$v^{\pm\pm} \to e^{\mp \lambda} v^{\pm\pm}$.

The spinor representations are the basic representations of the double
cover $\text{Spin}(1,1)$ with $T_{01} = \frac{1}{2}\rho_0\rho_1$ where
$\rho^\mu$ are $1+1$ dimensional Dirac matrices. Let $\rho_c = -\rho^0
\rho^1 = \rho_0\rho_1$ and define \emph{left-handed} and \emph{right-handed} spinors
$v^+$ and $v^-$ to satisfy $\rho_c v^\pm = \pm v^\pm$. We then have
$T_{01} = \frac{1}{2}\rho_c$ and $v^\pm$ transform as
\begin{equation}
g(\lambda)\cdot v^\pm = e^{\pm\frac{1}{2}\lambda}\, v^\pm\ .
\end{equation}
(Observe that the product $v^\pm w^\pm$ transforms in the same way as
$x^{\pm\pm}$.) We raise and lower the indices using the totally
antisymmetric $\varepsilon$-symbol with $\varepsilon_{+-} = +1 =
\varepsilon^{+-}$:
\begin{equation}
  v^+ = \varepsilon^{+-} v_- = v_-\ ,\quad v^- = \varepsilon^{-+} v_+ = -v_+\ .
\end{equation}
We thus conclude that an irreducible representation of $\text{SO}(1,1)$ is an object
with some number of $+$ signs $v^{++\cdots +}$ (left-moving) or some
number of $-$ signs $w^{--\cdots -}$ (right-moving).

\textbf{Note:} The Berezin differentials $\ud\theta^+$,
$\ud\w\theta{}^+$ transform as $\ud\theta^+ \to
e^{-\frac{1}{2}\lambda}\ud\theta^+$, $\ud\w\theta{}^+ \to
e^{-\frac{1}{2}\lambda}\ud\w\theta{}^+$. Thus, the most general
superspace action is of the form
\begin{equation}
\int \ud^2x\, \ud\theta^+ \ud\w\theta{}^+ K_{--} + \int \ud^2x \left(\ud\theta^+ \mc{W}_{-} - \text{h.c.}\right) \ ,
\end{equation}
where $K_{--}$ and $\mc{W}_-$ are functions of the various superfields
in the theory with $K_{--}$ unconstrained and $\w\uD_+\mc{W}_- =
0$. Equivalently, one can write
\begin{equation}
\int \ud^2x\, \uD_+ \w\uD_+ K_{--} + \int \ud^2x \left(\uD_+ \mc{W}_{-} - \text{h.c.}\right) \ ,
\end{equation}
upto total $\partial_{++}$ derivative terms.

\subsection{Chiral}
A chiral superfield $\Phi$ is a Lorentz scalar and satisfies $\w\uD_+
\Phi = 0$ and has components
\begin{equation}
\phi := \Phi_\bvert\ ,\quad \sqrt{2}\,\zeta_{\ndt+} :=
(\uD_+\Phi)_\bvert\ .
\end{equation}
The object $\w\uD_+\uD_+\Phi$ contains nothing new:
$(\w\uD_+\uD_+\Phi)_\bvert = 2i\partial_{++} \phi$. Thus, this
multiplet contains a scalar $\phi$ and a right-handed fermion
$\zeta$. The free action is
\begin{align}
\mc S_{\text{chiral}} &=
                        -\frac{i}{2}\int\ud^2x\,\uD_+\w\uD_+\
                        \w\Phi\, \partial_{--} \Phi = \int\ud^2x \left(-\w{\partial^{\mu}\phi}\,\partial_{\mu}\phi - i\w\zeta_{\ndt+} \partial_{--}\zeta_{\ndt+}\right)\ .
\end{align}

\subsection{Fermi}
A Fermi superfield $\Psi_-$ is a left-handed spinor and satisfies
$\w\uD_+ \Psi_- = \sqrt{2}E(\Phi)$ where $E(\Phi)$ is a holomorphic
function of the chiral multiplets $\Phi_i$ in the theory. $\Psi_-$ has
components
\begin{equation}
\psi_- := (\Psi_-)_\bvert\ ,\quad -\sqrt{2}\,G :=
(\uD_+\Psi_-)_\bvert\ ,\quad (\uD_+\w\uD_+\Psi_-)_\bvert = 2 \frac{\partial E}{\partial \phi_i}\, \zeta_{\ndt+,i}\ .
\end{equation}
The two-derivative action for $\Psi_-$ is
\begin{align}
\mc S_{\text{Fermi}} &= \frac{1}{2}\int \ud^2x\, \uD_+ \w\uD_+\ \w\Psi_-\Psi_-\ ,\nonumber\\
                     &= \int \ud^2x \left(-i\w\psi_- \partial_{++} \psi_- + |G|^2 - |E(\phi)|^2 + \w\psi_-\,\frac{\partial E}{\partial \phi_i}\ \zeta_{\ndt i+} + \frac{\partial \w E}{\partial \w\phi{}^i}\,\w\zeta{}^i_{\ndt+}\,\psi_-\right)\ .
\end{align} 
We see that the left-handed fermion $\psi_-$ satisfies
$\partial_{++}\psi_-=0$ for $E = 0$ and hence is right-moving
on-shell.

\subsection{Potential terms}
Let $\Phi_i$ collectively denote all the $(0,2)$ chiral multiplets in
the theory and $\Psi_a$ the $(0,2)$ Fermi multiplets (we suppress the
Lorentz index on $\Psi_a$). We have already seen the $E$-term
previously when we discussed kinetic terms. We can also write a
superpotential, also known as ``$J$-term'' in $(0,2)$ literature:
\begin{align}
  \mc S_{J} &=  -\frac{1}{\sqrt{2}}\int \ud^2x\, \uD_+ \left(J^{a}(\Phi_i) \Psi_a\right)  - \text{h.c.}\ ,\nonumber\\
            &= \int \ud^2x \left(J^a(\phi_i)G_a  + \w{G}{}^a \w{J}_a(\w{\phi}) - \frac{\partial J^a}{\partial \phi_j}\zeta_{j\ndt+}\psi_{a-} - \w\psi{}^a_{-}\frac{\partial \w{J}_a}{\partial \w{\phi}{}^j}\w\zeta{}^j_{+}\right)\ .
\end{align}
Invariance of the above term under $\n{0,2}$ supersymmetry requires
$\w\nabla_+ (\Psi_a J^a) = 0$. This implies
\begin{equation}
0 = E_a J^a =: E\cdot J\ .
\end{equation}
This constraint is necessary for the action to be $\n{0,2}$
supersymmetric. If the action for a theory can be written in $(0,2)$
superspace but the above constraint in violated, then the theory is
only $(0,1)$ supersymmetric.

\subsection{Vector}
Suppose we have some matter fields $\Upsilon$ transforming under a rigid
symmetry $\Upsilon \to e^{iK}\Upsilon$ with $K = K^aT_a$ an hermitian
parameter. We choose hermitian generators $T_a$ with $\tr(T_a T_b) =
\frac{1}{2}\delta_{ab}$ in the fundamental representation. We gauge
this symmetry by introducing gauge-covariant supercovariant
derivatives $\nabla_{+}$, $\w\nabla_{+}$ and $\nabla_{\pm\pm}$
which transform as $\nabla \longrightarrow e^{iK} \nabla e^{-iK}$
under gauge transformations. The superspace constraints are
\begin{equation}
\nabla_{+}^2 = 0\ ,\ \ \w\nabla{}^2_{+} = 0\ ,\ \ \text{and}\ \ \{\nabla_{+}\, ,\w\nabla_{+}\} =
2i\nabla_{++}\ .
\end{equation}
The non-trivial curvatures are given by
\begin{equation}
\mc{F} = [\nabla_{++},\nabla_{--}]\ ,\quad \mc{F}_- = [\w\nabla_{+}, \nabla_{--}]\ ,\quad \w{\mc{F}}_- = [\nabla_{+}, \nabla_{--}]\ .
\end{equation}
The Bianchi identities give
\begin{equation}
\nabla_+\mc{F} = \w\nabla_+\mc{F} = 0\ ,\quad \w\nabla_+\mc{F}_- = 0 = \nabla_+\w{\mc{F}}_-\ .
\end{equation}
The components of the above field strengths are given by
\begin{equation}
\lambda_- := -(\mc{F}_-)_\bvert\ ,\quad v_{01} := \mc{F}_{\,\bvert} \ ,\quad D + iv_{01} := \left(\nabla_+\mc{F}_-\right)_\bvert\ .
\end{equation}
The gauge action is given by
\begin{align}
\mc S_{\text{gauge}} &= \frac{1}{2g^2}\int \ud^2x\, \uD_+\w\uD_+\, \tr\, \w{\mc{F}}_-\mc{F}_-\ ,\nonumber\\
                     &= \frac{1}{g^2}\int \ud^2x\ \tr \left(\frac{1}{2} v_{01}^2 - i\w\lambda_- \uD_{++}\lambda_- + \frac{1}{2} D^2\right)\ .
\end{align}

The chirality constraint for a chiral superfield $\Phi$ in a complex
representation of the gauge group becomes $\w\nabla_+\Phi = 0$ and the
components are defined to be
\begin{equation}
\phi := \Phi_\bvert\ ,\quad \sqrt{2}\,\zeta_{\ndt+} :=
(\nabla_+\Phi)_\bvert\ .
\end{equation}
The minimally coupled action is
\begin{align}
\mc S_{\text{chiral}} &= -\frac{i}{2}\int\ud^2x\,\uD_+\w\uD_+\ \w\Phi\, \nabla_{--} \Phi\ ,\nonumber\\
                     &= \int\ud^2x \left(-\w{\uD^{\mu}\phi}\,\uD_{\mu}\phi - i\w\zeta_{\ndt+} \uD_{--}\zeta_{\ndt+} + i \sqrt{2}\w\phi \lambda_-\zeta_{\ndt+} - i\sqrt{2} \w\zeta_{\ndt+}\w\lambda_-\phi - \w\phi D \phi\right)\ .
\end{align}

Similarly, the constraint for a Fermi superfield $\Psi_{a-}$ in some
representation of the gauge group becomes $\w\nabla_+\Psi_a =
\sqrt{2}E_a(\Phi)$. The minimally coupled action is
\begin{align}
\mc S_{\text{Fermi}} &= \frac{1}{2}\int \ud^2x\, \uD_+ \w\uD_+\, \w\Psi{}^a\Psi_a\ ,\nonumber\\
                     &= \int \ud^2x \left(-i\w\psi{}^a_- \uD_{++} \psi_{a-} + \w{G}{}^aG_a - E_a \w{E}{}^a + \w\psi{}^a_{-}\,\frac{\partial E_a}{\partial \phi_j}\ \zeta_{j\ndt+} + \frac{\partial \w E{}^a}{\partial \w\phi{}^j}\,\w\zeta{}^j_{\ndt+}\,\psi_{a-}\right)\ .
\end{align} 

\subsection{Holomorphic representation}
The constraints $\nabla{}^2_+ = \w\nabla{}^2_+ = 0$ can be solved by
introducing a complex Lie algebra valued superfield $\Omega = \Omega^a
T_a$ called the \emph{prepotential}:
\begin{equation}
\nabla_+ = e^{-i\Omega} \uD_+ e^{i\Omega} := \uD_+ + i\Gamma_+\ ,\quad \w\nabla_+ =
e^{-i\w\Omega}\,\w\uD_+ e^{i\w\Omega} := \w\uD_+ - i\w\Gamma_+\ .
\end{equation}
where we have defined the spinor connections $\Gamma_+$ and
$\w\Gamma_+$. We also define
$\nabla_{\pm\pm} := \uD_{\pm\pm} + i\Gamma_{\pm\pm}$. The gauge
transformation $\nabla \to e^{iK} \nabla e^{-iK}$ can be reproduced by
assigning the following transformation rule for $\Omega$:
\begin{align}
e^{i\Omega} \to e^{i\Omega} e^{-iK}\ ,\quad e^{i\w\Omega} \to e^{i\w\Omega}
e^{-iK}\ .
\end{align}
The above solution has additional gauge invariances:
\begin{align}
&e^{i\Omega} \to e^{i\w\Lambda}e^{i\Omega}\ ,\quad e^{i\w\Omega} \to
  e^{i\Lambda} e^{i\w\Omega}\ .
\end{align}
where $\Lambda$ is a Lie algebra valued chiral superfield
$\w\uD_+\Lambda = 0$. One can use the hermitian $K$ to gauge away the
hermitian part of $\Omega$. Equivalently, we look at the $K$-inert
hermitian object
\begin{equation}
e^{V} := e^{i\Omega} e^{-i\w\Omega}\ ,\quad\text{with}\quad e^{V} \longrightarrow e^{i\w\Lambda}\,e^{V}\,e^{-i\Lambda}\ .
\end{equation}
(In the gauge where $\Omega = -\w\Omega$, we have $V =
2i\Omega$.)

One can go to the \emph{holomorphic representation} via a
(non-unitary) change of basis
$\nabla \to e^{i\w\Omega}\nabla e^{-i\w\Omega}$,
$\Upsilon \to e^{i\w\Omega}\Upsilon$ for a general
matter superfield $\Upsilon$. The spinor derivatives become
$\nabla_+ = e^{-V}\uD_+ e^{V}\ ,\ \w\nabla_+ = \w\uD_+$ which gives
\begin{equation}
  i\Gamma_+ = e^{-V}(\uD_+ e^V)\ ,\quad \w\Gamma_+ = 0\ ,
\end{equation}
thus justifying the name \emph{holomorphic}. In this representation,
the chirality constraint becomes $\w\uD_+\Upsilon = 0$. All the
derivatives are $K$-inert but transform under $\Lambda$ as
$\nabla \to e^{i\Lambda}\nabla e^{-i\Lambda}$ with
$\w\uD_+\Lambda = 0$ and the connections transform as
\begin{equation}
i\delta\Gamma_+ = -i\nabla_+\Lambda\ ,\quad \delta\w\Gamma_+ = 0\ .
\end{equation}
 The components
of $\Gamma_+$ are $\gamma_+ := (\Gamma_+)_\bvert$,
$v_{++} := \frac{1}{2}(\w\nabla_+\Gamma_+)_\bvert$ of which $\gamma_+$
can be set to zero using the gauge transformation above. The same
gauge freedom gives
\begin{equation}
\delta v_{++} = \frac{1}{2}(\w\nabla_+\delta\Gamma_+)_\bvert = -\frac{1}{2}(\{\w\nabla_+,\nabla_+\}\Lambda)_\bvert = -i\nabla_{++}\lambda\ ,
\end{equation}
which is the usual transformation for a non-abelian gauge field
$v_{++}$. The final constraint $\{\w\nabla_+,\nabla_+\} =
2i\nabla_{++}$ in fact gives $2i\Gamma_{++} = \w\nabla_+\Gamma_+$
whose bosonic part is $2v_{++}$. The curvatures are given by
\begin{equation}
\mc{F}_- = [\w\uD_+\, ,\nabla_{--}]\ ,\quad \wt{\mc{F}}_- = [\nabla_+,\nabla_{--}]\ .
\end{equation}
The superspace Lagrangians for the chiral, Fermi and vector multiplets
in the holomorphic representation are $\w\Phi e^{V}\nabla_{--} \Phi$ ,
$\w\Psi e^V \Psi$ and $\mc{F}_-\wt{\mc{F}}_-$ respectively.

\subsection{Duality exchanging \tp{$E \leftrightarrow J$}{E and J}}
Consider a Fermi superfield $\Psi_a$ satisfying $\w\nabla_+ \Psi_a =
\sqrt{2} E_a$. The most general action with $J$-term is
\begin{equation}
\mc{S}[\Psi_a] = -\frac{1}{2}\int \ud^2x\, \uD_+\w\uD_+\, \w\Psi{}^a\Psi_a  -\frac{1}{\sqrt{2}}\int \ud^2x\left\{\uD_+\, \Psi_a J^{a} + \w\uD_+ \w\Psi{}^a \w{J}_a\right\}\ .
\end{equation}
The kinetic term can be reproduced from the following first order
action for $\Psi_a$ by integrating out the unconstrained Grassmann
superfield $\Lambda_a$:
\begin{equation}
\mc{S}[\Lambda_a, \Psi_a] = \frac{1}{2}\int \ud^2x\,  \uD_+\w\uD_+ \left\{\w\Lambda{}^a\Lambda_a - \Psi_a\w\Lambda{}^a - \Lambda_a \w\Psi{}^a\right\} - \frac{1}{\sqrt{2}}\int \ud^2x\left\{\uD_+ \Psi_a J^{a} + \w\uD_+ \w\Psi{}^a \w{J}_a\right\}\ .
\end{equation}
Instead, we could integrate out $\Psi_a$. To do this, we push in
$\w\uD_+$ in the Lagrange multiplier term
$\Psi_a\w\Lambda{}^a$ (and appropriately for its complex conjugate) to
get
\begin{equation}
\mc{S}[\Lambda_a, \Psi_a] = \frac{1}{2}\int \ud^2x\, \uD_+ \left\{-\sqrt{2}E_a\w\Lambda{}^a + \Psi_a \w\nabla_+\w\Lambda{}^a - \sqrt{2}\Psi_a J^{a}\right\} - \text{h.c.}\ .
\end{equation}
Integrating out $\Psi_a$ gives $\w\nabla_+\w\Lambda{}^a = \sqrt{2}
J^a$. Relabelling $\w\Lambda{}^a = \Psi'^a$, we have
$\w\nabla_+\Psi'^a = \sqrt{2} J^a$ and the action
\begin{equation}
\mc{S}[\Psi'_a] = -\frac{1}{2}\int \ud^2x\,\uD_+\w\uD_+\, \w{\Psi}'{}_a\Psi'{}^a  - \frac{1}{\sqrt{2}}\int \ud^2x\left\{ \uD_+(\Psi'{}^a E_a) + \w\uD_+(\w\Psi'{}_a \w{E}{}^a)\right\}\ .
\end{equation}
\textbf{Note:} The new Fermi multiplet $\Psi'{}^a$ transforms in the
conjugate representation of the various symmetry groups in the theory
as compared to $\Psi_a$.

\subsection{Reduction to \tp{$\n{0,1}$}{N=(0,1)} superspace}
We study the reduction to $\n{0,1}$ superspace. The derivatives
$\uD_+, \w\uD_+$ are written as
\begin{equation}
  \uD_+ =: \frac{\uD + i\uQ}{\sqrt{2}}\ ,\quad  \w\uD_+ =: \frac{\uD - i\uQ}{\sqrt{2}}\ ,\quad\Longrightarrow\quad \uD = \frac{\uD_+ + \w\uD_+}{\sqrt{2}}\ ,\quad \uQ = \frac{\uD_+ - \w\uD_+}{\sqrt{2} i}\ ,
\end{equation}
with
\begin{equation}
  \uQ^2 = \uD^2 = i\partial_{++}\ ,\quad \{\uD, \uQ\} = 0\ .
\end{equation}
Here, $\uD$ is the $\n{0,1}$ real supercovariant derivative and $\uQ$
serves as the (non-manifest) generator of the extra supersymmetry.
The $(0,2)$ chiral and Fermi superfields (and their anti-chiral
counterparts) become $(0,1)$ scalar and Fermi superfields with
components
\begin{align}
  &(\Phi_i)_\bvert = \phi_i\ ,\quad (\uD\Phi_i)_\bvert = \zeta_{i+}\ ,\quad (\w\Phi{}^i)_\bvert = \w\phi{}^i\ ,\quad (\uD\w\Phi{}^i)_\bvert = -\w\zeta{}^i_+\ ,\nonumber\\
  &(\Psi_{a})_\bvert = \psi_{a-}\ ,\quad (\uD\Psi_{a})_\bvert = -(G_a - E_a)\ ,\quad (\w\Psi{}^a)_\bvert = \w\psi{}^a_-\ ,\quad (\uD\w\Psi{}^a)_\bvert = -(\w{G}{}^a - \w{E}{}^a)\ .
\end{align}
\emph{In particular, we see that the complex structure on the space of
  fields imposed by the superspace derivatives $\uD_+$ and $\w\uD_+$
  is lost upon reduction to $(0,1)$ superspace.} The above $(0,1)$
superfields transform as follows under the extra supersymmetry:
\begin{align}
  \uQ \Phi_i = -i\uD\Phi_i\ ,\quad \uQ\Psi_{a-} = -i\uD\Psi_{a-} + 2iE_a\ .
\end{align}
The $(0,1)$ actions are then given by
\begin{align}
  \mc{S}_{\text{chiral}} &= -i\int\ud^2x\, \uD\, ((\uD \w\Phi{}^i)\partial_{--}\Phi_i)\ ,\quad  \mc{S}_{\text{Fermi}} = \int\ud^2x\, \uD\, (\w\Psi{}^a\uD\Psi_{a} - \w{E}{}^a\Psi_{a} - \w\Psi{}^aE_a)\ .
\end{align}
The $J$-term becomes the $\n{0,1}$ superpotential
\begin{equation}
  \mc{S}_{J} = \int\ud^2x J^a E_a -\int\ud^2x\,\uD\, (J^a\Psi_a) + \text{h.c.}= -\int\ud^2x\,\uD\, (J^a\Psi_a) + \text{h.c.}\ .
\end{equation}
The first term would have prevented us from writing the $(0,2)$
superpotential in $(0,1)$ superspace but it vanishes due to the
identity $J\cdot E = 0$ that is required for $\n{0,2}$
supersymmetry. The total action for the Fermi superfield is then
\begin{equation}
\int\ud^2x\, \uD\, \left(\w\Psi{}^a\uD\Psi_{a} - (\w{E}{}^a + J^a) \Psi_{a} - \w\Psi{}^a(E_a + \w{J}_a)\right)\ .
\end{equation}
As we can see, the $E$- and $J$- terms are on equal footing in $(0,1)$
superspace and they can be written as a single $(0,1)$ superpotential
term $\uD\left((J^a + \w{E}{}^a)\Psi_a\right)$. The bosonic potential that arises from such a superpotential term is
\begin{equation}\label{01pot}
  \int\ud^2x \sum_a |J^a + \w{E}{}^a|^2\ .
\end{equation}

\subsection{\tp{${(2,2)} \to {(0,2)}$}{(2,2) to (0,2)}}
We shall be schematic here and details can be found in Section 6 of
\cite{Wphases}. A twisted chiral superfield $\Sigma$ satisfies
$\w\nabla_+\Sigma = \nabla_-\Sigma = 0$. The $(0,2)$
decomposition is then a chiral and a Fermi multiplet:
\begin{align}
\Sigma &:= \Sigma_{\,\bvert}\ ,\quad\text{with}\quad \w\nabla_+\Sigma = 0\ ,\nonumber\\
 \wt\Sigma_- &:= \frac{1}{\sqrt{2}}(\w\nabla_-\Sigma)_\bvert\ ,\quad\text{with}\quad \w\nabla_+\wt\Sigma_- = 0\ ,
\end{align}
where $\bvert$ indicates that we have set $\theta^- = \w\theta{}^-
= 0$. All other combinations of the supercovariant derivatives acting
on $\Sigma$ are either zero or derivatives of the above fields. The
(2,2) field strength is a special case of a twisted chiral multiplet:
$2\sqrt{2}\,\Sigma = \{\w\nabla_+, \nabla_-\}$. The complex scalar
$\sigma$ now sits in a separate (0,2) chiral multiplet $\Sigma$ and
$\wt\Sigma_-$ is the familiar (0,2) field strength $\mc{F}_-$ (upto a
factor of $-i/2$).

A (2,2) chiral superfield $\Phi$ satisfies $\w\nabla_+\Phi =
\w\nabla_-\Phi = 0$. The (0,2) decomposition is then
\begin{align}
  \Phi &:= \Phi_{\bvert}\ ,\quad\text{with}\quad \w\nabla_+\Phi = 0\ ,\nonumber\\
  \Phi_- &:= \frac{1}{\sqrt{2}}(\nabla_-\Phi)_\bvert\ ,\quad\text{with}\quad \w\nabla_+\Phi_- = \frac{1}{\sqrt{2}}\{\w\nabla_+,\nabla_-\}\Phi = 2\Sigma\Phi\ ,
\end{align}
where $\Sigma$ is the (0,2) chiral multiplet that contains the complex
scalar $\sigma$. Thus, a (2,2) chiral multiplet $\Phi$ splits into a
(0,2) chiral $\Phi$ and a Fermi $\Phi_-$ which has an $E$-term
$E_{\Phi_-} = \sqrt{2}\Sigma\Phi$.

A (2,2) superpotential ${W}(\Phi_i)$ gives rise to (0,2)
superpotential $\mc{W}(\Phi_i, \Phi_{i-})$ after the $\uD_-$ in the
measure has been pushed into the action:
\begin{equation}
\int \uD_+\uD_- {W}(\Phi_i) = \int \uD_+ \frac{\partial{W}}{\partial\Phi_i} \nabla_-\Phi_i = \sqrt{2} \int \uD_+ \frac{\partial{W}}{\partial\Phi_i} \Phi_{i-}\ ,
\end{equation}
giving a $J$-term $J^i = -2\frac{\partial{W}}{\partial\Phi_i}$. The
constraint $\w\nabla_+ (J^i \Phi_{i-}) = 0$ becomes
\begin{equation}
\frac{\partial{W}}{\partial\Phi_i}\Sigma\Phi_i = 0\ ,
\end{equation}
which is nothing but the condition of gauge invariance of $W(\Phi)$.
$\hfill\square$

\end{document}